\documentclass[twocolumn,prb,notitlepage,floats,superscriptaddress,amsmath,amssymb]{revtex4-2}

\usepackage[version=3]{mhchem} 
\usepackage{bm}
\usepackage[utf8]{inputenc}
\usepackage{textcomp}
\usepackage[T1]{fontenc}
\usepackage{graphicx}
\usepackage{upgreek}
\usepackage{color}
\usepackage{ulem}
\usepackage{hyperref} 
\hypersetup{colorlinks,citecolor=blue, filecolor=blue ,linkcolor=blue , urlcolor=blue}
\usepackage{sidecap}
\usepackage{amssymb}
\usepackage[caption=false]{subfig}
\usepackage{impnattypo}

\def \FUW{Faculty of Physics, University of Warsaw, Pasteura 5, 02-093 Warsaw, Poland}

\begin{document}

\title{The Spectroscopic Atlas of Excitonic Complexes in hBN-Encapsulated S-TMD Monolayers: From Excitons to Oxcitons and Beyond}

\author{Maciej R. Molas}
\email{maciej.molas@fuw.edu.pl}
\affiliation{\FUW}

\begin{abstract}

Hexagonal boron nitride (hBN) encapsulation has transformed the optical spectroscopy of semiconducting transition-metal dichalcogenide monolayers (MLs) by suppressing extrinsic disorder and enabling the resolution of an increasingly complex hierarchy of excitonic states. 
This Review provides a spectroscopy-oriented framework for understanding and identifying excitonic complexes in hBN-encapsulated Mo- and W-based MLs and their alloys. 
The discussion covers bright and dark excitons, Rydberg states, charged excitons, biexcitons, phonon-assisted replicas, and higher-order Fermi-sea many-body states, including hexcitons and oxcitons, with an emphasis on the experimental signatures that distinguish them. 
Particular attention is given to the combined use of photoluminescence and absorption-related spectroscopy, electrostatic gating, polarization analysis, excitation-power and temperature dependences, magnetic fields, and time-resolved measurements to establish microscopic assignments. 
The role of alloying as an additional degree of freedom for engineering the excitonic hierarchy, spin-valley configuration, and magneto-optical response is also highlighted. 
By organizing the rapidly expanding phenomenology around experimentally accessible fingerprints rather than individual material systems, this Review aims to provide a practical and broadly applicable guide for identifying, comparing, and engineering excitonic states in high-quality two-dimensional semiconductors.
\end{abstract}

\maketitle


\section{Introduction} 
\label{introduction}
Atomically thin Mo- and W-based semiconducting transition-metal dichalcogenides (\mbox{S-TMDs}) have become a model platform for exploring light-matter interactions in the two-dimensional (2D) limit~\cite{Mak2016,Koperski2017,Wurstbauer2017,Wang2018}. 
In contrast to their bulk counterparts, monolayer (ML) \mbox{S-TMDs}, including MoS$_2$, MoSe$_2$, MoTe$_2$, WS$_2$, and WSe$_2$, possess a direct band gap at the inequivalent $\mathrm{K}^{+}$ and $\mathrm{K}^{-}$ valleys of the Brillouin zone (BZ)~\cite{Mak2010,Arora2015,Arora_MoSe2,Lezama2015,Molas_Nanoscale}. 
Reduced dielectric screening and quantum confinement strongly enhance Coulomb interactions, giving rise to tightly bound electron-hole (\mbox{$e$-$h$}) pairs that dominate the optical response~\cite{Chernikov2014,Mak2016,Koperski2017,Wang2018}. 
At the same time, strong spin-orbit (SO) coupling, broken inversion symmetry, and valley-contrasting optical selection rules link the spin and valley degrees of freedom to the polarization of light, establishing \mbox{S-TMD} MLs as a particularly rich platform for excitonic, spin, and valley physics~\cite{Xiao2012,Cao2012,Mak2012,Jones2013,Xu2014,Schaibley2016}.

A major experimental advance was the development of van der Waals (vdW) heterostructures in which an \mbox{S-TMD} ML is encapsulated between thin flakes of hexagonal boron nitride (hBN). 
Such encapsulation suppresses extrinsic disorder and environmental inhomogeneity, reduces spatial fluctuations of the exciton transition energy, and produces remarkably narrow excitonic resonances in both photoluminescence (PL) and absorption-related spectroscopy, in some cases approaching the homogeneous linewidth limit~\cite{Cadiz2017,Ajayi2017,Wierzbowski2017,Shree2018,Martin2020,Ryu2024,Meshulam2026}. 
This improvement in spectral quality has changed the character of optical spectroscopy in \mbox{S-TMD} MLs: features that are broadened, merged, or completely hidden in conventional Si/SiO$_2$-supported MLs can become spectrally resolved in hBN-encapsulated structures.
As a result, the central experimental challenge has shifted from observing excitonic resonances to determining their microscopic origin.

This challenge is substantial because the optical response of \mbox{S-TMD} MLs contains a dense hierarchy of excitonic states. 
Beyond the neutral exciton and its non-hydrogenic Rydberg series~\cite{Chernikov2014,Stier2017,Goryca2019,Molas2019Energy}, the spin-valley band structure gives rise to optically bright, spin-forbidden, and momentum-forbidden excitons. 
The relative ordering of the lowest bright and dark states leads to the distinction between \textit{bright} and \textit{darkish} \mbox{S-TMD} MLs~\cite{Molas2017,Robert2017}.
The presence of excess charge carriers further enriches the spectrum through positively and negatively charged excitons, including singlet, triplet, bright, and dark trion configurations~\cite{Courtade2017,Vaclavkova2018,Zinkiewicz2020,Grzeszczyk2021}. 
More complex few-particle states, including neutral and charged biexcitons, emerge under optical excitation and electrostatic doping~\cite{Barbone2018,Li2018,Ye2018}, while phonon-assisted recombination generates characteristic replicas of spin- and momentum-forbidden excitons and trions~\cite{Li2019replica,Li2019momentum,LiuValley,He2020,Liu2020,Li2020phonon,Selig2020Sidebands}. 
At still higher carrier densities, interactions between a photoexcited $e$-$h$ pair and multiple distinguishable Fermi reservoirs lead to higher-order many-body states, including six-particle hexcitons and eight-particle oxcitons~\cite{Tuan2022}, as well as recently identified multivalley excitonic complexes involving an even larger number of occupied electronic reservoirs~\cite{Dijkstra2025}.

\begin{figure*}[t]
		\subfloat{}%
		\centering
		\includegraphics[width=1\linewidth]{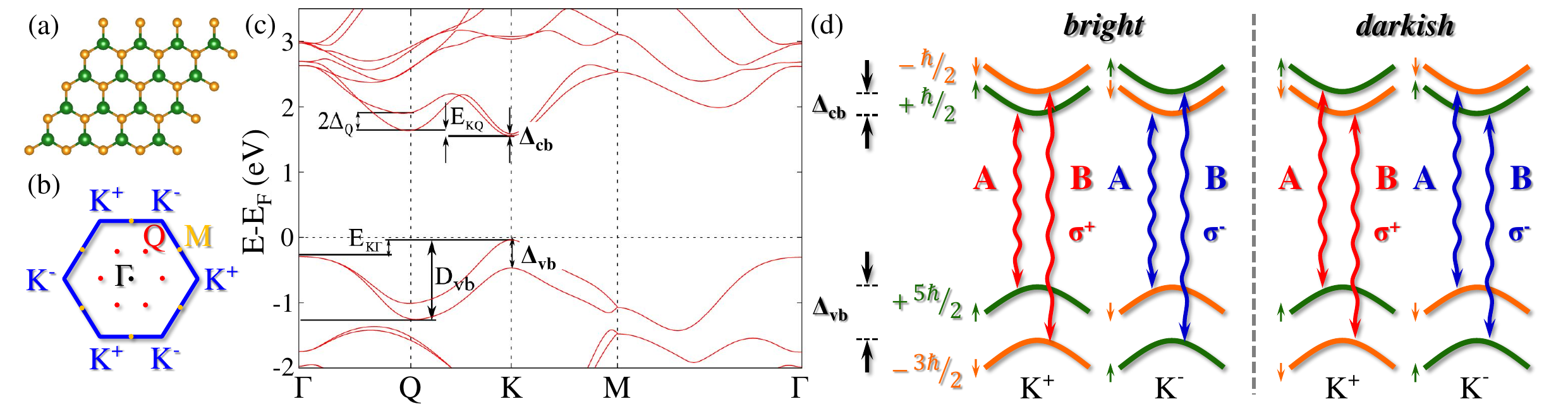}
        \caption{Electronic structure and optical selection rules of Mo- and W-based \mbox{S-TMD} MLs. 
        (a) Top view of the crystal structure of an \mbox{S-TMD} ML, showing the hexagonal arrangement of the transition-metal and chalcogen atoms.
        (b) Hexagonal Brillouin zone showing the inequivalent $\mathrm{K}^{+}$ and $\mathrm{K}^{-}$ valleys, the six $\mathrm{Q}$ valleys located along the $\Gamma$-$\mathrm{K}$ directions, and the high-symmetry $\Gamma$ and $\mathrm{M}$ points.
        (c) Representative electronic band dispersion along the $\Gamma$-$\mathrm{Q}$-$\mathrm{K}$-$\mathrm{M}$-$\Gamma$ direction, including SO coupling. 
        Relevant band-edge energy differences and the CB and VB spin splittings are indicated. 
        Panel (c) is adapted from Ref.~\cite{kormanyos2015} under a Creative Commons Attribution 3.0 (CC BY 3.0) license.
        (d) Schematic band-edge structure at the $\mathrm{K}^{+}$ and $\mathrm{K}^{-}$ valleys for \textit{bright} and \textit{darkish} \mbox{S-TMD} MLs. 
        Green and orange bands denote states with opposite angular-momentum projections, while the green and orange arrows indicate the corresponding spin projections. 
        At the $\mathrm{K}^{+}$ valley, the angular-momentum projections are $+\hbar/2$ and $-\hbar/2$ for the CB states, and $+5\hbar/2$ and $-3\hbar/2$ for the VB states, respectively; the signs are reversed at $\mathrm{K}^{-}$ by time-reversal symmetry.
        Red ($\sigma^{+}$) and blue ($\sigma^{-}$) arrows represent the valley-contrasting A and B optical transitions. 
        The opposite ordering of the spin-split CB states distinguishes the \textit{bright} and \textit{darkish} families.}
\label{fig:electronic_structure}
\end{figure*}

The growing number of closely spaced optical resonances makes their identification increasingly nontrivial. 
Transition energy alone is rarely sufficient for a reliable microscopic assignment. 
Instead, robust identification requires the combined use of several experimental observables, including PL and absorption-related spectroscopy, electrostatic gating, polarization selection rules, excitation-power and temperature dependences, magnetic-field response, diamagnetic shifts, and time-resolved measurements. 
Each of these probes constrains a different property of the excitonic state, such as its charge, spin-valley configuration, oscillator strength, spatial extent, particle number, or recombination dynamics. 
A central aim of this Review is therefore to organize the complex excitonic phenomenology in terms of such experimentally accessible spectroscopic fingerprints.

Alloying introduces an additional degree of freedom for controlling this hierarchy. 
In hBN-encapsulated alloy MLs, including WSSe and MoWSe$_2$, continuous tuning of the band gap, SO splittings, and spin-valley electronic structure has been shown to modify not only the exciton transition energies but also the ordering, fine structure, and magnetic response of excitonic states~\cite{Singh2020,Wu2021,Pucko2022,OlkowskaPucko2026}. 
Alloying can therefore be viewed not simply as a route to band-gap engineering, but as a means of designing the excitonic spectrum itself, including the relative ordering of bright and dark states and their magneto-optical properties.

Several excellent reviews have addressed the optical and excitonic properties of atomically thin \mbox{S-TMDs}, including their spin-valley physics, magneto-optical response, exciton dynamics, and many-body behavior~\cite{Koperski2017,Wang2018,Berkelbach2018,Mueller2018,Schaibley2016,Arora2021,Chen2023Excitonic,Li2025ExcitonDynamics,Roadmap2025}.
The present Review offers a complementary perspective by focusing specifically on high-quality hBN-encapsulated \mbox{S-TMD} MLs and their alloys, with the discussion organized around experimentally observed excitonic states and the spectroscopic criteria used to identify them.
The discussion begins with the electronic structure and neutral A and B excitons, before progressing to Rydberg states, bright and dark excitons, charged complexes, biexcitons, phonon-assisted states, and higher-order Fermi-sea many-body excitations.
Throughout, particular emphasis is placed on the connection between microscopic exciton configurations and experimentally measurable fingerprints, as well as on how electrostatic doping, magnetic fields, temperature, polarization, and alloying can be used to distinguish and control these states.
In this way, the Review aims to provide a practical spectroscopic atlas that bridges few-particle exciton physics and the increasingly complex many-body optical response of modern vdW heterostructures.

\section{Electronic structure and optical selection rules}
\label{sec:electronic_structure}

The characteristic optical properties of \mbox{S-TMD} MLs originate from their crystal symmetry and from the orbital composition of the electronic states forming the band edges. 
A ML with the chemical formula MX$_2$, where M = Mo or W and X = S, Se, or Te, consists of a plane of transition-metal atoms sandwiched 
between two planes of chalcogen atoms, as illustrated in Fig.~\ref{fig:electronic_structure}(a). 
The metal atoms are coordinated by six chalcogen atoms in a trigonal-prismatic geometry, forming the 1H hexagonal structure of 
an individual ML. 
The corresponding crystal belongs to the $D_{3h}$ point group and lacks inversion symmetry, while preserving a horizontal mirror plane with 
respect to the metal plane~\cite{Xiao2012,Cao2012,kormanyos2015}.

The corresponding BZ is hexagonal and contains two sets of inequivalent corners, denoted as $\mathrm{K}^{+}$ and $\mathrm{K}^{-}$, which are related by time-reversal symmetry [Fig.~\ref{fig:electronic_structure}(b)].
In addition to the $\mathrm{K}^{\pm}$ valleys, six symmetry-related $\mathrm{Q}$ valleys are located along the $\Gamma$-$\mathrm{K}$ directions.
These can be divided into two sets of three valleys, denoted as $\mathrm{Q}^{+}$ and $\mathrm{Q}^{-}$, which are related by time-reversal symmetry.
Figure~\ref{fig:electronic_structure}(c) shows a representative electronic band dispersion along the $\Gamma$-$\mathrm{K}$-$\mathrm{M}$-$\Gamma$ direction with SO coupling included, together with the relevant band-edge 
energy differences and spin splittings~\cite{kormanyos2015}. 
The fundamental direct optical transitions occur at the $\mathrm{K}^{\pm}$ valleys, whereas the energetically nearby $\mathrm{Q}$ minima and the VB states around $\Gamma$ become important for momentum-indirect excitonic configurations. 
The discussion of neutral momentum-forbidden excitons below focuses on the $\mathrm{K}^{+}$-$\mathrm{K}^{-}$ intervalley exciton $\mathrm{I}$.
Excitons involving the $\mathrm{Q}$ valleys, such as $\mathrm{K}$-$\mathrm{Q}$ configurations, are considered later in the context of phonon-assisted recombination, where scattering processes can supply the crystal momentum required for optical emission.

At the $\mathrm{K}^{\pm}$ points, the conduction-band (CB) edge is predominantly formed by the transition-metal $d_{z^2}$ orbital, whereas the valence-band (VB) edge is mainly composed of chiral combinations of the $d_{x^2-y^2}$ and $d_{xy}$ orbitals.
Introducing the valley index $\tau=+1$ ($-1$) for $\mathrm{K}^{+}$ ($\mathrm{K}^{-}$), the corresponding orbital parts can be approximately written as $\left|c,\tau\right\rangle \simeq \left|d_{z^2}\right\rangle$, and 
$\left|v,\tau\right\rangle \simeq \frac{1}{\sqrt{2}} \left( \left|d_{x^2-y^2}\right\rangle +i\tau\left|d_{xy}\right\rangle \right)$, respectively~\cite{Xiao2012,Cao2012,kormanyos2015}. 
Thus, the dominant atomic-orbital contribution is characterized by $m_l=0$ for the CB and $m_l=2\tau$ for the VB.

\begin{figure}[t]
    \centering
    \includegraphics[width=1\linewidth]{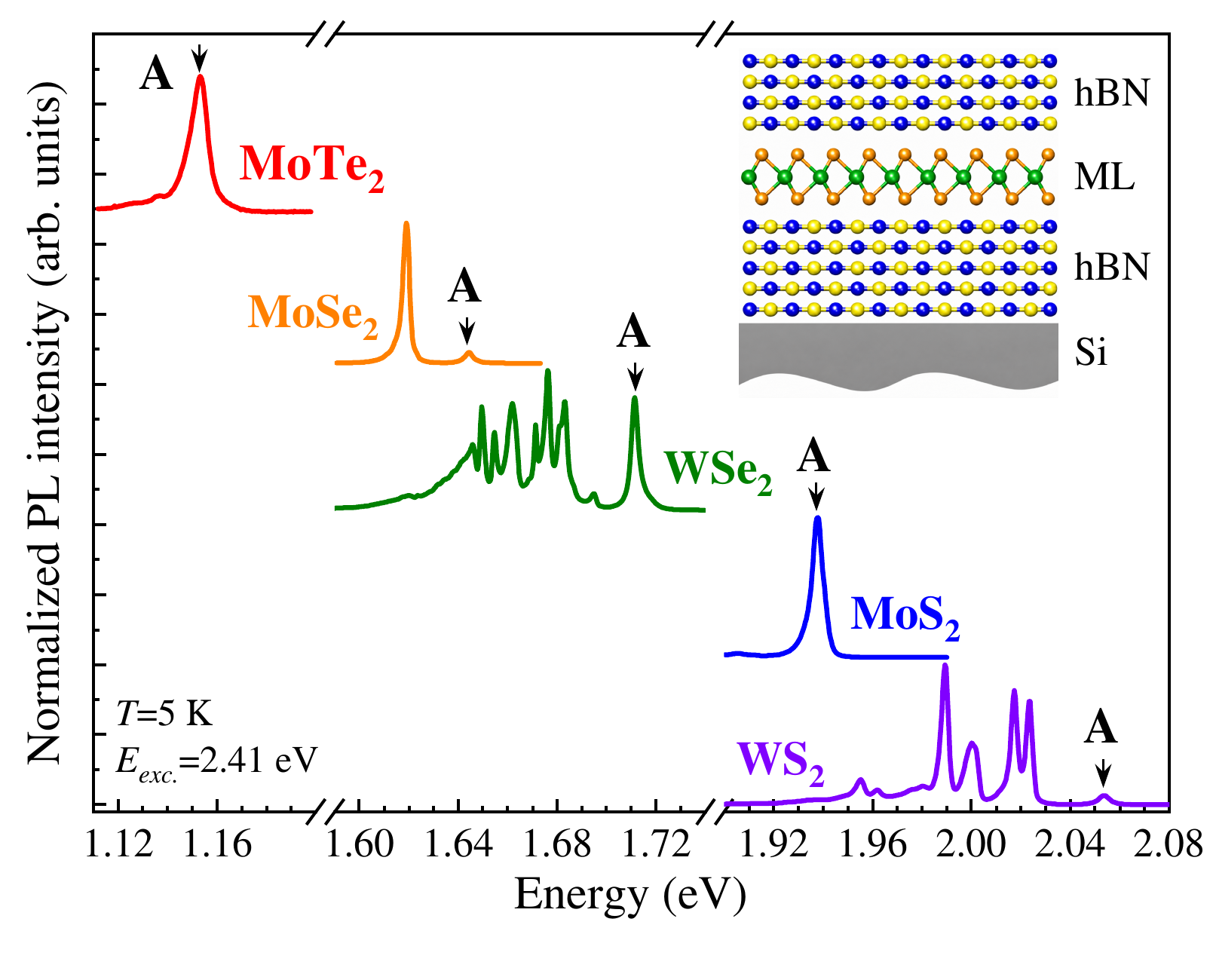}
    \caption{Low-temperature photoluminescence (PL) spectra of hBN-encapsulated MoTe$_2$, MoSe$_2$, WSe$_2$, MoS$_2$, and WS$_2$ MLs measured at $T=5$~K using an excitation energy of $E_{\mathrm{exc}}=2.41$~eV. 
    The spectra are normalized and vertically offset for clarity. 
    Black arrows indicate the emission energy of the neutral A exciton in each material. 
    The inset schematically illustrates the hBN/\mbox{S-TMD} ML/hBN structure on a Si substrate.
    The spectra were measured by the author using samples prepared by Dr Karol Nogajewski and Dr Miroslav Bartoš.
    }
    \label{fig:binary_TMDs}
\end{figure}

The strong SO interaction lifts the spin degeneracy of both bands. 
As shown in Fig.~\ref{fig:electronic_structure}(d), the resulting splittings are denoted as $\Delta_{\mathrm{cb}}$ and $\Delta_{\mathrm{vb}}$ for the CB and VB, respectively. 
The VB splitting is large because of the substantial orbital angular momentum of the $d_{x^2-y^2}\pm i d_{xy}$ states, whereas the CB splitting is considerably smaller because the dominant $d_{z^2}$ orbital has $m_l=0$. 
Time-reversal symmetry reverses both the valley and spin indices, producing the characteristic spin-valley locking of \mbox{S-TMD} MLs 
\cite{Xiao2012,Xu2014,Schaibley2016}.

The spin-split VB gives rise to two prominent series of direct optical transitions, conventionally denoted as A~and B 
[Fig.~\ref{fig:electronic_structure}(d)]. 
The A transition originates from the upper spin-split VB, whereas the B transition involves the lower VB. 
The ordering of the much more closely spaced CB states determines whether the lowest intravalley exciton is optically allowed. 
In \textit{bright} \mbox{S-TMD} MLs, the lowest CB state participates in the dipole-allowed A transition. 
In \textit{darkish} MLs, the ordering of the CB states is reversed, placing a spin-forbidden excitonic state below the optically active A exciton~\cite{Molas2017,Robert2017,Kapuscinski2021}. 
It should be noted that the bright-dark exciton splitting is not determined by $\Delta_{\mathrm{cb}}$ alone, since $e$-$h$ exchange and differences in exciton binding energies also contribute~\cite{Echeverry2016,Kapuscinski2021}.

A complete description of the optical selection rules requires consideration of the total crystal angular momentum of the band-edge Bloch states rather than of the spin projection alone. 
In Fig.~\ref{fig:electronic_structure}(d), the green and orange bands denote states with opposite angular-momentum projections, whereas the arrows indicate their spin projections. 
At the $\mathrm{K}^{+}$ valley, the relevant projections are $+1/2\hbar$ and $-1/2\hbar$ for the two CB states and $+5/2\hbar$ and $-3/2\hbar$ for the two VB states; the signs are reversed at $\mathrm{K}^{-}$ by time-reversal symmetry. 
Because of the threefold rotational symmetry, crystal angular momentum is conserved modulo $3\hbar$.
For circularly polarized light propagating perpendicular to the ML plane, the optical selection rule can be written as $j_z^{c}-j_z^{v}=\pm\hbar \quad (\mathrm{mod}\;3\hbar)$, where the sign determines the photon helicity. 
For the A transition at $\mathrm{K}^{+}$,
\begin{equation}
\frac{1}{2}\hbar-\frac{5}{2}\hbar = -2\hbar \equiv +\hbar \quad (\mathrm{mod}\;3\hbar),
\end{equation}
while for the B transition
\begin{equation}
-\frac{1}{2}\hbar-\left(-\frac{3}{2}\hbar\right) = +\hbar.
\end{equation}
Thus, both A and B transitions couple to $\sigma^{+}$ light at $\mathrm{K}^{+}$ and, by time-reversal symmetry, to $\sigma^{-}$ light at $\mathrm{K}^{-}$~\cite{Xiao2012,Cao2012,Mak2012}.

A frequently used simplified picture considers only the spin projections and treats transitions with $\Delta s_z=0$ as optically allowed. 
This approximation is useful for distinguishing bright and spin-forbidden excitons, but it does not determine the photon helicity or valley selectivity, which originate from the orbital and Bloch-phase contributions to the crystal angular momentum~\cite{Cao2012}. 
The angular-momentum representation in Fig.~\ref{fig:electronic_structure}(d) therefore provides the more complete description, while the spin-only picture serves as a convenient intuitive limit.

The same considerations explain the occurrence of optically dark excitonic states. 
Intravalley $e$-$h$ pairs formed from CB and VB states with incompatible spin projections are spin forbidden and may acquire optical activity through spin mixing, for example in an in-plane magnetic field~\cite{Slobodeniuk2016,Molas2017,Molas2019}.
By contrast, excitons composed of carriers residing in different valleys carry a finite center-of-mass momentum and are therefore momentum forbidden for direct photon emission. 
Both the $\mathrm{K}^{+}$-$\mathrm{K}^{-}$ intervalley exciton $\mathrm{I}$ and momentum-indirect configurations involving the $\mathrm{Q}$ valleys are discussed in the following sections.
Such momentum-forbidden excitons can become optically active through phonon-assisted recombination, which provides the required momentum, or through disorder-induced scattering, which relaxes crystal-momentum conservation~\cite{Li2019momentum,Liu2020,Bao2020}.

\section{Bright-exciton hierarchy}
\subsection{A and B excitons across the S-TMD family}
\label{subsec:AB_excitons}

The neutral A exciton represents the dominant low-energy optical resonance of Mo- and W-based \mbox{S-TMD} MLs and provides a natural starting point for comparing the different members of this material family. 
Figure~\ref{fig:binary_TMDs} compares representative low-temperature PL spectra of hBN-encapsulated MoTe$_2$, MoSe$_2$, WSe$_2$, MoS$_2$, and WS$_2$ MLs. 
For high-quality hBN-encapsulated \mbox{S-TMD} MLs, the low-temperature PL linewidth of the neutral A exciton can be reduced to a few meV or even below 2~meV. 
Reported linewidths are as low as $\sim2$~meV for MoS$_2$~\cite{Cadiz2017,Courtade2018}, below 1.5-1.7~meV for MoSe$_2$~\cite{Ajayi2017,Han2018,Courtade2018}, $\sim3$~meV for MoTe$_2$~\cite{Han2018}, $\sim2$-4~meV for WS$_2$~\cite{Cadiz2017,Zuberek2022}, and $\sim4$~meV for WSe$_2$~\cite{Cadiz2017,Wierzbowski2017,Li2018}. 
More recently, a systematic comparative study of hBN-encapsulated MoS$_2$, MoSe$_2$, WS$_2$, and WSe$_2$ MLs examined the temperature evolution of the exciton linewidth and oscillator strength over a broad temperature range~\cite{Meshulam2026}. 
The A-exciton emission spans nearly 0.9~eV across the family, from the near-infrared spectral range in MoTe$_2$ to 2~eV in WS$_2$, reflecting the pronounced chemical dependence of the quasiparticle band gap and exciton binding energy~\cite{Arora2015,Arora_MoSe2,Molas_Nanoscale,Goryca2019}.
The precise transition energy can nevertheless be shifted by the dielectric environment, strain, carrier density, and local disorder, as demonstrated experimentally in MoSe$_2$, WS$_2$, and WSe$_2$ MLs~\cite{Borghardt2017,Aslan2018,VanTuan2019,Cadiz2017,Wierzbowski2017, Vaclavkova2018, Raja2017}.

Temperature-dependent PL provides an additional methods of distinguishing excitonic states based on their distinct thermal behavior.
In hBN-encapsulated WSe$_2$, for example, the relative spectral weights of excitonic resonances change markedly with temperature, with some features progressively weakening or disappearing upon heating~\cite{Jadczak2021Upconversion}.
More generally, temperature-dependent absorption and PL measurements on hBN-encapsulated \mbox{S-TMD} MLs reveal pronounced differences in the thermal behavior of neutral, charged, bright, and dark excitonic states, reflecting changes in their populations and redistribution among different configurations~\cite{arora2019dark}.
Thus, the temperature-induced disappearance, enhancement, or redistribution of individual resonances provides a useful complementary fingerprint for identifying their microscopic origin.

\begin{table}[t]
\centering
\caption{Representative band-edge and low-temperature excitonic parameters of binary \mbox{S-TMD} MLs.
$E_\mathrm{A}$ and $E_\mathrm{B}$ denote the energies of the neutral A and B excitons, respectively, $\Delta_{\mathrm{cb}}$ is the CB SO splitting, and
$\Delta_{AB}=E_\mathrm{B}-E_\mathrm{A}$ is the A-B exciton energy separation.
The sign convention for $\Delta_{\mathrm{cb}}$ follows Ref.~\cite{Koperski2017}.
Representative values are compiled from Refs.~\cite{kormanyos2015,Koperski2017,Arora2015,Arora_MoSe2,Molas_Nanoscale,Goryca2019,Robert2020}.}
\label{tab:binary_TMDs}

\begin{tabular}{lcccc}
\hline
Material &
$E_\mathrm{A}$ (eV) &
Family &
$\Delta_{\mathrm{cb}}$ (meV) &
$\Delta_{\mathrm{AB}}$ (meV) \\
\hline
MoS$_2$  & $\sim1.94$ & \textit{darkish} & $\sim +3$ & 150 - 160 \\

MoSe$_2$ & $\sim1.64$ & \textit{bright}  & $20$ - $22$ & $\sim200$ \\

MoTe$_2$ & $\sim1.17$ & \textit{bright}  & $32$ - $36$ & $\sim250$ \\

WS$_2$   & $\sim2.05$ & \textit{darkish} & $-29$ - $-32$ & 380 - 400 \\

WSe$_2$  & $\sim1.72$ & \textit{darkish} & $-36$ - $-37$ & 400  -430 \\
\hline
\end{tabular}
\end{table}

Despite these large differences in transition energy, the near-band-edge optical response of all five compounds is characterized by two principal intravalley excitons, conventionally denoted as A and B. 
As discussed in Sec.~\ref{sec:electronic_structure}, they originate predominantly from optical transitions involving the upper and lower SO-split valence bands at the $\mathrm{K}^{\pm}$ valleys. 
The A exciton is readily observed in low-temperature PL, whereas the higher-energy B exciton is generally more clearly resolved in absorption-related techniques, such as reflectance contrast (RC) or transmission spectroscopy, because carriers excited into the B-exciton branch rapidly relax toward lower-energy states.

The A-B separation increases substantially across the S-TMD family, from about 150-160 meV in MoS$_2$ to above 400 meV in WSe$_2$, as summarized in Table~\ref{tab:binary_TMDs}.
The experimentally measured A-B exciton separation should not, however, be identified directly with the single-particle VB splitting, since the finite CB splitting and differences between the binding energies of the A and B excitons also contribute.

The ordering of the spin-split CB states provides a useful starting point for distinguishing between \textit{bright} and \textit{darkish} \mbox{S-TMD} MLs.
MoSe$_2$ and MoTe$_2$ are conventionally classified as \textit{bright} MLs, whereas WS$_2$ and WSe$_2$ belong to the \textit{darkish} family.
MoS$_2$ represents a notable exception to a classification based solely on the single-particle CB splitting.
Although its small positive $\Delta_{\mathrm{cb}}$ historically suggested a bright excitonic ground state, magneto-optical experiments have demonstrated that the spin-forbidden dark exciton lies below the bright A exciton~\cite{Molas2017,Robert2020,Kapuscinski2021}.
This discrepancy arises because \mbox{$e$-$h$} exchange interactions and differences in exciton binding energies also contribute to the excitonic fine structure, with effects comparable to the small CB splitting in MoS$_2$~\cite{Echeverry2016,Robert2020}.
The resulting ordering of the lowest excitonic states strongly affects their thermal population and, consequently, the low-temperature PL response.

\begin{figure}[t]
    \centering
    \includegraphics[width=1\linewidth]{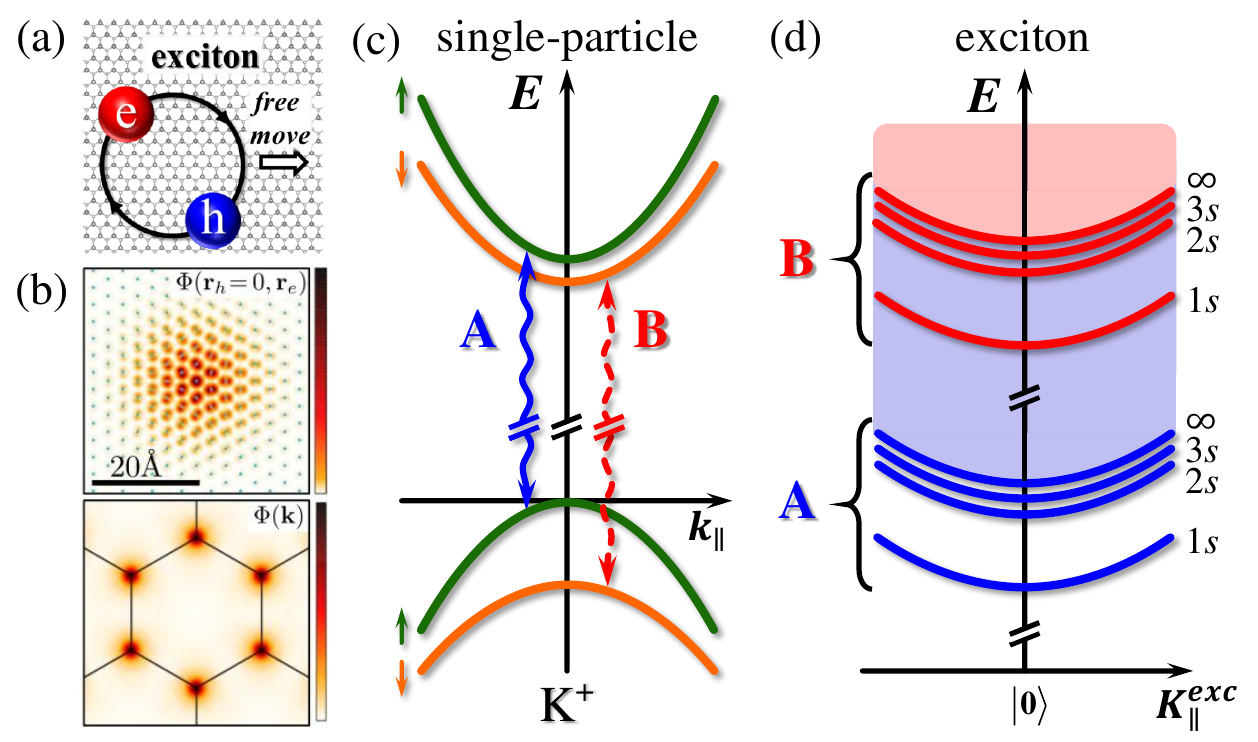}
    \caption{
    Schematic representation of excitons and their Rydberg series in \mbox{S-TMD} MLs. 
    (a) Schematic illustration of a Coulomb-bound $e$-$h$ pair forming an exciton whose center of mass can move freely within the ML plane.
    (b) Calculated A-exciton wave function in ML MoS$_2$: the upper panel shows its real-space distribution with the hole position fixed at the center, 
    whereas the lower panel shows the corresponding distribution in momentum space, with dominant contributions around the $\mathrm{K}^{\pm}$ valleys. 
    Panel (b) is adapted from Ref.~\cite{Druppel2017} under a Creative Commons Attribution 4.0 (CC BY 4.0) license.
    (c) Schematic single-particle band structure at the $\mathrm{K}^{+}$ valley, showing the interband transitions giving rise to the A and B excitons. 
    (d) Corresponding excitonic dispersion, illustrating the A and B Rydberg series composed of the $1s$, $2s$, $3s$, $\ldots$ states converging toward their respective free $e$-$h$ continua ($n\rightarrow\infty$). 
    $K_{\parallel}^{\mathrm{exc}}$ denotes the in-plane center-of-mass wave vector of the exciton.
    }
    \label{fig:Rydberg_scheme}
\end{figure}

The main band-edge and excitonic parameters discussed above are summarized in Table~\ref{tab:binary_TMDs}.

\subsection{Exciton Rydberg series}
\label{subsec:Rydberg}

The strong Coulomb attraction between electrons and holes leads to the formation of tightly bound excitons in \mbox{S-TMD} MLs, as illustrated schematically in 
Fig.~\ref{fig:Rydberg_scheme}(a). 
Their relative motion is confined by the $e$-$h$ interaction, whereas the exciton center of mass can propagate within the ML plane. The microscopic character of such a state is illustrated in Fig.~\ref{fig:Rydberg_scheme}(b) for the A exciton in ML MoS$_2$: the wave function extends over several lattice constants in real space, while the corresponding momentum-space distribution is concentrated around the $\mathrm{K}^{\pm}$ valleys~\cite{Druppel2017,Wang2018}.

\begin{figure*}[t]
    \centering
    \includegraphics[width=0.9\linewidth]{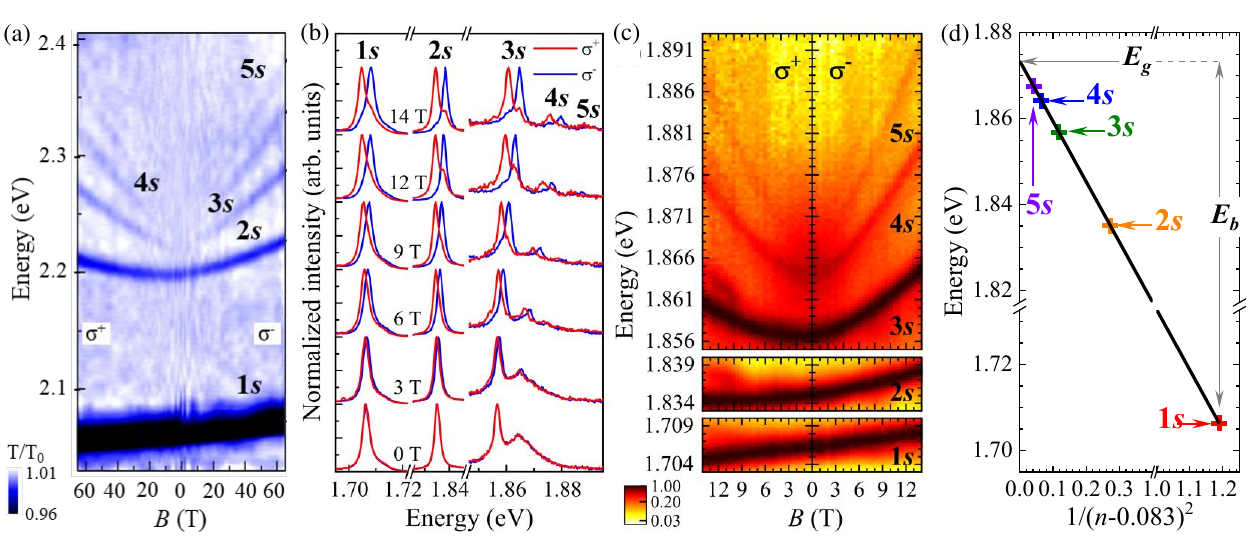}
    \caption{
    Experimental observation and analysis of excitonic Rydberg states in hBN-encapsulated \mbox{S-TMD} MLs. 
    (a) False-color map of the circularly polarized transmission spectra of a WS$_2$ ML measured in magnetic fields up to 65~T, revealing the $1s$, $2s$, 
    $3s$, $4s$, and $5s$ states of the neutral A-exciton Rydberg series. 
    Panel (a) adapted from Ref.~\cite{Goryca2019} under a Creative Commons Attribution 4.0 (CC BY 4.0) license.
    (b) Helicity-resolved ($\sigma^{+}$ and $\sigma^{-}$) PL spectra of an hBN-encapsulated WSe$_2$ ML at selected magnetic fields up to 14~T, showing 
    the field evolution of the $1s$-$5s$ excitonic states. 
    (c) Corresponding false-color representation of the magneto-PL spectra, highlighting the progressively stronger magnetic-field response of the 
    higher-lying Rydberg states. 
    (d) Zero-field energies of the $1s$-$5s$ states plotted as a function of $1/(n-0.083)^2$. The linear dependence extrapolates to the free 
    $e$-$h$ continuum at $E_{\mathrm{g}}$, allowing the A-exciton binding energy $E_{\mathrm{b}}=E_{\mathrm{g}}-E_{1s}$ to be determined. 
    Panels (b)-(d) are adapted from Ref.~\cite{Molas2019Energy}.
    }
    \label{fig:Rydberg_exp}
\end{figure*}

The A and B interband transitions discussed in Sec.~\ref{sec:electronic_structure} are schematically illustrated in Fig.~\ref{fig:Rydberg_scheme}(c).
When the $e$-$h$ interaction is included, each transition gives rise to a series of bound excitonic states, with the $1s$, $2s$, $3s$, $\ldots$ states converging toward their respective free $e$-$h$ continua, as illustrated in Fig.~\ref{fig:Rydberg_scheme}(d).
The $1s$ state is the most strongly bound and most localized in real space, whereas higher-lying states become progressively less bound and more extended~\cite{Chernikov2014,Stier2017,Goryca2019,Molas2019Energy}.
The energy difference between the $1s$ state and the corresponding continuum defines the exciton binding energy, $E_{\mathrm{b}}$.
In hBN-encapsulated \mbox{S-TMD} MLs, $E_{\mathrm{b}}$ typically remains on the order of 0.15-0.3~eV~\cite{Wang2018,Molas2019spec,Goryca2019}.

For an ideal 2D hydrogen atom interacting through a purely Coulombic potential, the bound-state energies follow
\begin{equation}
E_n =
E_{\mathrm{g}}-
\frac{\mathrm{Ry}^{*}}
{\left(n-\frac{1}{2}\right)^2},
\qquad n=1,2,\ldots ,
\label{eq:2D_hydrogen}
\end{equation}
where $E_{\mathrm{g}}$ is the quasiparticle band gap and $\mathrm{Ry}^{*}=\mathrm{Ry}\mu/(\varepsilon^{2}m_{0})$ is the effective Rydberg energy, with $\mathrm{Ry}=13.6$~eV, $\mu$ the reduced $e$-$h$ effective mass, $m_{0}$ the free-electron mass, and $\varepsilon$ the effective dielectric constant of the surrounding medium. 
Experimentally observed excitonic spectra in \mbox{S-TMD} MLs, however, deviate strongly from this ideal 2D hydrogenic sequence because dielectric screening 
in an atomically thin polarizable layer is intrinsically nonlocal~\cite{Chernikov2014,Wang2018}.

This behavior is commonly described using the Rytova-Keldysh interaction~\cite{Rytova1967,Keldysh1979,Cudazzo2011,Berkelbach2013,Slobodeniuk2023}, which in momentum 
space can be written as
\begin{equation}
V(q)= -\frac{e^2}
{2\varepsilon_0 q\left(\kappa+r_0 q\right)},
\label{eq:RK}
\end{equation}
where $\kappa=(\varepsilon_{\mathrm{top}}+\varepsilon_{\mathrm{bottom}})/2$ describes the dielectric environment and $r_0=2\pi\chi_{\mathrm{2D}}$ is the intrinsic screening length determined by the 2D polarizability $\chi_{\mathrm{2D}}$ of the ML. 
The corresponding reduced screening length is $r_0^{*}=r_0/\kappa$. 
At large $e$-$h$ separations, the interaction approaches the Coulomb potential screened predominantly by the surrounding medium, whereas at short 
distances the intrinsic polarizability of the ML becomes increasingly important. 
Consequently, the compact $1s$ exciton and the more extended excited states experience different effective screening, producing the characteristic 
non-hydrogenic spacing of the Rydberg series~\cite{Chernikov2014,Raja2017,Wang2018,Slobodeniuk2023}.

More recently, sum-frequency generation spectroscopy has enabled the simultaneous observation of both $s$- and $p$-type Rydberg excitons in hBN-encapsulated MoS$_2$, MoSe$_2$, WS$_2$, and WSe$_2$ MLs, providing an additional experimental probe of the nonlocal dielectric screening and excitonic level structure~\cite{Takahashi2024}.

While the Rytova-Keldysh interaction generally requires a numerical treatment, a useful analytical approximation can be obtained by replacing it, outside the 
short-range core region, with a modified Kratzer potential~\cite{Molas2019Energy}. 
For the $s$-type excitonic states, this approximation leads to
\begin{equation}
\begin{aligned}
& E_{ns} = E_{\mathrm{g}} - \frac{\mathrm{Ry}^{*}}
{\left(n+\delta\right)^2}, \\
& \delta = g_{\mathrm{K}}\lambda-\frac{1}{2},
\qquad
\lambda^2=\frac{2r_0^{*}}{a_{\mathrm{B}}^{*}}.
\end{aligned}
\label{eq:rydberg_modified}
\end{equation}
where $g_{\mathrm{K}}$ is a dimensionless parameter of the modified-Kratzer approximation and $a_{\mathrm{B}}^{*}=a_{\mathrm{B}}\varepsilon m_{0}/\mu$ is the effective excitonic Bohr radius with $a_{\mathrm{B}}$ the Bohr radius.
The subscript "K" is introduced here to distinguish $g_{\mathrm{K}}$ from the exciton $g$ factor discussed below. 
The effective Rydberg energy $\mathrm{Ry}^{*}$ is governed mainly by the reduced $e$-$h$ mass and the dielectric environment, whereas $\delta$ predominantly 
accounts for the effect of the intrinsic ML polarizability. 
Thus, the modified-Kratzer model provides a convenient analytical approximation to the non-hydrogenic Rydberg sequence generated by the full Rytova-Keldysh interaction.

Figure~\ref{fig:Rydberg_exp} illustrates the experimental observation and analysis of high-lying Rydberg states. 
In hBN-encapsulated WS$_2$, states up to $5s$ can be resolved in high-field transmission spectroscopy [Fig.~\ref{fig:Rydberg_exp}(a)], while Figs.~\ref{fig:Rydberg_exp}(b)-(c) show the evolution of the $1s$-$5s$ A-exciton states in magneto-PL of hBN-encapsulated WSe$_2$~\cite{Goryca2019,Molas2019Energy}. 

The magnetic-field evolution of an excitonic resonance contains both Zeeman and diamagnetic contributions. 
The valley-Zeeman effect produces a splitting between the two circularly polarized components ($E_n^{\sigma^\pm}$),
\begin{equation}
\Delta E_n^{\mathrm{Z}} = E_n^{\sigma^+}-E_n^{\sigma^-} = g_n\mu_{\mathrm{B}}B ,
\label{eq:zeeman_rydberg}
\end{equation}
where $g_n$ is the exciton $g$ factor of the $n$th Rydberg state, $n$ is the principal quantum number, and $\mu_{\mathrm{B}}$ is the Bohr magneton.
In contrast, the mean exciton energy, $\langle E_n(B)\rangle =[{E_n^{\sigma^+}(B)+E_n^{\sigma^-}(B)}]/{2}$, undergoes, in the perturbative regime, a quadratic diamagnetic shift,
\begin{equation}
\langle E_n(B)\rangle-E_n(0) = \sigma_n B^2 = \frac{e^2}{8\mu} \left\langle r^2\right\rangle_n B^2 , 
\label{eq:diamagnetic}
\end{equation}
where $E_n(0)$ is the zero-field energy of the $n$th Rydberg state and $\langle r^2\rangle_n$ is the mean-square in-plane $e$-$h$ separation.
The Zeeman contribution is linear and odd in $B$, whereas the diamagnetic contribution is quadratic and even in $B$. 
Their simultaneous presence is directly visible in the magnetic-field evolution of the Rydberg states in Figs.~\ref{fig:Rydberg_exp}(a)-(c). 
The exciton radius increases rapidly with the principal quantum number, leading to progressively larger diamagnetic coefficients for higher Rydberg states.
This makes magneto-optical spectroscopy particularly powerful for identifying excited excitonic states~\cite{Stier2017,Goryca2019,Molas2019Energy}.

For the WSe$_2$ ML, the zero-field energies exhibit an approximately linear dependence on $1/(n+\delta)^2$ [Fig.~\ref{fig:Rydberg_exp}(d)]. 
Extrapolation to $1/(n+\delta)^2\rightarrow0$ determines the quasiparticle band-gap energy $E_{\mathrm{g}}$, corresponding to the free $e$-$h$ continuum, 
and hence the exciton binding energy $E_{\mathrm{b}}=E_{\mathrm{g}}-E_{1s}$. 

Selected experimental values of the \mbox{A-exciton} binding energy in hBN-encapsulated binary \mbox{S-TMD} MLs obtained at low temperatures are summarized in Table~\ref{tab:binding_energies}.
Only measurements performed below 20~K are included.

\begin{table}[b]
\centering
\caption{
Selected A-exciton binding energies, $E_{\mathrm{b}}$, obtained from low-temperature ($T<20$~K) spectroscopy of hBN-encapsulated binary \mbox{S-TMD} MLs. 
The reported values retain some model dependence through the description of the screened $e$-$h$ interaction and the determination of the free $e$-$h$ continuum.
PLE and SFG denote PL-excitation and sum-frequency generation spectroscopy, respectively.
}
\label{tab:binding_energies}

\begin{tabular}{lcccl}
\hline
Material & Year & $E_{\mathrm{b}}$ (meV) & Spectroscopy & Ref. \\
\hline

MoS$_2$ & 2018 & $\sim222$ & reflectance/PLE & \cite{Robert2018} \\
        & 2019 & $\sim217$ & PL & \cite{Molas2019Energy} \\
        & 2019 & $221\pm3$ & magneto-absorption & \cite{Goryca2019} \\
        & 2020 & $261\pm5$ & photocurrent & \cite{Vaquero2020} \\
        & 2024 & $239.5\pm0.6$ & SFG & \cite{Takahashi2024} \\[1mm]

MoSe$_2$ & 2018 & $\sim186$ & reflectance/upconversion PL & \cite{Han2018} \\
         & 2019 & $\sim216$ & PL & \cite{Molas2019Energy} \\
         & 2019 & $231\pm3$ & magneto-absorption & \cite{Goryca2019} \\
         & 2023 & $\sim214$ & reflectance & \cite{Kipczak2023} \\
         & 2024 & $\sim211$ & reflectance & \cite{Golyshkov2024} \\
         & 2024 & $207\pm3$ & SFG & \cite{Takahashi2024} \\[1mm]

MoTe$_2$ & 2018 & $\sim156$ & reflectance/upconversion PL & \cite{Han2018} \\
         & 2019 & $177\pm3$ & magneto-absorption & \cite{Goryca2019} \\[1mm]

WS$_2$ & 2019 & $\sim174$ & PL & \cite{Molas2019Energy} \\
       & 2019 & $180\pm3$ & magneto-absorption & \cite{Goryca2019} \\
       & 2024 & $187\pm3$ & SFG & \cite{Takahashi2024} \\[1mm]

WSe$_2$ & 2017 & $\sim161$ & magneto-absorption & \cite{Stier2017} \\
        & 2019 & $\sim167$ & magneto-PL & \cite{Molas2019Energy} \\
        & 2024 & $\sim168$ & reflectance & \cite{Golyshkov2024} \\
        & 2024 & $164\pm1$ & SFG & \cite{Takahashi2024} \\
\hline
\end{tabular}
\end{table}

Overall, the reported binding energies show a reasonably consistent material-dependent picture.
The values are particularly stable for WSe$_2$ ($\sim0.16$-$0.17$~eV) and WS$_2$ ($\sim0.17$-$0.19$~eV), whereas a larger spread is found for MoSe$_2$ ($\sim0.19$-$0.23$~eV) and MoS$_2$ ($\sim0.22$-$0.26$~eV).
The available values for MoTe$_2$ lie in the $\sim0.16$-$0.18$~eV range.
No clear systematic evolution with publication year is apparent.
The remaining variations likely reflect differences in sample geometry and dielectric environment, as well as in the experimental and theoretical procedures used to determine the free $e$-$h$ continuum.

Although most experimental studies have focused on the intense $1s$ excitonic resonances, recent advances in optical spectroscopy increasingly enable the investigation and control of higher-lying Rydberg states. 
For example, ultrafast spectroscopy of hBN-encapsulated WSe$_2$ ML has revealed distinct dynamics of the excited $2s$ exciton, including a longer decay time than that of the $1s$ state and the formation of optically induced charged $2s$ excitonic complexes~\cite{Genco2025}.

\section{Optically forbidden excitons}

\begin{figure*}[t]
    \centering
    \includegraphics[width=1\linewidth]{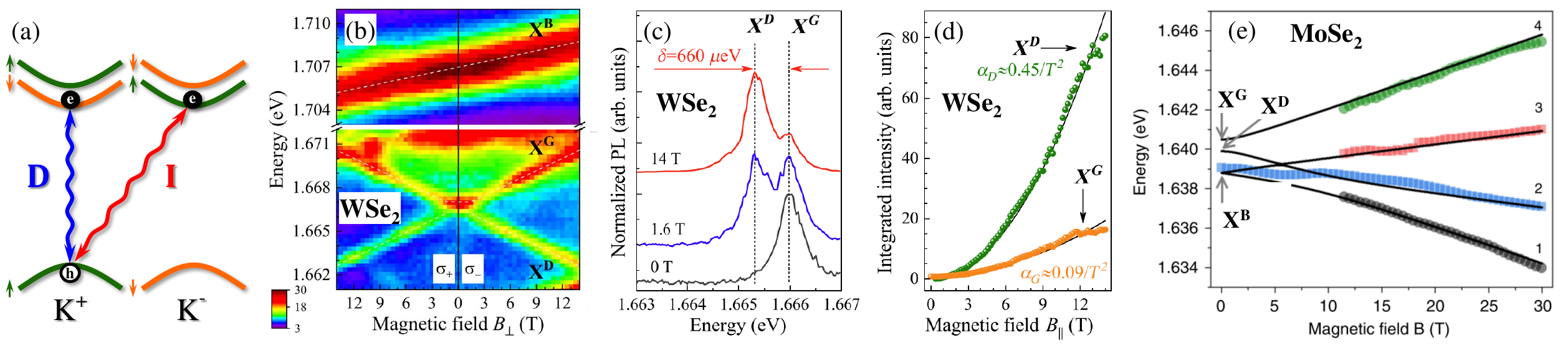}
    \caption{
    Neutral dark excitons and their magneto-optical response in \mbox{S-TMD} MLs.
    (a) Schematic representation of an intravalley spin-forbidden exciton ($\mathrm{D}$) and an intervalley momentum-dark exciton ($\mathrm{I}$), in which the 
    electron and hole occupy the same and opposite $\mathrm{K}$ valleys, respectively.
    (b) False-color magneto-PL map of an hBN-encapsulated WSe$_2$ ML in a perpendicular magnetic field $B_{\perp}$, showing the bright exciton 
    $X^{B}$ and the lower-energy grey and dark excitons, $X^{G}$ and $X^{D}$.
    (c) Selected PL spectra illustrating the magnetic-field evolution of the $X^{G}$-$X^{D}$ doublet and its zero-field splitting $\delta\approx660~\mu$eV.
    (d) Integrated PL intensities of $X^{D}$ and $X^{G}$ as a function of the in-plane magnetic field $B_{\parallel}$, demonstrating the approximately 
    quadratic magnetic brightening of both states.
    Panels (b)-(d) are adapted from Ref.~\cite{Molas2019}. 
   (e) Magnetic-field evolution of the four mixed neutral-exciton states in an hBN-encapsulated MoSe$_2$ ML in a magnetic field tilted by $45^\circ$ 
    with respect to the ML plane. Solid lines represent the calculated field dependence of the exciton energies.
    Panel (e) is adapted from Ref.~\cite{Robert2020} under a Creative Commons Attribution 4.0 (CC BY 4.0) license.
    }
     \label{fig:dark_excitons}
\end{figure*}

The excitonic spectrum of \mbox{S-TMD} MLs extends well beyond the optically bright states discussed above. 
Depending on the spin and valley configurations of the constituent electron and hole, an exciton may become optically inactive because of spin-selection rules or crystal-momentum conservation. 
Such dark states are particularly important in the \textit{darkish} \mbox{S-TMD}s, where they form the lowest-energy excitonic states and strongly influence low-temperature PL, exciton relaxation, and spin-valley dynamics~\cite{Koperski2017,Wang2018,Molas2017,Zhang2017,Robert2020,Zinkiewicz2020,Zinkiewicz2021}.

Figure~\ref{fig:dark_excitons} summarizes the relevant neutral dark-exciton configurations and their characteristic magneto-optical response. 
In Fig.~\ref{fig:dark_excitons}(a), $\mathrm{D}$ denotes an intravalley spin-forbidden configuration, whereas $\mathrm{I}$ denotes an intervalley momentum-dark configuration. 
Within the spin-forbidden exciton manifold, $e$-$h$ exchange gives rise to two states: the grey exciton ($X^{\mathrm{G}}$), which possesses a weak out-of-plane optical dipole, and the strictly dark exciton ($X^{\mathrm{D}}$), whose electric-dipole transition is forbidden at zero magnetic field, as illustrated in Figs.~\ref{fig:dark_excitons}(b)--(d).

\subsection{Spin-forbidden excitons}

A spin-forbidden exciton is formed by an electron and a hole residing in the same valley but associated with bands of opposite spin character. 
Its direct optical recombination is therefore strongly suppressed by the spin selection rule. 
The relative ordering of the bright and spin-forbidden excitons is material dependent and cannot be inferred solely from the single-particle SO splitting of the CB, since differences in exciton binding energies and $e$-$h$ exchange interactions also contribute to the excitonic fine structure~\cite{Echeverry2016,Robert2017,Robert2020}.

It is useful to define the bright-dark splitting as $\Delta_{\mathrm{BD}}=E_{\mathrm{X}^{\mathrm{B}}}-E_{\mathrm{X}^{\mathrm{D}}}$, such that $\Delta_{\mathrm{BD}}>0$ corresponds to a dark excitonic ground state. 
In WSe$_2$, the spin-forbidden exciton is consistently found below the bright exciton, with a typical bright-dark splitting of approximately 41~meV~\cite{Wang2017,Robert2017,Li2018,Ye2018,Barbone2018,Molas2019,Schneider2020,Förste2020,Kapuscinski2021,Jadczak2021Upconversion,Zinkiewicz2022,Li2023Screening,Ren2023,Kipczak2024}.
A dark excitonic ground state is likewise observed in WS$_2$, with a typical bright-dark splitting of approximately 40~meV~\cite{Zinkiewicz2020,Zinkiewicz2021}.
For MoS$_2$, magneto-PL measurements have established a dark excitonic ground state with $\Delta_{\mathrm{BD}}\approx14$~meV~\cite{Robert2020}. 
In contrast, MoSe$_2$ exhibits a bright excitonic ground state, with the spin-forbidden exciton located only about 1.3-1.5~meV above the bright state~\cite{Lu2019,Robert2020}.

The spin-forbidden nature of these excitons strongly suppresses their coupling to light in conventional optical geometries, making their direct experimental identification challenging. 
A particularly powerful approach to access these states is magnetic brightening induced by an in-plane magnetic field. 
The field mixes CB states with opposite spin projections, thereby admixing an optically allowed component into the nominally spin-forbidden exciton and enabling its observation in PL. 
This approach has provided a direct means of establishing the bright-dark exciton ordering in several members of the \mbox{S-TMD} family~\cite{Molas2017,Robert2020}.

\subsection{Dark-grey exciton fine structure}

The spin-forbidden exciton manifold exhibits an additional fine structure arising from the short-range $e$-$h$ exchange interaction. 
This interaction couples the two spin-forbidden excitonic configurations associated with the $\mathrm{K}^{+}$ and $\mathrm{K}^{-}$ valleys, and their symmetric and antisymmetric coherent superpositions form two eigenstates conventionally referred to as the grey exciton $\mathrm{X}^{\mathrm{G}}$ and the strictly dark exciton $\mathrm{X}^{\mathrm{D}}$. 
The $\mathrm{X}^{\mathrm{G}}$ state carries a weak out-of-plane optical dipole and can therefore couple to light polarized perpendicular to the ML plane, whereas $\mathrm{X}^{\mathrm{D}}$ remains dipole forbidden at zero magnetic field~\cite{Wang2017,Robert2017,Molas2019}.

In analogy to the bright-dark splitting $\Delta_{\mathrm{BD}}$, the grey-dark exchange splitting is defined as
$\delta_{\mathrm{GD}}=E_{\mathrm{X}^{\mathrm{G}}}-E_{\mathrm{X}^{\mathrm{D}}}$.
This fine structure is particularly well resolved in high-quality hBN-encapsulated WSe$_2$ MLs, for which reported values of $\delta_{\mathrm{GD}}$ lie in the narrow range of approximately $0.60$--$0.66$~meV [Fig.~\ref{fig:dark_excitons}(c)]~\cite{Robert2017,Molas2019}.
A comparable splitting of approximately $0.53$~meV has been reported for hBN-encapsulated WS$_2$~\cite{Zinkiewicz2020}. 
In MoS$_2$ and MoSe$_2$, by contrast, the $\mathrm{X}^{\mathrm{D}}$ and $\mathrm{X}^{\mathrm{G}}$ components have not been spectrally resolved due to their small energy separation relative to the linewidths.
Consequently, the grey-dark splitting could not be determined directly, although magneto-PL analysis suggests that it is small, on the order of 1~meV or less~\cite{Robert2020}.

Time-resolved spectroscopy provides an additional fingerprint of the bright or dark character of an excitonic state. 
Owing to their strong oscillator strength, bright excitons exhibit very short intrinsic radiative lifetimes, typically on the picosecond scale~\cite{Robert2016}. 
In contrast, the strongly reduced optical coupling of spin-forbidden states leads to substantially longer recombination times.
For hBN-encapsulated WSe$_2$, time-resolved PL revealed a decay time of only $\sim2$~ps for the bright exciton, compared with $\sim110$~ps for the grey exciton at 12~K~\cite{Robert2017}.
The lifetime of the lower-energy, strictly dark state was not determined in these measurements.
The exciton lifetime therefore provides a useful complementary signature of optical darkness, although it can also be affected by localization, non-radiative processes, and population transfer between excitonic states.

The valley composition of the $\mathrm{X}^{\mathrm{D}}$ and $\mathrm{X}^{\mathrm{G}}$ states can be directly modified by an out-of-plane magnetic field. 
The valley-Zeeman term lifts the degeneracy between the underlying $\mathrm{K}^{+}$ and $\mathrm{K}^{-}$ configurations and competes with the exchange interaction that defines the zero-field eigenstates. 
As a result, $\mathrm{X}^{\mathrm{D}}$ and $\mathrm{X}^{\mathrm{G}}$ become progressively mixed and evolve toward valley-polarized states with increasing $B_{\perp}$, as illustrated in Fig.~\ref{fig:dark_excitons}(b)~\cite{Molas2019}.

An in-plane magnetic field provides complementary access to this fine structure by optically activating the spin-forbidden manifold. 
In particular, polarization-resolved measurements reveal distinct linear-polarization selection rules for $\mathrm{X}^{\mathrm{D}}$ and $\mathrm{X}^{\mathrm{G}}$, reflecting their different exchange-defined coherent valley superpositions~\cite{Molas2019}. 
The magnetic-brightening mechanism and the corresponding field dependence of these states are discussed in a later subsection.

\subsection{Momentum-forbidden intervalley excitons}
\label{momentum-forbidden}

Optical inactivity can also originate from crystal-momentum conservation rather than from the spin selection rule. 
The simplest example considered here is the intervalley exciton $\mathrm{I}$, formed by an electron and a hole residing in opposite $\mathrm{K}^{+}$ and $\mathrm{K}^{-}$ valleys, as schematically illustrated in Fig.~\ref{fig:dark_excitons}(a). 
In contrast to the spin-forbidden $\mathrm{X}^{\mathrm{D}}$ and $\mathrm{X}^{\mathrm{G}}$ states discussed above, the $e$-$h$ pair forming $\mathrm{I}$ can be spin allowed. 
Nevertheless, the exciton carries a large finite center-of-mass momentum corresponding to the separation between the two valleys and therefore lies far outside the photon light cone. 
Since the in-plane momentum carried by a photon is negligible on the scale of the BZ, radiative recombination of the $\mathrm{K}^{+}$-$\mathrm{K}^{-}$ intervalley exciton is forbidden by momentum conservation in an ideal translationally invariant crystal~\cite{Selig2016,Selig2020}.

Despite its momentum-forbidden character, a weak emission line associated with the $\mathrm{I}$ exciton has been experimentally observed in ML WSe$_2$~\cite{He2020,Liu2020}. 
Such emission can become optically accessible when translational symmetry is locally broken, for example by disorder or defects, which relax the requirement of crystal-momentum conservation and enable a weak zero-phonon recombination channel. 
A distinct recombination pathway involves the emission of a zone-edge phonon that supplies the momentum required to connect the $\mathrm{K}^{+}$ and $\mathrm{K}^{-}$ valleys. 
This phonon-assisted process gives rise to characteristic replicas of the $\mathrm{I}$ exciton at lower energies~\cite{He2020,Liu2020}.

Recent experiments have extended the spectroscopic access to momentum-forbidden intervalley excitons beyond low-temperature PL. 
Time-resolved momentum microscopy directly resolved a long-lived $\mathrm{K}^{-}$-$\mathrm{K}^{+}$ exciton in ML WS$_2$, while room-temperature dual-action spectroscopy revealed a spectroscopic signature of the $\mathrm{K}^{+}$-$\mathrm{K}^{-}$ intervalley exciton in hBN-encapsulated WSe$_2$~\cite{Zhu2025,Wragg2025}.

The coexistence of weak zero-phonon emission and phonon-assisted replicas provides characteristic optical signatures of the momentum-forbidden intervalley exciton. 
In particular, the energies, polarization properties, and magnetic-field response of the phonon replicas provide information on both the underlying excitonic state and the phonons mediating its recombination. 
These exciton-phonon processes are discussed in detail in a subsequent section.

\subsection{Magnetic brightening and optical fingerprints}

Magnetic fields provide powerful methods of accessing optically forbidden excitons and probing their spin-valley composition. 
An in-plane magnetic field $B_{\parallel}$ couples to the carrier spin through the Zeeman interaction and mixes the SO-split CB states with opposite spin projections. 
As a result, an optically allowed component is admixed into the nominally spin-forbidden exciton, enabling its radiative recombination~\cite{Slobodeniuk2016,Molas2017,Zhang2017}. 
In the perturbative regime, the mixing amplitude is linear in $B_{\parallel}$, and the corresponding oscillator strength therefore increases approximately as $B_{\parallel}^{2}$~\cite{Slobodeniuk2016,Molas2017,Zhang2017}. 
Under fixed experimental conditions, the integrated PL intensity can consequently be described phenomenologically as
\begin{equation}
I_j(B_{\parallel})
=
I_j(0)+\alpha_j B_{\parallel}^{2},
\label{eq:dark_brightening}
\end{equation}
where $j=\mathrm{D},\mathrm{G}$, $I_j(0)$ denotes the integrated intensity of the corresponding excitonic emission at zero in-plane magnetic field, and
$\alpha_j$ is the magnetic-brightening coefficient.
The latter depends on the population of the corresponding excitonic state, the strength of magnetic-field-induced mixing between bright
and dark states, and the radiative and nonradiative recombination processes~\cite{Kapuscinski2021}.
Consequently, $\alpha_j$ is not an intrinsic material constant and, in WSe$_2$, exhibits a strong dependence on temperature~\cite{Kipczak2024}
and electrostatic doping~\cite{Krasucki2026Doping}.
For an ideal strictly dark neutral state, $I_{\mathrm{D}}(0)$ vanishes, whereas a finite zero-field contribution may occur experimentally due to residual symmetry breaking or weak state mixing; for the grey exciton, a finite $I_{\mathrm{G}}(0)$ is also allowed because of its weak out-of-plane optical dipole. 
The approximately quadratic field dependence observed for both $\mathrm{X}^{\mathrm{D}}$ and $\mathrm{X}^{\mathrm{G}}$ in WSe$_2$ is shown in Fig.~\ref{fig:dark_excitons}(d)~\cite{Molas2019}.

A perpendicular magnetic field $B_{\perp}$ probes the valley degree of freedom through the valley-Zeeman effect. 
For the dark-grey doublet, the resulting valley-Zeeman term competes with the short-range $e$-$h$ exchange interaction and mixes the zero-field $\mathrm{X}^{\mathrm{D}}$ and $\mathrm{X}^{\mathrm{G}}$ eigenstates, as discussed above and illustrated in Fig.~\ref{fig:dark_excitons}(b). 
Thus, the two field orientations provide complementary information: $B_{\parallel}$ activates the spin-forbidden states optically, whereas $B_{\perp}$ modifies their valley composition.

\begin{figure}[t]
    \centering
    \includegraphics[width=1\linewidth]{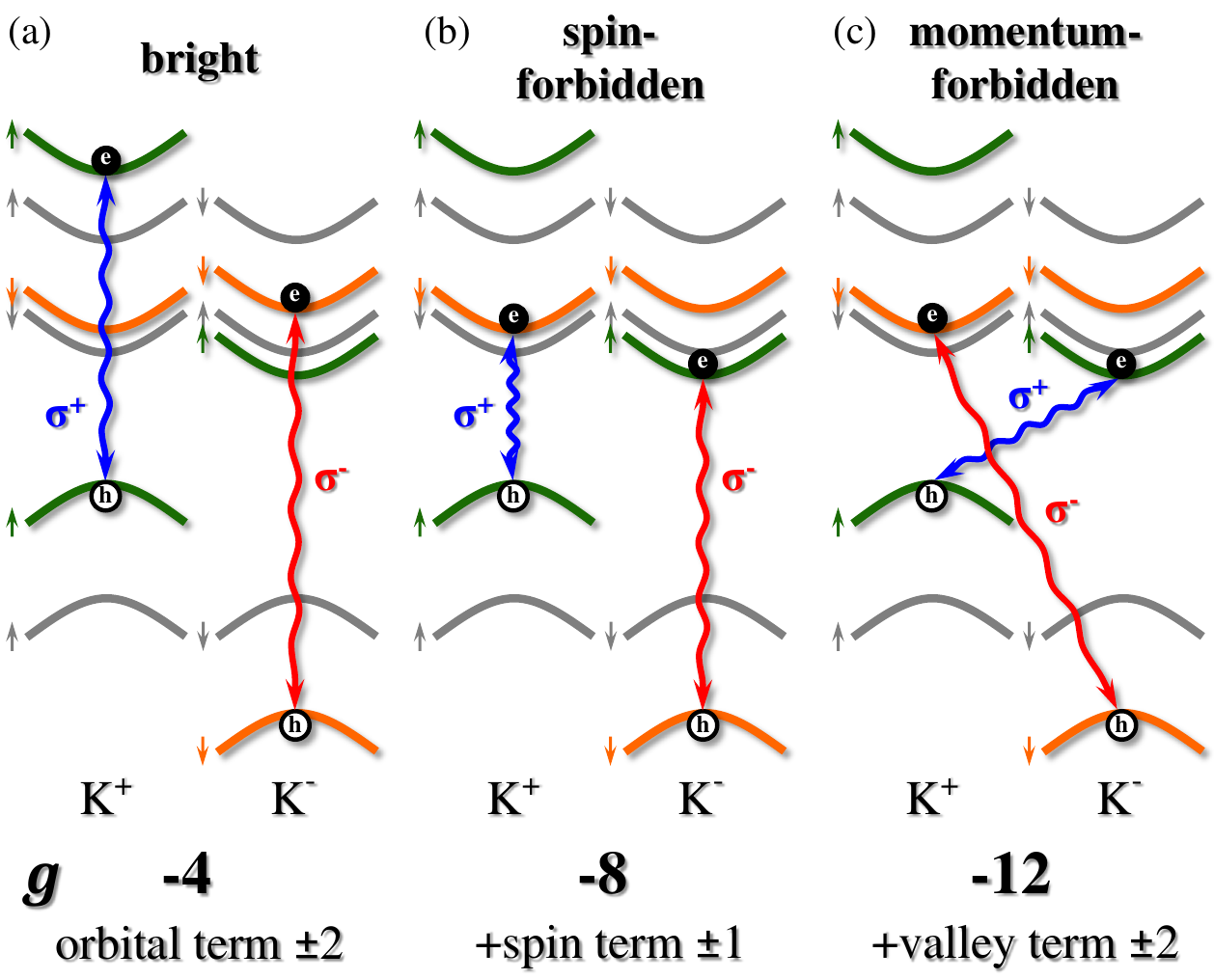}
    \caption{Schematic illustration of the magnetic-moment contributions to the exciton $g$ factor for (a) bright intravalley, (b) spin-forbidden intravalley, and (c) momentum-forbidden intervalley excitons in \mbox{S-TMD} MsL. 
    In the simplified single-particle picture, the bright exciton exhibits a characteristic $g$ factor close to $-4$, mainly determined by the atomic-orbital contribution. 
    For the spin-forbidden exciton, an additional spin contribution increases the magnitude to approximately $g=-8$, whereas for the momentum-forbidden $\mathrm{K}^{+}$-$\mathrm{K}^{-}$ exciton, the additional valley magnetic moment leads to values around $g=-12$. 
    The indicated values represent characteristic trends rather than universal constants, as the precise exciton $g$ factor depends on the material and the specific band states involved.}
     \label{fig:zeeman}
\end{figure}

Both effects can be exploited simultaneously in a tilted magnetic field. 
The in-plane component brightens the spin-forbidden states, while the out-of-plane component lifts the valley degeneracy and produces their Zeeman splitting. 
Figure~\ref{fig:dark_excitons}(e) illustrates this approach for an hBN-encapsulated MoSe$_2$ ML. 
Four excitonic branches become resolved at high magnetic fields and can be quantitatively described by a model incorporating the bright-dark splitting $\Delta_{\mathrm{BD}}$, the grey-dark exchange splitting $\delta_{\mathrm{GD}}$, and the interaction with the external magnetic field. 
Since the excitonic states become mixed at finite field, the labels $\mathrm{X}^{\mathrm{B}}$, $\mathrm{X}^{\mathrm{G}}$, and $\mathrm{X}^{\mathrm{D}}$ strictly refer to their zero-field limits~\cite{Robert2020}.

\begin{figure*}[t]
    \centering
    \includegraphics[width=1\linewidth]{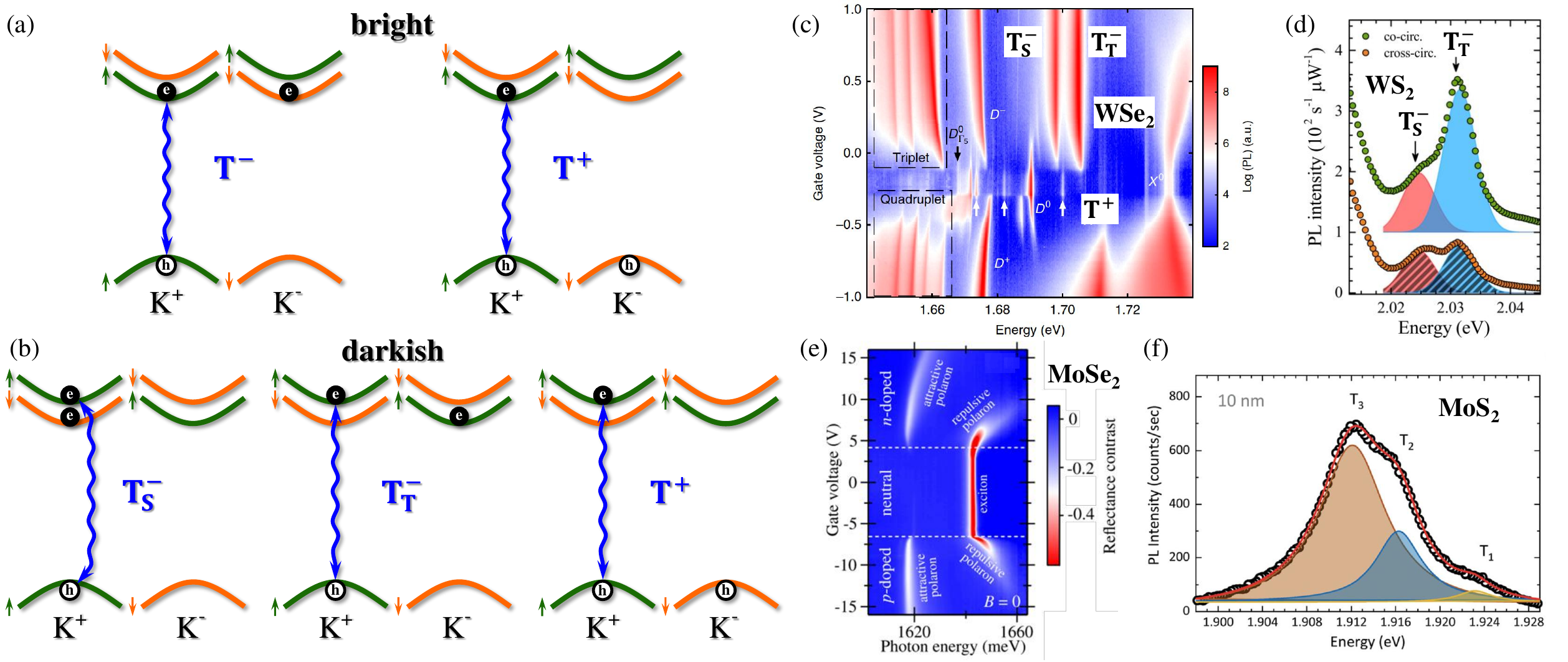}
    \caption{Charged excitons and their spectroscopic manifestations in hBN-encapsulated \mbox{S-TMD} MLs.
    (a) Schematic spin-valley configurations in a \textit{darkish} \mbox{S-TMD} ML, showing the intravalley singlet $\mathrm{T}_{\mathrm{S}}^{-}$ and intervalley triplet $\mathrm{T}_{\mathrm{T}}^{-}$ negative trions, together with the positive trion $\mathrm{T}^{+}$.
    (b) Corresponding configurations of the negative and positive trions, $\mathrm{T}^{-}$ and $\mathrm{T}^{+}$, in a \textit{bright} \mbox{S-TMD} ML.
    (c) Gate-dependent PL of an hBN-encapsulated WSe$_2$ ML, illustrating the evolution of the charged-exciton spectrum with carrier density, including the
    bright singlet and triplet negative trions, $\mathrm{T}_{\mathrm{S}}^{-}$ and $\mathrm{T}_{\mathrm{T}}^{-}$, the positive trion $\mathrm{T}^{+}$, and lower-energy dark excitonic complexes.
    Panel (c) adapted from Ref.~\cite{He2020} under a Creative Commons Attribution 4.0 (CC BY 4.0) license.
    (d) Polarization-resolved PL of an hBN-encapsulated WS$_2$ ML, resolving the singlet and triplet negative trions, $\mathrm{T}_{\mathrm{S}}^{-}$ and $\mathrm{T}_{\mathrm{T}}^{-}$.
    Panel (d) adapted from Ref.~\cite{Vaclavkova2018} with permission from IOP Publishing Ltd.
    (e) Gate-dependent reflectance contrast of an hBN-encapsulated MoSe$_2$ ML, showing the evolution of the excitonic response upon electron and hole doping
    and the attractive- and repulsive-polaron branches.
    Panel (e) adapted from Ref.~\cite{Smolenski2019} under a Creative Commons Attribution 4.0 (CC BY 4.0) license.
    (f) Low-temperature PL spectrum of an hBN-encapsulated MoS$_2$ ML, revealing three components of the negative-trion manifold, $\mathrm{T}_{1}$, $\mathrm{T}_{2}$, and $\mathrm{T}_{3}$. 
    Panel (f) adapted from Ref.~\cite{Grzeszczyk2021} published by the Royal Society of Chemistry under a Creative Commons Attribution-NonCommercial 3.0 Unported (CC BY-NC 3.0) license.
}
\label{fig:charged_excitons}
\end{figure*}

\subsection{Exciton $g$ factors as spectroscopic fingerprints}

The exciton $g$ factor provides a particularly useful spectroscopic fingerprint of the spin and valley configuration of an excitonic state. 
Within a simple single-particle picture, the magnetic-field-induced shifts of the band-edge states can be rationalized in terms of spin, atomic-orbital, and valley
contributions~\cite{Koperski2019}.
This phenomenological decomposition provides an intuitive explanation for the characteristic progression of exciton $g$ factors from approximately $-4$ for bright intravalley excitons, through values around $-8$ for spin-forbidden intravalley excitons, to values
of order $-12$ for momentum-forbidden intervalley excitons, as schematically illustrated in Fig.~\ref{fig:zeeman}.

For the bright intravalley exciton, the electron and hole occupy the same valley and have the same spin character. The spin contributions to the transition energy therefore cancel. 
In the simplest orbital picture, the CB edge is predominantly derived from the $d_{z^2}$ orbital and carries essentially no atomic orbital angular momentum, whereas the VB edge is formed predominantly by the $d_{x^2-y^2}\pm i d_{xy}$ orbitals with opposite orbital angular momenta in the
$\mathrm{K}^{+}$ and $\mathrm{K}^{-}$ valleys. 
The resulting difference between the two valley-selective optical transitions gives the characteristic $g\approx-4$ of the bright exciton~\cite{Koperski2019}.

For a spin-forbidden intravalley exciton, the electron and hole have opposite spin character. 
Consequently, their spin Zeeman contributions no longer cancel. 
Taking the free-electron-like spin contribution as approximately $\pm\mu_{\mathrm{B}}B$ for an individual band state, this produces an additional contribution of approximately $-4$ to the excitonic $g$ factor, leading to the simple estimate $g\approx-8$. 
Experimentally, dark-exciton $g$ factors of approximately $-9$ to $-10$ have been reported for WSe$_2$~\cite{Robert2017,Li2019replica,Förste2020}, while values of $-8.6$ and $-6.5$ have been obtained for MoSe$_2$ and MoS$_2$, respectively~\cite{Robert2020}.

A still larger magnetic response is expected for the spin-allowed  momentum-forbidden $\mathrm{K}^{+}$-$\mathrm{K}^{-}$ intervalley exciton.
Because the electron and hole occupy opposite valleys, the valley-dependent magnetic moments that cancel for an intravalley bright exciton instead
contribute to the Zeeman splitting. 
Within the simplified phenomenological picture, this additional valley contribution increases the magnitude of the
$g$ factor from approximately $-4$ to values of order $-12$~\cite{Koperski2019}. 
Consistent with this picture, values around $g=-12.5$ to $-13$ have been reported for the momentum-forbidden $\mathrm{I}$ exciton in ML WSe$_2$~\cite{He2020,Liu2020}.

The separation into independent spin, atomic-orbital, and valley terms should, however, be regarded as an intuitive description rather than a complete microscopic decomposition.
In a full band-structure treatment, the orbital magnetic moment of a Bloch state contains contributions arising from coupling to other bands and is intrinsically linked to its valley character.
Thus, the orbital and valley magnetic responses are not strictly independent quantities.
First-principles calculations evaluate the magnetic moments via the summation-over-bands method directly from the electronic wave functions and include coupling between the relevant bands, providing quantitative exciton $g$ factors without relying on the simple additive picture~\cite{Wozniak2020}.
Such calculations reproduce the characteristic hierarchy of magnetic responses.
For WSe$_2$, the calculated $g$ factors are $g^{\mathrm{calc}}=-4.0$, $|g^{\mathrm{calc}}|=10.1$, and $|g^{\mathrm{calc}}|=13.6$ for bright, spin-forbidden, and momentum-forbidden intervalley excitons, respectively~\cite{Förste2020}.
The corresponding values for WS$_2$ are $g^{\mathrm{calc}}=-3.56$, $|g^{\mathrm{calc}}|=8.73$, and $|g^{\mathrm{calc}}|=12.20$~\cite{Zinkiewicz2021}.
In this approach, the exciton $g$ factor is commonly approximated by the difference between the $g$ factors of the electron and hole Bloch states involved in the corresponding optical transition.
More recently, a generalized approach to excitonic $g$ factors has been developed that explicitly incorporates many-body interactions and the excitonic wave function obtained from Bethe-Salpeter equation calculations~\cite{PauloPRB2026}.
This treatment was shown to account for the nonmonotonic behavior of the $g$ factors of the X$^\mathrm{B}$ and X$^\mathrm{D}$ excitons in ML WSe$_2$.
Although the approximate sequence $g\sim-4$, $-8$, and $-12$ is not universal, the systematic increase in $|g|$ from bright to spin-forbidden and momentum-forbidden excitons provides a useful spectroscopic fingerprint of their distinct spin-valley configurations, particularly when several resonances occur within a narrow spectral range.

\section{Charged excitons: from trions to exciton polarons}
\label{sec:charged_excitons}

The presence of resident charge carriers introduces another level of complexity into the excitonic spectrum of \mbox{S-TMD} MLs. 
In the dilute-carrier limit, charged excitonic resonances are conveniently described within the few-particle trion picture, in which a neutral exciton binds an additional electron or hole~\cite{Mak2013,Ross2013,Courtade2017}. 
The resulting negatively and positively charged complexes are conventionally denoted as $\mathrm{T}^{-}$ and $\mathrm{T}^{+}$, respectively. 
Because the additional carrier can occupy different spin and valley states, several trion configurations may coexist, producing a characteristic fine structure governed by the underlying spin-valley electronic structure~\cite{Courtade2017,Vaclavkova2018,Grzeszczyk2021}. 
Figure~\ref{fig:charged_excitons} summarizes representative charged-exciton configurations in panels (a,b) and their experimental manifestations in panels (c)-(f). 
This few-particle description is most appropriate at low carrier densities; with increasing doping, the optical response gradually evolves toward the more general exciton-polaron regime discussed below.

Charged excitonic resonances are not restricted to the $1s$ exciton manifold but also emerge for excited Rydberg states. 
Excited $2s$ trions have been identified in ML WS$_2$ and WSe$_2$~\cite{Arora2019Trion,Wagner2020}, while optically induced charged $2s$ excitonic complexes have recently been observed in hBN-encapsulated WSe$_2$~\cite{Genco2025}. 
More generally, gate-dependent spectroscopy of hBN-encapsulated MoSe$_2$ and WSe$_2$ has revealed exciton-polaron Rydberg series associated with excited excitonic states, demonstrating that many-body interactions with the Fermi sea extend beyond the $1s$ exciton to higher Rydberg states~\cite{Liu2021Polaron}.

\subsection{Charged excitons in darkish MLs}

In \textit{darkish} \mbox{S-TMD} MLs, the charged-exciton spectrum directly reflects the spin-valley structure of the conduction band and is therefore substantially richer than a simple pair of positive- and negative-trion resonances.
In the few-particle picture, a negative trion $\mathrm{T}^{-}$ consists of two electrons and one hole, whereas a positive trion $\mathrm{T}^{+}$ contains one electron and two holes.
For the negative trion, two optically bright configurations are particularly important: the intravalley singlet $\mathrm{T}_{\mathrm{S}}^{-}$ and the intervalley triplet $\mathrm{T}_{\mathrm{T}}^{-}$, schematically illustrated in Fig.~\ref{fig:charged_excitons}(a)~\cite{Courtade2017,Vaclavkova2018}.

Electrostatic gating provides a direct means of controlling the charge state and resolving this fine structure.
A representative example is hBN-encapsulated WSe$_2$ [Fig.~\ref{fig:charged_excitons}(c)], where the neutral exciton dominates close to charge neutrality, while hole doping gives rise to positively charged excitonic states and electron doping reveals the singlet and triplet negative trions~\cite{He2020}.
The same spectrum also contains lower-energy charged dark states, demonstrating that the distinction between optically bright and spin-forbidden excitons extends naturally to charged excitonic complexes.

The singlet-triplet fine structure is particularly clearly resolved in hBN-encapsulated WS$_2$ [Fig.~\ref{fig:charged_excitons}(d)].
Polarization-resolved PL distinguishes the two negative-trion configurations: the intervalley triplet preserves the excitation helicity more efficiently than the intravalley singlet, providing an additional spectroscopic fingerprint of their distinct valley character~\cite{Vaclavkova2018}.

\begin{figure*}[t]
    \centering
    \includegraphics[width=1\linewidth]{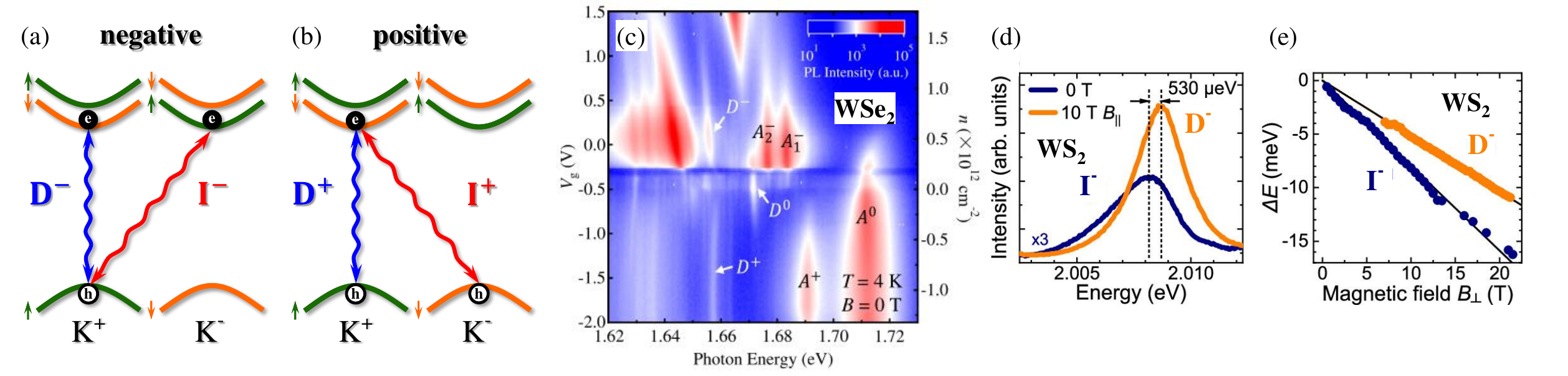}
    \caption{Dark charged excitons and their spin-valley fingerprints in hBN-encapsulated S-TMD MLs.
    (a,b) Schematic spin-valley configurations and possible recombination pathways of negatively and positively charged dark excitons, including intravalley spin-forbidden ${\mathrm D}^{\mp}$ and intervalley momentum-forbidden ${\mathrm I}^{\mp}$ trions.
    (c) Gate-dependent PL of an hBN-encapsulated WSe$_2$ ML, showing the evolution from the neutral dark exciton $\mathrm{D}^{0}$ to negatively and positively charged dark trions, $\mathrm{D}^{-}$ and $\mathrm{D}^{+}$, upon electron and hole doping, respectively.
    Panel (c) is reproduced from Ref.~\cite{Liu2019} with permission from the American Physical Society, \href{https://doi.org/10.1103/PhysRevLett.123.027401} {https://doi.org/10.1103/PhysRevLett.123.027401}. Copyright 2019 American Physical Society.
    (d) Low-temperature PL spectra of an hBN-encapsulated WS$_2$ ML measured at $B_{\parallel}=0$ and 10~T, resolving the intervalley momentum-forbidden ${\mathrm I}^{-}$ and magnetically brightened intravalley spin-forbidden ${\mathrm D}^{-}$ trions, separated by approximately $530~\mu$eV.
    (e) Zeeman splitting of ${\mathrm D}^{-}$ and ${\mathrm I}^{-}$ as a function of the out-of-plane magnetic-field
    component, yielding $g=-8.9$ and $g=-13.7$, respectively. 
    Panels (d) and (e) are adapted from Ref.~\cite{Zinkiewicz2021} under a Creative Commons Attribution 4.0 (CC BY 4.0) license.
}
\label{fig:dark_trions}
\end{figure*}

\subsection{Charged excitons in bright MLs and the trion-polaron crossover}

In \textit{bright} \mbox{S-TMD} MLs, the lowest-energy intravalley exciton is optically allowed, and at low doping levels the corresponding charged excitonic states can be conveniently described as positive and negative trions, as schematically illustrated in Fig.~\ref{fig:charged_excitons}(b).
A representative example is hBN-encapsulated MoSe$_2$, where electrostatic doping leads to a gradual evolution from the neutral-exciton resonance to lower-energy charged-exciton features.

An exciton polaron is a many-body quasiparticle formed when an exciton interacts with a Fermi sea of charge carriers, becoming dressed by particle-hole excitations~\cite{Efimkin2017,Sidler2017}.
The distinction between a trion and an exciton polaron is controlled primarily by the ratio of the Fermi energy to the trion binding energy rather than by a universal carrier-density threshold.
In the dilute limit, $E_{\mathrm{F}}\ll E_{\mathrm{tr}}$, the few-particle trion and many-body Fermi-polaron approaches converge and provide essentially equivalent descriptions of the optical response~\cite{Glazov2020}.
When $E_{\mathrm{F}}$ becomes comparable to $E_{\mathrm{tr}}$, however, Pauli blocking and many-body correlations between the exciton and the Fermi sea become increasingly important, making the polaron description more appropriate~\cite{Efimkin2017,Chang2018}.

Within this many-body picture, the optical spectrum contains a lower-energy attractive-polaron branch and a higher-energy repulsive-polaron branch~\cite{Efimkin2017,Sidler2017}.
This evolution has been observed experimentally in doped \mbox{S-TMD} MLs using several complementary optical techniques~\cite{Sidler2017,Smolenski2019,Goldstein2020,Wagner2020,Liu2021Polaron}.
In particular, Fig.~\ref{fig:charged_excitons}(e) shows the gate-dependent reflectance contrast of hBN-encapsulated MoSe$_2$, where increasing electron or hole doping produces a lower-energy attractive-polaron branch and a higher-energy repulsive-polaron branch, accompanied by a redistribution of oscillator strength~\cite{Smolenski2019}.

The conventional trion and exciton-polaron pictures should therefore not be regarded as mutually exclusive descriptions.
Rather, they represent complementary limits of the same exciton-carrier interaction problem: the trion language remains particularly intuitive in the dilute-carrier regime, whereas the Fermi-polaron description becomes essential once many-body correlations with a developed Fermi sea dominate the optical response~\cite{Glazov2020,Chen2023Excitonic}.

\subsection{MoS$_2$: an exceptional charged-exciton fine structure}

MoS$_2$ represents a particularly interesting case among the \mbox{S-TMD} MLs.
Within the single-particle band picture, MoS$_2$ would be classified as a \textit{bright} ML because the ordering of the spin-split CB states favors the spin-allowed transition.
However, the CB SO splitting is exceptionally small, with $\Delta_{\mathrm{cb}}\sim+3$~meV (Table~\ref{tab:binary_TMDs})
\cite{kormanyos2015}.
Consequently, excitonic contributions that are relatively weak in other \mbox{S-TMDs}, in particular, differences in the binding energies of bright and dark excitons and the short-range $e$-$h$ exchange interaction, can compete with the single-particle SO splitting and reverse the bright-dark ordering.
Magneto-PL experiments have indeed demonstrated that the spin-forbidden exciton in hBN-encapsulated MoS$_2$ lies approximately 14~meV below the bright exciton, placing MoS$_2$ in the \textit{darkish} family at the excitonic level~\cite{Robert2020}.

This unusual band and excitonic structure also leads to an exceptionally rich spectrum of negatively charged excitons. 
High-resolution low-temperature spectroscopy of hBN-encapsulated MoS$_2$ revealed three negative-trion components, $\mathrm{T}_{1}$, $\mathrm{T}_{2}$, and $\mathrm{T}_{3}$, as shown in Fig.~\ref{fig:charged_excitons}(f)~\cite{Grzeszczyk2021}.
Within a few-particle spin-valley picture, the $\mathrm{T}_{1}$ and $\mathrm{T}_{2}$ resonances were assigned to intravalley and intervalley
spin-singlet negative trions, respectively, whereas the lowest-energy $\mathrm{T}_{3}$ feature, observed under photodoping conditions, was
tentatively attributed to an intervalley spin-triplet negative trion~\cite{Grzeszczyk2021}.

A more microscopic interpretation was subsequently provided by gate-dependent high-field magneto-PL combined with \textit{ab initio} many-body calculations. 
The three resonances were shown to have mixed intra- and intervalley character arising from three-particle correlations and $e$-$h$ exchange interactions, rather than representing pure spin-valley configurations~\cite{Klein2022}.
The small CB SO splitting in MoS$_2$ brings several charged-exciton configurations close in energy, thereby enhancing their hybridization. 
At higher electron densities, these resonances gradually evolve toward many-body polaronic states~\cite{Klein2021}.

\begin{figure*}[t]
    \centering
    \includegraphics[width=1\linewidth]{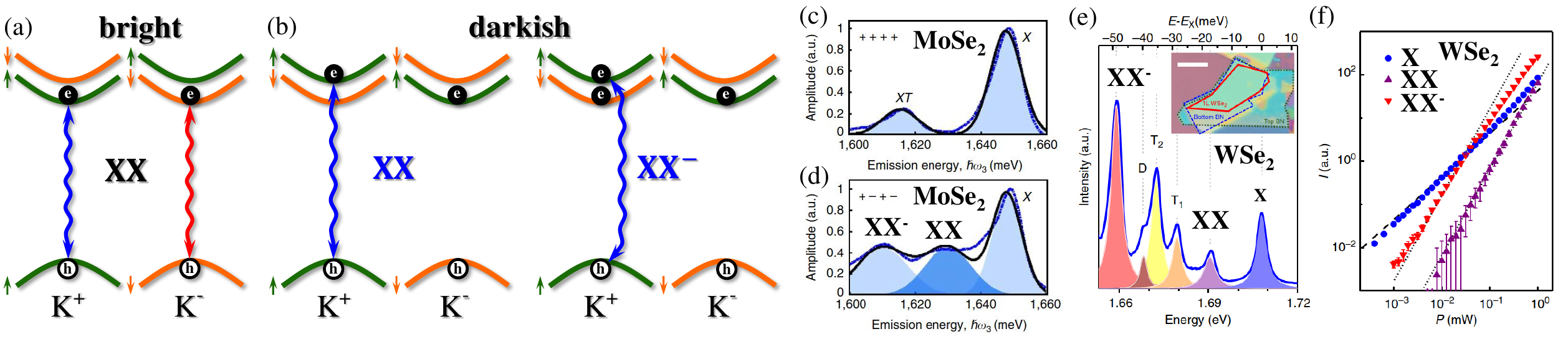}
    \caption{Neutral and charged biexcitons in S-TMD MLs.
    (a) Schematic spin-valley configuration of a neutral biexciton $\mathrm{XX}$ in a \textit{bright} ML.
    (b) Representative configurations of the neutral biexciton $\mathrm{XX}$ and negatively charged biexciton $\mathrm{XX}^{-}$
    in a \textit{darkish} ML.
    (c,d) Polarization-resolved 2D coherent spectroscopy of ML MoSe$_2$: emission-energy slices measured under
    (c) co-circular and (d) cross-circular excitation.
    The cross-circular configuration reveals the neutral biexciton $\mathrm{XX}$ and the charged biexciton $\mathrm{XX}^{-}$. 
    Panels (c) and (d) are adapted from Ref.~\cite{Hao2017}  under a Creative Commons Attribution 4.0 (CC BY 4.0) license.
    (e) Low-temperature PL spectrum of an hBN-encapsulated WSe$_2$ ML, showing the neutral exciton $\mathrm{X}$, trion resonances, the dark
    exciton, and the neutral and negatively charged biexcitons, $\mathrm{XX}$ and $\mathrm{XX}^{-}$.
    (f) Integrated PL intensities of $\mathrm{X}$, $\mathrm{XX}$, and $\mathrm{XX}^{-}$ as a function of excitation power, demonstrating the
    superlinear power dependence of the two biexcitonic resonances. 
    Panels (e) and (f) are adapted from Ref.~\cite{Li2018} under a Creative Commons Attribution 4.0 (CC BY 4.0) license.
    }
    \label{fig:biexcitons}
\end{figure*}

\subsection{Dark trions}

The distinction between bright and optically forbidden neutral excitons extends naturally to charged excitonic complexes.
In \textit{darkish} \mbox{S-TMD} MLs, an additional carrier can bind to a dark exciton, giving rise to negatively or positively charged dark trions.
The corresponding intravalley spin-forbidden and intervalley momentum-forbidden recombination pathways for negatively and positively charged
dark excitons are schematically illustrated in Figs.~\ref{fig:dark_trions}(a) and \ref{fig:dark_trions}(b), respectively.
Dark trions have been identified in several hBN-encapsulated S-TMD MLs, with their energetic ordering relative to bright trions depending strongly on the material~\cite{arora2019dark}.

Electrostatic gating provides a direct method of controlling their charge state, as illustrated for hBN-encapsulated WSe$_2$ in
Fig.~\ref{fig:dark_trions}(c).
Starting from the charge-neutral dark exciton, electron and hole doping give rise to negatively and positively charged dark-trion resonances, respectively~\cite{Liu2019,Li2019,He2020,Zinkiewicz2022}.
Their characteristic large Zeeman splittings, with $g$ factors close to $-9$, provide an important fingerprint of their spin-forbidden character
\cite{Liu2019,Li2019}.
The charged dark excitons additionally exhibit binding energies on the order of 15~meV and substantially longer lifetimes than their bright counterparts~\cite{Liu2019,Li2019}.

Spin-forbidden negative dark trions in hBN-encapsulated WS$_2$ ML were identified through their magnetic brightening, large $g$ factor, and long radiative lifetime~\cite{Zinkiewicz2020}.
The designation "\textit{dark trion}", however, does not uniquely specify the optical recombination pathway.
A more complete picture is provided by hBN-encapsulated WS$_2$ ML, where the same negative-trion carrier configuration can recombine through either an intravalley spin-forbidden channel, denoted ${\mathrm D}^{-}$, or an intervalley momentum-forbidden channel, ${\mathrm I}^{-}$~\cite{Zinkiewicz2021}.
As shown in Fig.~\ref{fig:dark_trions}(d), the ${\mathrm I}^{-}$ emission dominates at zero magnetic field, whereas
the ${\mathrm D}^{-}$ state becomes optically active upon application of an in-plane magnetic field.
The two resonances are separated by only about $530~\mu$eV~\cite{Zinkiewicz2021}.
The finite zero-field emission of the momentum-forbidden ${\mathrm I}^{-}$ state was suggested to arise from
momentum-relaxing processes, most likely involving defect-assisted recombination~\cite{Zinkiewicz2021}.

Beyond conventional recombination pathways, Coulomb interactions can activate additional radiative channels of dark trions.
Intervalley electron-electron scattering was theoretically shown to mix bright and dark trion configurations in W-based \mbox{S-TMD} MLs, giving rise to optically active semi-dark trions~\cite{Danovich2017}.
Experimental evidence for such a state was subsequently reported in hBN-encapsulated WS$_2$, where an additional emission line, denoted T$'$, was attributed to the recombination of a semi-dark negative trion~\cite{Zinkiewicz2021}.
More recently, a related Coulomb-assisted brightening mechanism was proposed for hBN-encapsulated WSe$_2$, involving second-order radiative recombination of negatively charged dark trions accompanied by an intervalley transition of the remaining electron~\cite{Jindal2025Dark}.
A hole-hole-assisted emission channel of positively charged dark trions was also identified, producing a near-infrared emission line approximately 478~meV below the conventional dark-trion resonance~\cite{Jindal2025Dark}.

The microscopic nature of the different dark-trion recombination channels can be established from their Zeeman response.
In an out-of-plane magnetic-field, the intravalley spin-forbidden ${\mathrm D}^{-}$ and intervalley momentum-forbidden ${\mathrm I}^{-}$ transitions exhibit markedly different $g$ factors, $g=-8.9$ and $g=-13.7$, respectively [Fig.~\ref{fig:dark_trions}(e)]~\cite{Zinkiewicz2021}.
These values closely follow the characteristic hierarchy previously discussed for neutral excitons, with $|g|$ values of approximately 8-9 for spin-forbidden intravalley states and 13-14 for momentum-forbidden intervalley states~\cite{Zinkiewicz2021}.
Interestingly, the semi-dark trion emission T$'$ in WS$_2$ exhibits a much smaller $g$ factor of $-3.3$, comparable to those of bright excitonic complexes, reflecting its distinct recombination mechanism~\cite{Zinkiewicz2021}.
Similarly, the Coulomb-assisted emission of positively charged dark trions in WSe$_2$ exhibits a positive $g$ factor of about $+4$, opposite in sign to that of the conventional bright positive trion.
This unusual magnetic response supports a recombination mechanism involving the $\Gamma$ valley of the valence band~\cite{Jindal2025Dark}.

\section{Biexcitons and charged biexcitons}
\label{sec:biexcitons}

The strong Coulomb interaction in S-TMD MLs enables the formation of few-particle complexes beyond neutral and charged excitons.
A neutral biexciton, denoted as $\mathrm{XX}$, is a four-particle complex consisting of two bound $e$-$h$ pairs, whereas a negatively charged biexciton $\mathrm{XX}^{-}$ is a five-particle state containing three electrons and two holes and can equivalently be viewed as a bound exciton-trion complex.
Representative configurations are schematically illustrated in Figs.~\ref{fig:biexcitons}(a) and (b) for bright and darkish MLs,
respectively.
In darkish W-based MLs, the low-energy dark-exciton states further enrich the biexciton manifold and enable complexes involving both bright and dark excitonic constituents~\cite{Barbone2018,Li2018,Ye2018,Steinhoff2018,Zinkiewicz2021,Zhang2026Biexciton}.

\begin{figure*}[t]
    \centering
    \includegraphics[width=1\linewidth]{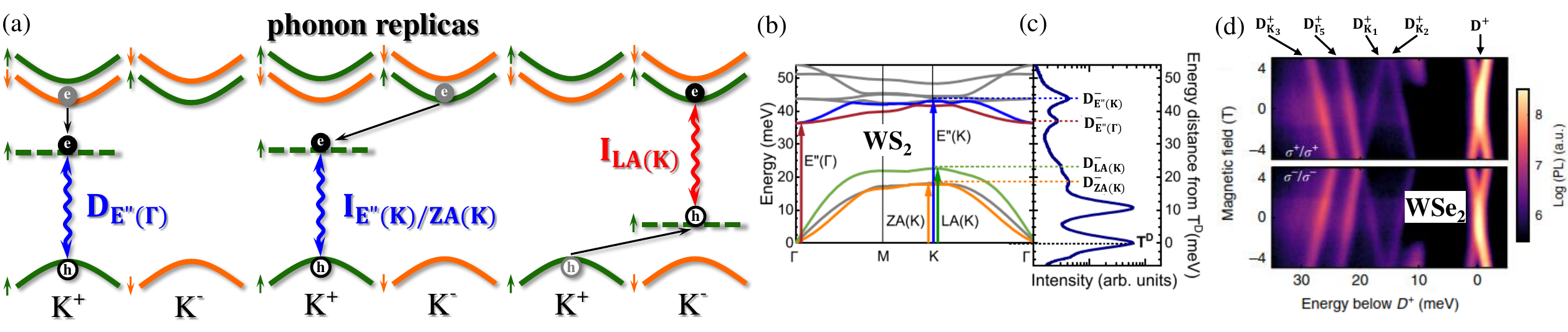}
    \caption{Phonon-assisted recombination and phonon replicas of dark excitonic complexes in S-TMD MLs.
    (a) Schematic phonon-assisted recombination pathways of neutral dark excitons. 
    The zone-center $E''(\Gamma)$ phonon activates an intravalley spin-forbidden transition, whereas $K$-point phonons provide the momentum
    required for intervalley recombination; representative channels involving $E''(K)$, ZA$(K)$, and LA$(K)$ phonons are shown.
    (b) Calculated phonon dispersion of ML WS$_2$, with the modes relevant to dark-trion phonon replicas highlighted.
    (c) Low-temperature PL spectrum of an hBN-encapsulated WS$_2$ ML showing the negative dark trion and its $E''(\Gamma)$, $E''(K)$, LA$(K)$, and
    ZA$(K)$ phonon replicas.
    Panels (b) and (c) are adapted from Ref.~\cite{Zinkiewicz2021} under a Creative Commons Attribution 4.0 (CC BY 4.0) license.
    (d) Circular-polarization-resolved magneto-PL of an hBN-encapsulated WSe$_2$ ML showing the positively charged dark trion $\mathrm{D}^{+}$
    and its $K_3$, $\Gamma_5$, $K_1$, and $K_2$ phonon replicas.
    Panel (d) Adapted from Ref.~\cite{He2020} under a Creative Commons Attribution 4.0 (CC BY 4.0) license.
    }
    \label{fig:phonon_replicas}
\end{figure*}

A particularly direct identification of neutral and charged biexcitons was obtained in ML MoSe$_2$ using polarization-resolved 2D coherent spectroscopy~\cite{Hao2017, Nagler2018}.
For co-circular excitation [Fig.~\ref{fig:biexcitons}(c)], the spectrum is dominated by the exciton and exciton-trion coupling resonance.
In contrast, cross-circular excitation [Fig.~\ref{fig:biexcitons}(d)] reveals additional resonances assigned to the
neutral biexciton $\mathrm{XX}$ and the charged biexciton $\mathrm{XX}^{-}$.
The appearance of these additional features only under cross-circular excitation identifies both complexes as intervalley states, with binding energies of approximately 20~meV and 5~meV for $\mathrm{XX}$ and $\mathrm{XX}^{-}$, respectively~\cite{Hao2017}.

The complementary PL picture is illustrated for an hBN-encapsulated WSe$_2$ ML in Figs.~\ref{fig:biexcitons}(e) and (f).
In addition to the neutral exciton and trion resonances, the spectrum contains the neutral biexciton $\mathrm{XX}$ and a lower-energy negatively charged biexciton $\mathrm{XX}^{-}$~\cite{Li2018}.
Both features exhibit nearly quadratic excitation-power dependences, with fitted power-law exponents of 1.94 and 1.82 for $\mathrm{XX}$ and $\mathrm{XX}^{-}$, respectively, compared with 1.29 for the neutral exciton [Fig.~\ref{fig:biexcitons}(f)]~\cite{Li2018}.
This characteristic power dependence provides an important fingerprint of their multi-exciton origin.

Excitation-power dependence alone, however, is not sufficient for an unambiguous assignment of the charge state.
Gate-dependent PL of hBN-encapsulated WSe$_2$ ML showed that the neutral $\mathrm{XX}$ resonance is restricted to the charge-neutral regime.
In contrast, $\mathrm{XX}^{-}$ emerges in the weakly electron-doped regime, where neutral excitons and negative trions coexist~\cite{Li2018,Barbone2018}.
The combination of power- and gate-dependent spectroscopy therefore distinguishes the four-particle neutral biexciton from the five-particle
charged biexciton, or exciton-trion complex.

Magneto-optical studies of hBN-encapsulated WSe$_2$ and WS$_2$ MLs reveal that biexcitonic emission generally exhibits $g$ factors characteristic of optically bright recombination.
In WSe$_2$ ML, the neutral biexciton was shown to contain bright and dark excitonic constituents, with radiative recombination proceeding through its bright component; correspondingly, its spectral $g$ factor is close to that of the bright exciton, with reported values around $-4$~\cite{Li2018,Barbone2018}.
The negatively charged biexciton in WSe$_2$ ML exhibits $g_{\mathrm{XX}^{-}}\approx-3.9$ to $-5.3$~\cite{Li2018,Barbone2018}, while in hBN-encapsulated WS$_2$ two $\mathrm{XX}^{-}$ emission lines were reported with $g=-4.1$~\cite{Zinkiewicz2021}.
These results highlight that the Zeeman response of biexcitonic emission primarily reflects the radiative recombination channel rather than the full spin configuration of the complex.
Beyond such recombination processes, additional low-energy emission features can arise from phonon-assisted transitions, as discussed in the following section.

\section{Exciton-phonon interaction and phonon replicas}
\label{sec:phonon_replicas}

The rich low-energy PL spectra of S-TMD MLs do not arise exclusively from distinct Coulomb-bound excitonic complexes.
Optically forbidden excitons and trions can also recombine through the emission of a~phonon, producing satellite lines, commonly referred to as phonon replicas, below the corresponding excitonic resonance~\cite{Li2019replica,Li2019momentum,He2020,Liu2020,Selig2020Sidebands, Zinkiewicz2021}.
For a phonon-emission process, the replica is red-shifted from the zero-phonon transition approximately by the energy of the participating phonon,
\begin{equation}
E_{\mathrm{rep}}=E_{\mathrm{ZPL}}-\hbar\omega_{\nu}(\mathbf{q}),
\end{equation}
where $E_{\mathrm{ZPL}}$ is the zero-phonon-line energy, while $\nu$ and $\mathbf{q}$ denote the phonon branch and momentum, respectively.
Consequently, a reliable identification of a phonon replica requires not only the appropriate energy separation but also consistency with the polarization, carrier-density dependence, and magnetic-field response of the associated excitonic state.

\subsection{Phonon-assisted recombination of dark excitons and trions}

The basic phonon-assisted recombination channels of neutral dark excitons are schematically illustrated in Fig.~\ref{fig:phonon_replicas}(a).
The same mechanisms apply to charged dark excitons, for which the additional carrier acts as a spectator during radiative recombination.
A zone-center phonon can activate an intravalley spin-forbidden transition, whereas zone-edge phonons provide the momentum required for intervalley
recombination~\cite{Li2019replica,LiuValley,He2020,Liu2020}.
In particular, the $\mathrm{E}"(\Gamma)$ mode mediates an intravalley spin-flip process, while $\mathrm{K}$-point phonons enable intervalley electron- and hole-scattering channels.

This correspondence is particularly clear for hBN-encapsulated WS$_2$ ML.
Comparison of the calculated phonon dispersion [Fig.~\ref{fig:phonon_replicas}(b)] with the PL spectrum [Fig.~\ref{fig:phonon_replicas}(c)] identifies four replicas of the negative dark trion, red-shifted by approximately 37, 44, 23, and 18~meV and assigned to $\mathrm{E}"(\Gamma)$, $\mathrm{E}"(\mathrm{K})$, LA$(\mathrm{K})$, and ZA$(\mathrm{K})$ phonons, respectively~\cite{Zinkiewicz2021}.
Their magneto-optical response provides an independent fingerprint: the $\mathrm{E}"(\Gamma)$ replica exhibits $g=-8.9$, identical to the intravalley
dark trion, whereas the $\mathrm{K}$-point replicas have $|g|\approx13$, characteristic of intervalley recombination~\cite{Zinkiewicz2021}.

A richer set of phonon replicas is observed for the positively charged dark trion in hBN-encapsulated WSe$_2$ ML [Fig.~\ref{fig:phonon_replicas}(d)].
Four lines appear approximately 26, 22, 18, and 13~meV below $\mathrm{D}^{+}$ and are assigned to the $K_3$, $\Gamma_5$, $K_1$, and
$K_2$ phonons, respectively~\cite{He2020}.
Their $g$ factors, $-13.4$, $-9.7$, $+12.2$, and $-13.0$, together with their helicity selection rules, distinguish the underlying scattering
channels: the $\mathrm{K}_3$ mode provides the dominant spin-conserving intervalley
electron-scattering channel, whereas $\mathrm{K}_1$ mediates intervalley hole scattering~\cite{He2020}.
The $\mathrm{K}_3$ mode is particularly notable because it carries chiral lattice motion and couples the phonon pseudo-angular momentum to the valley degree of
freedom.

Phonon replicas have consequently been identified for neutral spin-forbidden and momentum-forbidden excitons as well as for positively and negatively
charged dark trions in WSe$_2$~\cite{Li2019replica,Li2019momentum,LiuValley,He2020,Liu2020,Li2020phonon}.
Thus, charging changes the energy and many-body configuration of the excitonic state but preserves the underlying spin-valley constraints on
phonon-assisted recombination.
Moreover, recent strain-dependent studies of WS$_2$ and WSe$_2$ have further revealed strongly localized phonon-dressed dark-exciton states with extended lifetimes and pronounced exciton-phonon coupling~\cite{Chand2026}.

\begin{figure*}[t]
    \centering
    \includegraphics[width=0.845\linewidth]{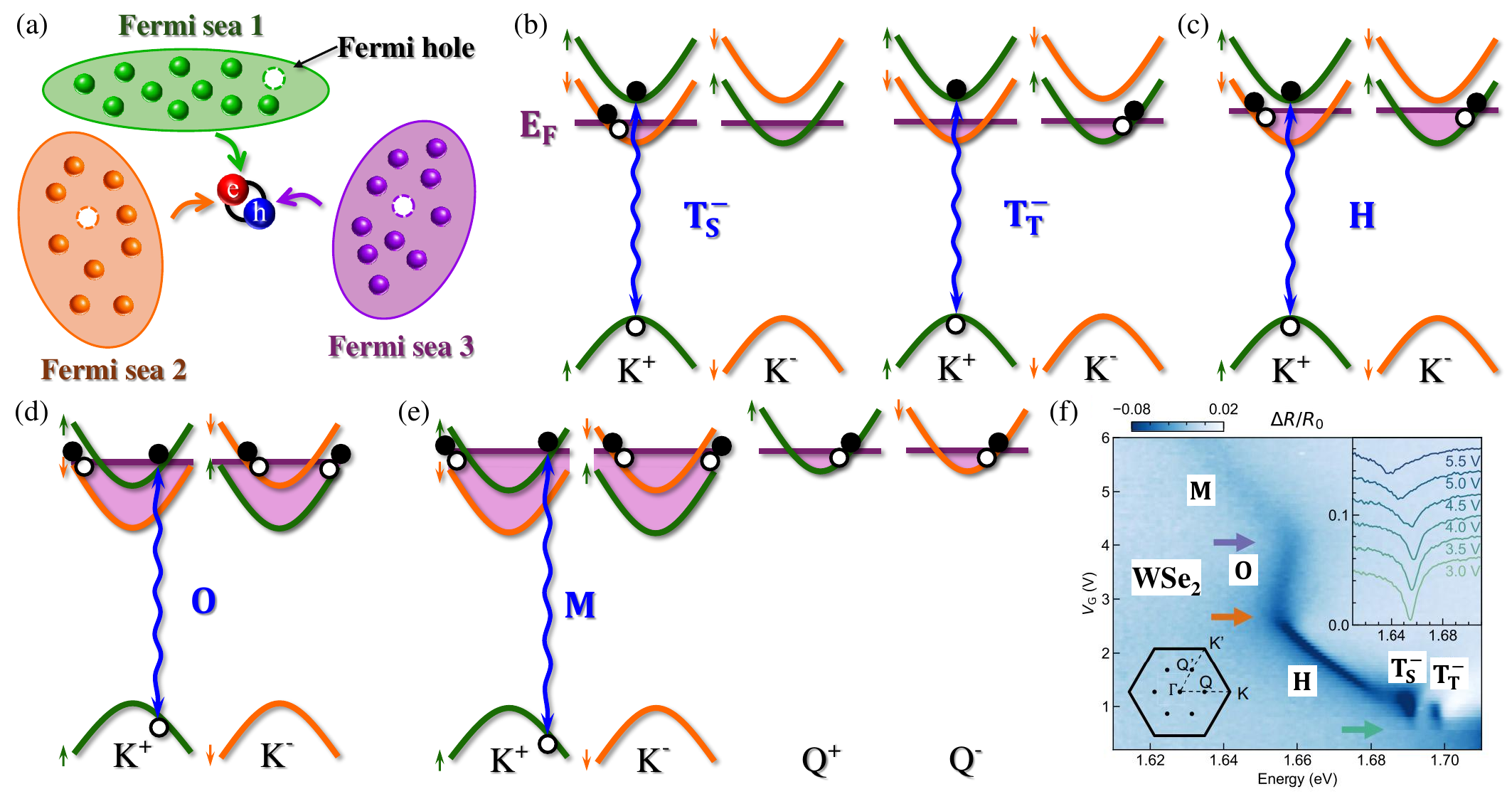}
    \caption{
    Evolution from few-body excitonic complexes to Fermi-sea many-body states.
    (a) Schematic illustration of a photoexcited $e$-$h$ pair interacting with several quantum-mechanically distinguishable Fermi reservoirs. 
    Each participating reservoir contributes a Fermi particle-hole excitation, where the Fermi hole represents a vacancy in an initially occupied CB state, rather than a photoexcited VB hole.
    (b)-(e) Schematic configurations of the negatively charged singlet and triplet excitons, $\mathrm{T}_{\mathrm{S}}^{-}$ and $\mathrm{T}_{\mathrm{T}}^{-}$, the six-particle hexciton ($\mathrm{H}$), the eight-particle oxciton ($\mathrm{O}$), and the larger multivalley complex $\mathrm{M}$ in electron-doped WSe$_2$ ML. 
    Progressive occupation of additional spin-valley states increases the number of distinguishable Fermi reservoirs available for correlation with the photoexcited $e$-$h$ pair. 
    At the highest electron densities, occupation of the $\mathrm{Q}^{\pm}$ valleys enables the formation of the multivalley $\mathrm{M}$ state.
    (f) Gate-dependent reflection-contrast spectrum of an hBN-encapsulated WSe$_2$ ML showing the evolution from $\mathrm{T}_{\mathrm{S}}^{-}$ and $\mathrm{T}_{\mathrm{T}}^{-}$ to $\mathrm{H}$, $\mathrm{O}$, and finally $\mathrm{M}$ with increasing electron density. 
    Panel (f) adapted from Ref.~\cite{Dijkstra2025} under a Creative Commons Attribution 4.0 (CC BY 4.0) license.
    }
\label{fig:many_body}
\end{figure*}

\subsection{Momentum-indirect Q-K excitons and multiphonon replicas}

Phonon-assisted recombination is not restricted to excitons formed from the two inequivalent $\mathrm{K}$ valleys.
The secondary CB minima at the $\mathrm{Q}$ points (often denoted as $\Lambda$ in the literature) enable momentum-indirect $\mathrm{Q}$-$\mathrm{K}$ excitons composed of an electron in a $\mathrm{Q}$ valley and a hole at $\mathrm{K}$.
Because such an exciton carries a finite center-of-mass momentum, its direct radiative recombination is forbidden by momentum conservation and can be activated by disorder or by phonons carrying the required momentum.

Evidence for such processes was obtained in ML WS$_2$ using near-resonant PLE spectroscopy, where scattering by zone-edge phonons was shown to populate momentum-indirect excitonic states involving the $\mathrm{Q}$ ($\Lambda$) valleys~\cite{Bao2020}.
Subsequent PL studies identified a $\mathrm{Q}$-$\mathrm{K}$ momentum-dark exciton together with a series of phonon-assisted features involving zone-edge acoustic phonons, including higher-order replicas extending to multiple-phonon processes~\cite{Adler2024}.

The importance of the $\mathrm{Q}$ valleys extends beyond WS$_2$. 
In ML WSe$_2$, theoretical studies have predicted low-energy $\mathrm{Q}$-valley trions, while experiments have revealed many-body complexes emerging upon filling the $\mathrm{Q}$ valleys~\cite{Nielsen2025Q,Dijkstra2025}. 
Ultrafast spectroscopy further shows that phonon-assisted intervalley scattering plays a central role in exciton relaxation and valley dynamics in this material~\cite{Dogadov2026}.

\section{From few-body complexes to Fermi-sea many-body states}

The discussion of charged excitons in the preceding section introduced the crossover from the few-particle trion picture to the exciton-polaron regime. 
This provides a~natural starting point for a richer class of many-body states that emerges when a photoexcited $e$-$h$ pair interacts simultaneously with several quantum-mechanically distinguishable Fermi seas.

In contrast to biexcitons and charged biexcitons, which remain discrete few-particle complexes, hexcitons, oxcitons, and related states involve particle-hole excitations of the resident carrier system. 
As illustrated schematically in Fig.~\ref{fig:many_body}(a), binding of a resident CB electron to the photoexcited  $e$-$h$ pair leaves a vacancy in the initially occupied states of the Fermi sea, referred to as a Fermi hole.
Unlike the photoexcited hole in the VB, the Fermi hole represents a missing electron in the CB and carries a positive charge relative to the unperturbed Fermi sea.
The resulting Fermi particle-hole excitation forms part of the correlated many-body complex~\cite{Tuan2022theory,Dijkstra2025}.

When two distinguishable Fermi reservoirs participate, the resulting six-particle state is referred to as a hexciton, $\mathrm{H}$, whereas coupling to three reservoirs gives an eight-particle oxciton, $\mathrm{O}$~\cite{Li2021,Tuan2022}.
The relevant quantity is therefore not simply the carrier density but also the number of distinguishable spin-valley reservoirs available for correlation~\cite{Tuan2022,Dery2025}.
This connection between band filling and many-body excitonic states is illustrated for electron-doped WSe$_2$ in Figs.~\ref{fig:many_body}(b)-(e).

A particularly clear experimental demonstration was recently obtained in an hBN-encapsulated WSe$_2$ ML [Fig.~\ref{fig:many_body}(f)]~\cite{Dijkstra2025}. 
With increasing electron density, the optical response evolves from the exchange-split singlet and triplet negatively charged excitons, $\mathrm{T}_{\mathrm{S}}^{-}$ and $\mathrm{T}_{\mathrm{T}}^{-}$, to $\mathrm{H}$ and subsequently to $\mathrm{O}$, accompanied by a transfer of oscillator strength between the successive resonances. 
At still higher density, an additional resonance denoted $\mathrm{M}$ emerges when the Fermi level reaches the $\mathrm{Q}^{\pm}$ CB valleys.

The occupation of the multivalley $\mathrm{Q}^{\pm}$ manifold greatly increases the number of distinguishable carrier reservoirs. 
Within the composite-exciton picture, the $\mathrm{M}$ state may involve the photoexcited $e$-$h$ pair together with as many as nine Fermi particle-hole excitations, corresponding to up to 20 correlated quasiparticles~\cite{Dijkstra2025}. 
Its detailed microscopic structure remains to be established, and it is therefore more appropriately regarded as a highly correlated multivalley excitonic state than simply as a larger analogue of a biexciton or oxciton.

The progression from charged excitons to $\mathrm{H}$, $\mathrm{O}$, and $\mathrm{M}$ thus represents a conceptual crossover from few-body excitonic complexes to optical excitations in which the multivalley Fermi sea itself becomes an integral part of the correlated state.

\section{Alloy monolayers - tunable exciton hierarchy}

The preceding sections have discussed the rich family of excitonic states and complexes emerging in chemically well-defined S-TMD MLs.
Alloying introduces an additional degree of freedom to this already complex excitonic landscape by enabling the electronic and optical properties to be continuously modified through chemical composition. 
Recent studies have demonstrated that such compositional engineering can lead not only to continuous shifts of excitonic resonances, but also to qualitatively new and intriguing changes in their energetic hierarchy and optical response.

Both the transition-metal sublattice, as in Mo$_{x}$W$_{1-x}$X$_2$, and the chalcogen sublattice, as in M(S$_{x}$Se$_{1-x}$)$_2$, can be continuously varied, where $x$ denotes the fractional concentration of the corresponding alloy constituent.
This compositional degree of freedom enables systematic tuning of the optical band gap and excitonic transition energies over a broad spectral range~\cite{Chen2013,Tongay2014,Zhang2014,Mann2014,Li2014,Feng2014}.
Early studies demonstrated continuous band-gap tuning in both transition-metal-alloyed Mo$_{1-x}$W$_x$X$_2$ and chalcogen-alloyed M(S$_x$Se$_{1-x}$)$_2$ MLs~\cite{Chen2013,Tongay2014,Zhang2014,Mann2014,Li2014,Feng2014}.
More recently, systematic spectroscopy of Mo$_{1-x}$W$_x$S$_2$ and Mo(S$_{1-x}$Se$_x$)$_2$ MLs has shown that the A-, B-, and higher-energy excitonic transitions, as well as their temperature dependence, can be controlled through composition~\cite{Ciesiolkiewicz2026}.
However, the evolution of the band gap with alloy composition can deviate from linear interpolation (band-gap bowing), while the SO splitting can also change substantially~\cite{Tongay2014,Wang2015Alloy}.

A characteristic consequence of alloying is intrinsic compositional disorder. 
Even for macroscopically homogeneous crystals, random atomic substitution produces nanoscale fluctuations of the local electronic structure, which can broaden and localize excitonic states~\cite{Chen2013,Tongay2014,Zhang2014,Masenda2021}. 
Temperature-dependent PL of Mo$_{0.3}$W$_{0.7}$Se$_2$ ML, for example, revealed non-monotonic variations of the exciton energy and linewidth associated with relaxation in a disorder potential~\cite{Masenda2021}. 
The distinction between extrinsic and intrinsic disorder becomes particularly clear after hBN encapsulation. 
For WSSe MLs, encapsulation reduces the neutral-exciton linewidth only from about 23 to 18~meV, whereas the corresponding linewidths of WSe$_2$ and WS$_2$ MLs decrease from 13 to 3.5~meV and from 16 to 5.2~meV, respectively~\cite{Pucko2022}. 
Thus, while hBN efficiently suppresses fluctuations associated with the substrate and dielectric environment, the residual linewidth of an alloy can remain limited by compositional disorder.

\begin{figure*}[t]
    \centering
    \includegraphics[width=1\linewidth]{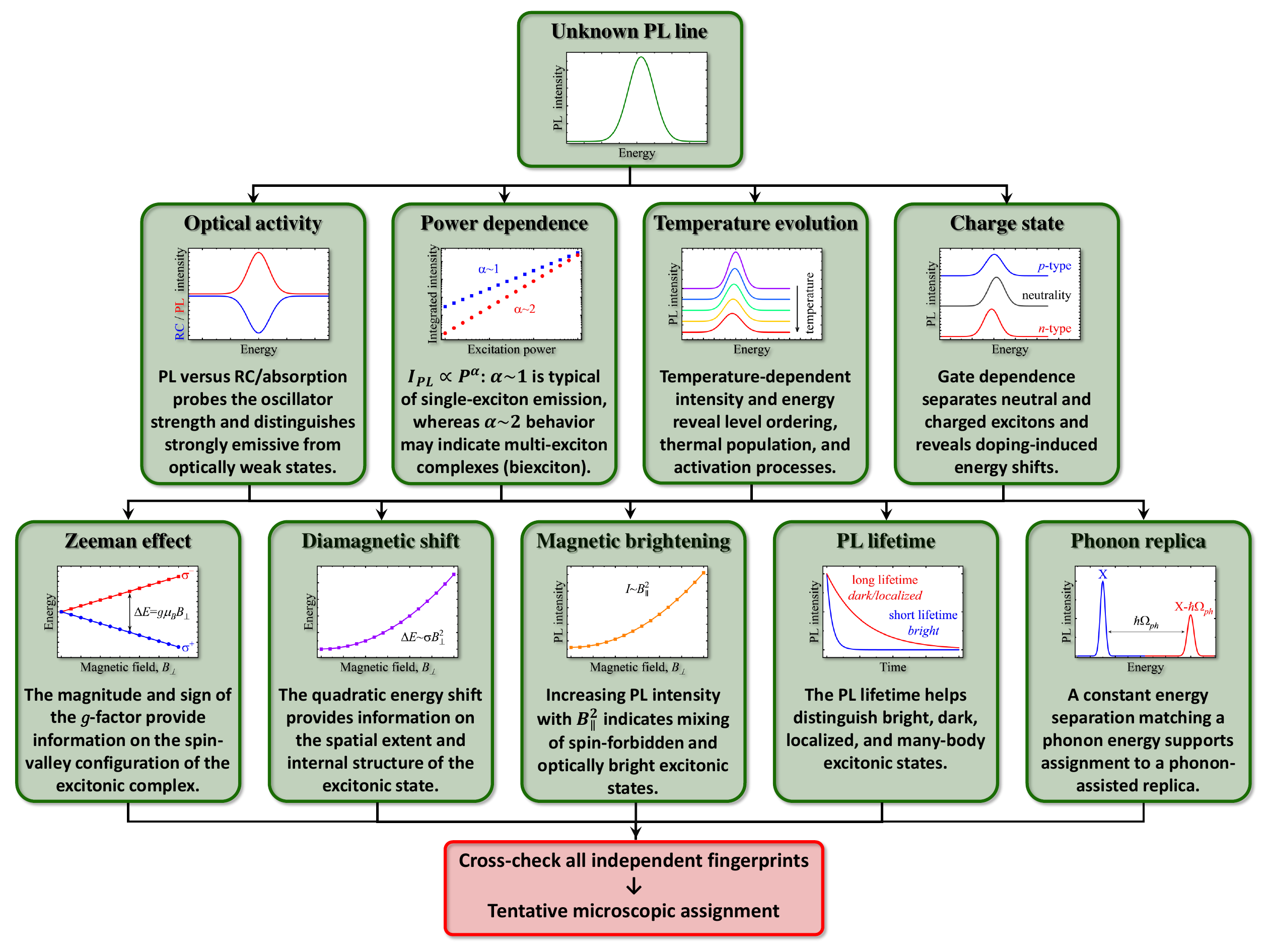}
    \caption{
    Experimental decision tree for identifying excitonic complexes in hBN-encapsulated S-TMD MLs.
    An unknown PL resonance is classified using complementary fingerprints provided by PL versus RC/absorption, gate-voltage dependence, excitation-power scaling, in-plane and out-of-plane magnetic fields, polarization selection rules, phonon-energy offsets, temperature evolution, diamagnetic shifts, and time-resolved PL. Each observable constrains a different property of the underlying state, including charge, particle number, spin and valley configuration, oscillator strength, spatial extent, and recombination dynamics. The final microscopic assignment should be based on the mutual consistency of several independent observables rather than on a single spectroscopic signature.
    }
\label{fig:how}
\end{figure*}

Alloying also provides direct access to the spin-valley structure of the band edges. 
In Mo$_{1-x}$W$_x$Se$_2$ ML, the CB spin splitting changes substantially with composition and evolves between the bright excitonic ground state characteristic of MoSe$_2$ ML and the darkish ordering of WSe$_2$ ML~\cite{Wang2015Alloy,Braganca2019}. 
The same evolution strongly affects valley polarization and exciton relaxation~\cite{Wang2015Alloy,Braganca2019}. 
Composition-dependent valley polarization has also been observed in Mo$_{1-x}$W$_x$S$_2$ MLs, where the polarization increases toward the W-rich limit~\cite{Jeong2026}. 
For chalcogen alloys, magneto-optical spectroscopy of hBN-encapsulated WSSe established a dark excitonic ground state, analogous to those of its WS$_2$ and WSe$_2$ parent compounds~\cite{Pucko2022}. 
These results demonstrate that alloying controls not only the optical transition energy but also the internal spin-valley hierarchy of excitonic states.

Magneto-optical spectroscopy provides an especially sensitive probe of this alloy-modified electronic structure. 
Enhanced excitonic Zeeman responses were already reported for Mo$_{0.5}$W$_{0.5}$Se$_2$ MLs~\cite{Wu2021}. 
More recently, systematic measurements of hBN-encapsulated Mo$_{x}$W$_{1-x}$Se$_2$ MLs revealed a strikingly non-monotonic composition dependence of the neutral-exciton $g$ factor~\cite{OlkowskaPucko2026}. 
While the parent MoSe$_2$ and WSe$_2$ MLs exhibit the conventional value $g_{\mathrm{X}}\approx-4$, the alloys reach $|g_{\mathrm{X}}|\approx10$ around $x\approx0.2$. 
First-principles calculations identify alloy-induced admixing of CB states originating from different valleys as the microscopic origin of this enhancement~\cite{OlkowskaPucko2026}. 
Alloying therefore does not simply interpolate between the properties of the two parent compounds, but can generate excitonic states with valley and orbital character that is absent in either binary endpoint.

Importantly, the complex excitonic hierarchy characteristic of binary S-TMDs is largely preserved in alloy MLs. 
In hBN-encapsulated WSSe, magneto-PL allowed bright, spin-forbidden, and momentum-forbidden negatively charged excitons to be distinguished, together with a phonon replica of the spin-dark trion~\cite{Pucko2022}. 
The same material also hosts localized single-photon emitters, showing that alloying can extend the range of excitonic phenomena beyond simple band-gap tuning~\cite{Pucko2022}. 
Combined with the strong composition dependence of exciton energies, SO splitting, valley polarization, and Zeeman response, these observations establish alloy MLs as a platform for engineering excitonic states with properties inaccessible in the binary end members~\cite{Wang2015Alloy,Wu2021,Pucko2022,OlkowskaPucko2026}.

\section{Identification of excitonic complexes}
\label{sec}

The high spectral quality of hBN-encapsulated S-TMD MLs has revealed a dense manifold of excitonic resonances, including bright and dark excitons, trions, biexcitons, higher-order complexes, Rydberg states, localized excitons, and phonon replicas. 
Since several of these transitions may occur within only a few meV of one another, assignment based solely on the emission energy is generally unreliable. 
A robust identification should instead rely on several complementary experimental fingerprints, as summarized in Figure~\ref{fig:how}.

Comparison of PL with RC or absorption provides information on the optical activity of a resonance and helps distinguish strongly emissive states from transitions with significant direct oscillator strength~\cite{Cadiz2017,Ajayi2017,Han2018,Goryca2019}. 
Excitation-power dependence offers complementary information on the number of photoexcited quasiparticles involved: approximately linear behavior is typical of single-exciton emission, whereas a superlinear dependence may support a biexcitonic or multiexcitonic assignment~\cite{Li2018,Barbone2018,Ye2018}. 
Such power-law analysis, however, should not be used as a unique identification criterion because saturation, exciton-exciton annihilation, and non-radiative processes can substantially modify the observed exponent.
Temperature-dependent spectroscopy probes thermal population, level ordering, and activation processes~\cite{Jadczak2021Upconversion,Meshulam2026,arora2019dark}. 
Electrostatic gating, in turn, provides one of the most direct probes of the charge state, separating neutral from positively and negatively charged excitonic complexes~\cite{He2020,Smolenski2019,Liu2019,Li2018}.

Magneto-optical measurements provide particularly powerful fingerprints. In an out-of-plane magnetic field, $B_{\perp}$, the valley Zeeman splitting, $\Delta E_{\mathrm{Z}}=g_{\mathrm{X}}\mu_{\mathrm{B}}B_{\perp}$, yields an effective $g$ factor that constrains the spin-valley configuration of the transition~\cite{Robert2017,Molas2019,Zinkiewicz2021,OlkowskaPucko2026}. 
The diamagnetic shift probes the spatial extent of the $e$-$h$ wave function and is especially useful for identifying excited Rydberg states~\cite{Stier2017,Goryca2019,Molas2019Energy}. 
An in-plane magnetic field, $B_{\parallel}$, can mix states with opposite spin projections and brighten spin-forbidden excitons, thereby providing a direct signature of spin-dark character~\cite{Molas2017,Zinkiewicz2020,Robert2020,Pucko2022}. 
Time-resolved measurements provide another complementary fingerprint, with PL lifetimes helping to distinguish rapidly recombining bright states from longer-lived dark, localized, or momentum-forbidden states~\cite{Robert2016,Robert2017,Liu2019,Zinkiewicz2020}. 
Finally, phonon replicas should be identified not only from an energy separation matching a phonon mode, but also from correlated gate, magnetic-field, polarization, or temperature dependences with the proposed parent excitonic state~\cite{Li2019replica,Li2019momentum,He2020,Zinkiewicz2021}.

Thus, the identification of an excitonic resonance should be treated as a multidimensional problem. 
The most reliable assignments are those for which several independent observables consistently support the same microscopic configuration, rather than those based on a single energy position, power exponent, or magnetic-field response.

\section{Summary and Outlook}

The development of high-quality hBN-encapsulated \mbox{S-TMD} MLs has transformed exciton spectroscopy from the observation of a limited number of broadened optical resonances into the study of a dense hierarchy of well-resolved excitonic states. 
This progress has also changed the central experimental challenge: the observation of a new spectral feature is no longer sufficient, and its microscopic origin must be established through a consistent set of independent spectroscopic fingerprints. 
As emphasized throughout this Review, transition energy alone rarely provides an unambiguous assignment. 
Instead, reliable identification increasingly relies on the combined use of emission and absorption-related spectroscopy, electrostatic gating, polarization selection rules, excitation-power and temperature dependences, magnetic-field response, diamagnetic shifts, and time-resolved measurements.

A major emerging direction is the evolution from conventional few-particle excitonic complexes toward highly correlated optical excitations involving the Fermi sea itself. 
The observation of hexcitons, oxcitons, and higher-order multivalley states demonstrates that increasing carrier density can qualitatively change the microscopic nature of an optically created excitation rather than simply shift or broaden an existing excitonic resonance. 
Establishing the internal structure, formation pathways, dynamics, coherence, and stability of these many-body states will require increasingly close interplay between high-resolution spectroscopy and microscopic many-body theory.

Alloying provides a complementary route toward controlling the excitonic hierarchy without relying solely on electrostatic doping. 
By modifying the band structure, SO splittings, valley composition, and relative ordering of bright and dark states, alloying can be used to engineer not only exciton transition energies but also their fine structure and magneto-optical response. 
Combining compositional control with hBN encapsulation, electrostatic tuning, strain, and external magnetic fields therefore offers a particularly promising strategy for designing excitonic states with targeted optical and spin-valley properties.

Looking forward, the combination of cleaner samples with ultrafast, nonlinear, multidimensional, and magneto-optical spectroscopy should extend the field beyond the identification of stationary excitonic resonances toward direct access to their formation, relaxation, coherence, correlations, and interactions. 
The next stage of exciton spectroscopy in 2D semiconductors will therefore be defined not only by resolving increasingly complex excitonic states, but by learning how to create, distinguish, and deliberately control their microscopic configurations. 


\section*{Acknowledgment}
The author would like to thank Artur Slobodeniuk for valuable discussions, insightful suggestions, and careful review of the scientific content of this article. 
The author also thanks Adam Babiński, Arka Karmakar, Wajid Ali, Tomasz Woźniak, Katarzyna Olkowska-Pucko, and Grzegorz Krasucki for their constructive feedback and critical reading of the manuscript.
The work was supported by the National Science Centre, Poland (grants no. 2022/46/E/ST3/00166 and 2025/57/B/ST5/03288).








\bibliographystyle{apsrev4-2}
\bibliography{biblio}

@Article{Mak2016,
  Title                    = {Photonics and optoelectronics of 2D semiconductor transition metal dichalcogenides},
  Author                   = {Mak, Kin Fai and Shan, Jie},
  Journal                  = {Nature Photonics},
  Year                     = {2016},
  Number                   = {4},
  Pages                    = {216--226},
  Volume                   = {10},
  ISSN                     = {1749-4893},
  Refid                    = {Mak2016},
  Url                      = {https://doi.org/10.1038/nphoton.2015.282}
}

@article{Wurstbauer2017,
  author  = {Ursula Wurstbauer and Bastian Miller and Eric Parzinger and Alexander W. Holleitner},
  title   = {Light-matter interaction in transition metal dichalcogenides and their heterostructures},
  journal = {Journal of Physics D: Applied Physics},
  volume  = {50},
  number  = {17},
  pages   = {173001},
  year    = {2017},
  doi     = {10.1088/1361-6463/aa5f81}
}

@Article{Schaibley2016,
  Title                    = {Valleytronics in 2D materials},
  Author                   = {Schaibley, John R. and Yu, Hongyi and Clark, Genevieve and Rivera, Pasqual and Ross, Jason S. and Seyler, Kyle L. and Yao, Wang and Xu, Xiaodong},
  Journal                  = {Nature Reviews Materials},
  Year                     = {2016},
  Number                   = {11},
  Pages                    = {16055--},
  Volume                   = {1},

  ISSN                     = {2058-8437},
  Refid                    = {Schaibley2016},
  Url                      = {https://doi.org/10.1038/natrevmats.2016.55}
}

@article{Molas2017,
    author={M R Molas and C Faugeras and A O Slobodeniuk and K Nogajewski and M Bartos and D M Basko and M Potemski},
    title={Brightening of dark excitons in monolayers of semiconducting transition metal dichalcogenides},
    journal={2D Materials},
    volume={4},
    number={2},
    pages={021003},
    doi={10.1088/2053-1583/aa5521},
    year={2017}
}

@Article{Zhang2017,
author={Zhang, Xiao-Xiao and Cao, Ting and Lu, Zhengguang and Lin, Yu-Chuan and Zhang, Fan and Wang, Ying and Li, Zhiqiang and Hone, James C. and Robinson, Joshua A. and Smirnov, Dmitry and Louie, Steven G. and Heinz, Tony F.},
title={Magnetic brightening and control of dark excitons in monolayer WSe2},
journal={Nature Nanotechnology},
year={2017},
month={Sep},
day={01},
volume={12},
number={9},
pages={883-888},
doi={10.1038/nnano.2017.105},
url={https://doi.org/10.1038/nnano.2017.105}
}

@article{Robert2017,
  title = {Fine structure and lifetime of dark excitons in transition metal dichalcogenide monolayers},
  author = {Robert, C. and Amand, T. and Cadiz, F. and Lagarde, D. and Courtade, E. and Manca, M. and Taniguchi, T. and Watanabe, K. and Urbaszek, B. and Marie, X.},
  journal = {Phys. Rev. B},
  volume = {96},
  issue = {15},
  pages = {155423},
  numpages = {8},
  year = {2017},
  month = {Oct},
  publisher = {American Physical Society},
  doi = {10.1103/PhysRevB.96.155423},
  url = {https://link.aps.org/doi/10.1103/PhysRevB.96.155423}
}

@article{Wang2017,
  title = {In-Plane Propagation of Light in Transition Metal Dichalcogenide Monolayers: Optical Selection Rules},
  author = {Wang, G. and Robert, C. and Glazov, M. M. and Cadiz, F. and Courtade, E. and Amand, T. and Lagarde, D. and Taniguchi, T. and Watanabe, K. and Urbaszek, B. and Marie, X.},
  journal = {Phys. Rev. Lett.},
  volume = {119},
  issue = {4},
  pages = {047401},
  numpages = {7},
  year = {2017},
  month = {Jul},
  publisher = {American Physical Society},
  doi = {10.1103/PhysRevLett.119.047401},
  url = {https://link.aps.org/doi/10.1103/PhysRevLett.119.047401}
}

@Article{Barbone2018,
author={Barbone, Matteo and Montblanch, Alejandro R.-P. and Kara, Dhiren M. and Palacios-Berraquero, Carmen and Cadore, Alisson R.
and De Fazio, Domenico and Pingault, Benjamin and Mostaani, Elaheh
and Li, Han and Chen, Bin and Watanabe, Kenji and Taniguchi, Takashi
and Tongay, Sefaattin and Wang, Gang and Ferrari, Andrea C.
and Atat{\"u}re, Mete},
title={Charge-tuneable biexciton complexes in monolayer WSe2},
journal={Nature Communications},
year={2018},
month={Sep},
day={13},
volume={9},
number={1},
pages={3721},
doi={10.1038/s41467-018-05632-4},
url={https://doi.org/10.1038/s41467-018-05632-4}
}

@Article{Li2018,
author={Li, Zhipeng and Wang, Tianmeng and Lu, Zhengguang and Jin, Chenhao and Chen, Yanwen and Meng, Yuze and Lian, Zhen and Taniguchi, Takashi and Watanabe, Kenji and Zhang, Shengbai and Smirnov, Dmitry
and Shi, Su-Fei},
title={Revealing the biexciton and trion-exciton complexes in BN encapsulated WSe2},
journal={Nature Communications},
year={2018},
month={Sep},
day={13},
volume={9},
number={1},
pages={3719},
doi={10.1038/s41467-018-05863-5},
url={https://doi.org/10.1038/s41467-018-05863-5}
}

@article{Liu2020,
  title = {Multipath Optical Recombination of Intervalley Dark Excitons and Trions in Monolayer ${\mathrm{WSe}}_{2}$},
  author = {Liu, Erfu and van Baren, Jeremiah and Liang, Ching-Tarng and Taniguchi, Takashi and Watanabe, Kenji and Gabor, Nathaniel M. and Chang, Yia-Chung and Lui, Chun Hung},
  journal = {Phys. Rev. Lett.},
  volume = {124},
  issue = {19},
  pages = {196802},
  numpages = {7},
  year = {2020},
  month = {May},
  publisher = {American Physical Society},
  doi = {10.1103/PhysRevLett.124.196802},
  url = {https://link.aps.org/doi/10.1103/PhysRevLett.124.196802}
}

@article{Molas2019,
  title = {Probing and Manipulating Valley Coherence of Dark Excitons in Monolayer ${\mathrm{WSe}}_{2}$},
  author = {Molas, M. R. and Slobodeniuk, A. O. and Kazimierczuk, T. and Nogajewski, K. and Bartos, M. and Kapu\ifmmode \acute{s}\else \'{s}\fi{}ci\ifmmode \acute{n}\else \'{n}\fi{}ski, P. and Oreszczuk, K. and Watanabe, K. and Taniguchi, T. and Faugeras, C. and Kossacki, P. and Basko, D. M. and Potemski, M.},
  journal = {Phys. Rev. Lett.},
  volume = {123},
  issue = {9},
  pages = {096803},
  numpages = {6},
  year = {2019},
  month = {Aug},
  publisher = {American Physical Society},
  doi = {10.1103/PhysRevLett.123.096803},
  url = {https://link.aps.org/doi/10.1103/PhysRevLett.123.096803}
}

@Article{He2020,
author={He, Minhao and Rivera, Pasqual and Van Tuan, Dinh and Wilson, Nathan P. and Yang, Min and Taniguchi, Takashi and Watanabe, Kenji and Yan, Jiaqiang and Mandrus, David G. and Yu, Hongyi and Dery, Hanan
and Yao, Wang and Xu, Xiaodong},
title={Valley phonons and exciton complexes in a monolayer semiconductor},
journal={Nature Communications},
year={2020},
month={Jan},
day={30},
volume={11},
number={1},
pages={618},
issn={2041-1723},
doi={10.1038/s41467-020-14472-0},
url={https://doi.org/10.1038/s41467-020-14472-0}
}

@article{LiuValley,
  title = {Valley-selective chiral phonon replicas of dark excitons and trions in monolayer $\mathrm{WS}{\mathrm{e}}_{2}$},
  author = {Liu, Erfu and van Baren, Jeremiah and Taniguchi, Takashi and Watanabe, Kenji and Chang, Yia-Chung and Lui, Chun Hung},
  journal = {Phys. Rev. Research},
  volume = {1},
  issue = {3},
  pages = {032007},
  numpages = {7},
  year = {2019},
  month = {Oct},
  publisher = {American Physical Society},
  doi = {10.1103/PhysRevResearch.1.032007},
  url = {https://link.aps.org/doi/10.1103/PhysRevResearch.1.032007}
}

@article{Li2019,
author = {Li, Zhipeng and Wang, Tianmeng and Lu, Zhengguang and Khatoniar, Mandeep and Lian, Zhen and Meng, Yuze and Blei, Mark and Taniguchi, Takashi and Watanabe, Kenji and McGill, Stephen A. and Tongay, Sefaattin and Menon, Vinod M. and Smirnov, Dmitry and Shi, Su-Fei},
title = {Direct Observation of Gate-Tunable Dark Trions in Monolayer WSe2},
journal = {Nano Letters},
volume = {19},
number = {10},
pages = {6886-6893},
year = {2019},
doi = {10.1021/acs.nanolett.9b02132}
}

@Article{Robert2020,
author={Robert, C. and Han, B. and Kapuscinski, P. and Delhomme, A. and Faugeras, C.
and Amand, T. and Molas, M. R. and Bartos, M. and Watanabe, K. and Taniguchi, T. and Urbaszek, B. and Potemski, M. and Marie, X.},
title={Measurement of the spin-forbidden dark excitons in MoS2 and MoSe2 monolayers},
journal={Nature Communications},
year={2020},
month={Aug},
day={12},
volume={11},
number={1},
pages={4037},
doi={10.1038/s41467-020-17608-4},
url={https://doi.org/10.1038/s41467-020-17608-4}
}

@article{arora2019dark,
 title = {Dark trions govern the temperature-dependent optical absorption and emission of doped atomically thin semiconductors},
  author = {Arora, Ashish and Wessling, Nils Kolja and Deilmann, Thorsten and Reichenauer, Till and Steeger, Paul and Kossacki, Piotr and Potemski, Marek and Michaelis de Vasconcellos, Steffen and Rohlfing, Michael and Bratschitsch, Rudolf},
  journal = {Phys. Rev. B},
  volume = {101},
  issue = {24},
  pages = {241413},
  numpages = {6},
  year = {2020},
  month = {Jun},
  publisher = {American Physical Society},
  doi = {10.1103/PhysRevB.101.241413},
  url = {https://link.aps.org/doi/10.1103/PhysRevB.101.241413}
}

@article{Koperski2017,
  author  = {Maciej Koperski and Maciej R. Molas and Ashish Arora and Karol Nogajewski and Artur O. Slobodeniuk and Cl{\'e}ment Faugeras and Marek Potemski},
  title   = {Optical properties of atomically thin transition metal dichalcogenides: observations and puzzles},
  journal = {Nanophotonics},
  volume  = {6},
  number  = {6},
  pages   = {1289--1308},
  year    = {2017},
  doi     = {10.1515/nanoph-2016-0165}
}

@article{Koperski2019,
	author={Maciej Koperski and Maciej R Molas and Ashish Arora and Karol Nogajewski and Miroslav Bartos and Jan Wyzula and Diana
	Vaclavkova and Piotr Kossacki and Marek Potemski},
	title={Orbital, spin and valley contributions to Zeeman splitting of excitonic resonances in $\textrm{MoSe}_2$, $\textrm{WSe}_2$ and $\textrm{WS}_2$ Monolayers},
	journal={2D Materials},
	volume={6},
	number={1},
	pages={015001},
	url={},
	year={2019},
	doi={10.1088/2053-1583/aae14b}
}

@Article{Arora2015,
author ="Arora, Ashish and Koperski, Maciej and Nogajewski, Karol and Marcus, Jacques and Faugeras, Clement and Potemski, Marek",
title  ="Excitonic resonances in thin films of $\textrm{WSe}_2$: from monolayer to bulk material",
journal  ="Nanoscale",
year  ="2015",
volume  ="7",
issue  ="23",
pages  ="10421",
doi  ="10.1039/C5NR01536G"
}

@article{Courtade2017,
    title = {{Charged excitons in monolayer ${\mathrm{WSe}}_{2}$: Experiment and theory}},
    author = {Courtade, E. and Semina, M. and Manca, M. and Glazov, M. M. and Robert, C. and Cadiz, F. and Wang, G. and Taniguchi, T. and Watanabe, K. and Pierre, M. and Escoffier, W. and Ivchenko, E. L. and Renucci, P. and Marie, X. and Amand, T. and Urbaszek, B.},
    journal = {Phys. Rev. B},
    volume = {96},
    issue = {8},
    pages = {085302},
    numpages = {12},
    year = {2017},
    month = {Aug},
    publisher = {American Physical Society},
    doi = {10.1103/PhysRevB.96.085302}
}

@Article{Zinkiewicz2020,
author ="Zinkiewicz, M. and Slobodeniuk, A. O. and Kazimierczuk, T. and Kapuściński, P. and Oreszczuk, K. and Grzeszczyk, M. and Bartos, M. and Nogajewski, K. and Watanabe, K. and Taniguchi, T. and Faugeras, C. and Kossacki, P. and Potemski, M. and Babiński, A. and Molas, M. R.",
title  ="Neutral and charged dark excitons in monolayer WS2",
journal  ="Nanoscale",
year  ="2020",
volume  ="12",
issue  ="35",
pages  ="18153-18159",
publisher  ="The Royal Society of Chemistry",
doi  ="10.1039/D0NR04243A",
url  ="http://dx.doi.org/10.1039/D0NR04243A"
}

@article{Zinkiewicz2021,
author = {Zinkiewicz, Małgorzata and Woźniak, Tomasz and Kazimierczuk, Tomasz and Kapuscinski, Piotr and Oreszczuk, Kacper and Grzeszczyk, Magdalena and Bartoš, Miroslav and Nogajewski, Karol and Watanabe, Kenji and Taniguchi, Takashi and Faugeras, Clement and Kossacki, Piotr and Potemski, Marek and Babiński, Adam and Molas, Maciej R.},
title = {Excitonic Complexes in n-Doped WS2 Monolayer},
journal = {Nano Letters},
volume = {21},
number = {6},
pages = {2519-2525},
year = {2021},
doi = {10.1021/acs.nanolett.0c05021}
}

@Article{Grzeszczyk2021,
author ="Grzeszczyk, Magdalena and Olkowska-Pucko, Katarzyna and Nogajewski, Karol and Watanabe, Kenji and Taniguchi, Takashi and Kossacki, Piotr and Babiński, Adam and Molas, Maciej R.",
title  ="Exposing the trion{'}s fine structure by controlling the carrier concentration in hBN-encapsulated MoS2",
journal  ="Nanoscale",
year  ="2021",
volume  ="13",
issue  ="44",
pages  ="18726-18733",
publisher  ="The Royal Society of Chemistry",
doi  ="10.1039/D1NR03855A",
url  ="http://dx.doi.org/10.1039/D1NR03855A"
}

@Article{Raja2017,
author={Raja, Archana and Chaves, Andrey and Yu, Jaeeun and Arefe, Ghidewon and Hill, Heather M. and Rigosi, Albert F. and Berkelbach, Timothy C. and Nagler, Philipp and Sch{\"u}ller, Christian and Korn, Tobias and Nuckolls, Colin
and Hone, James and Brus, Louis E. and Heinz, Tony F. and Reichman, David R.
and Chernikov, Alexey},
title={Coulomb engineering of the bandgap and excitons in two-dimensional materials},
journal={Nature Communications},
year={2017},
month={May},
day={04},
volume={8},
number={1},
pages={15251},
doi={10.1038/ncomms15251},
url={https://doi.org/10.1038/ncomms15251}
}

@article{Chernikov2014,
  title = {Exciton Binding Energy and Nonhydrogenic Rydberg Series in Monolayer ${\mathrm{WS}}_{2}$},
  author = {Chernikov, Alexey and Berkelbach, Timothy C. and Hill, Heather M. and Rigosi, Albert and Li, Yilei and Aslan, Ozgur Burak and Reichman, David R. and Hybertsen, Mark S. and Heinz, Tony F.},
  journal = {Phys. Rev. Lett.},
  volume = {113},
  issue = {7},
  pages = {076802},
  numpages = {5},
  year = {2014},
  month = {Aug},
  publisher = {American Physical Society},
  doi = {10.1103/PhysRevLett.113.076802},
  url = {https://link.aps.org/doi/10.1103/PhysRevLett.113.076802}
}

@Article{Goryca2019,
author={Goryca, M. and Li, J. and Stier, A. V. and Taniguchi, T. and Watanabe, K. and Courtade, E. and Shree, S. and Robert, C. and Urbaszek, B. and Marie, X. and Crooker, S. A.},
title={Revealing exciton masses and dielectric properties of monolayer semiconductors with high magnetic fields},
journal={Nature Communications},
year={2019},
month={Sep},
day={13},
volume={10},
number={1},
pages={4172},
doi={10.1038/s41467-019-12180-y},
url={https://doi.org/10.1038/s41467-019-12180-y}
}

@article{Vaclavkova2018,
	doi = {10.1088/1361-6528/aac65c},
	url = {https://doi.org/10.1088/1361-6528/aac65c},
	year = 2018,
	month = {jun},
	publisher = {{IOP} Publishing},
	volume = {29},
	number = {32},
	pages = {325705},
	author = {D Vaclavkova and J Wyzula and K Nogajewski and M Bartos and A O Slobodeniuk and C Faugeras and M Potemski and M R Molas},
	title = {Singlet and triplet trions in {WS}2 monolayer encapsulated in hexagonal boron nitride},
	journal = {Nanotechnology}
}

@Article{Kapuscinski2021,
author={Kapu{\'{s}}ci{\'{n}}ski, Piotr and Delhomme, Alex and Vaclavkova, Diana and Slobodeniuk, Artur O. and Grzeszczyk, Magdalena
and Bartos, Miroslav and Watanabe, Kenji
and Taniguchi, Takashi and Faugeras, Cl{\'e}ment and Potemski, Marek},
title={Rydberg series of dark excitons and the conduction band spin-orbit splitting in monolayer WSe2},
journal={Communications Physics},
year={2021},
month={Aug},
day={19},
volume={4},
number={1},
pages={186},
doi={10.1038/s42005-021-00692-3},
url={https://doi.org/10.1038/s42005-021-00692-3}
}

@article{Wang2018,
  title = {{Colloquium: Excitons in atomically thin transition metal dichalcogenides}},
  author = {Wang, Gang and Chernikov, Alexey and Glazov, Mikhail M. and Heinz, Tony F. and Marie, Xavier and Amand, Thierry and Urbaszek, Bernhard},
  journal = {Rev. Mod. Phys.},
  volume = {90},
  issue = {2},
  pages = {021001},
  numpages = {25},
  year = {2018},
  month = {Apr},
  publisher = {American Physical Society},
  doi = {10.1103/RevModPhys.90.021001}
}

@article{Molas2019Energy,
  title = {Energy Spectrum of Two-Dimensional Excitons in a Nonuniform Dielectric Medium},
  author = {Molas, M. R. and Slobodeniuk, A. O. and Nogajewski, K. and Bartos, M. and Bala, \L{}. and Babi\ifmmode \acute{n}\else \'{n}\fi{}ski, A. and Watanabe, K. and Taniguchi, T. and Faugeras, C. and Potemski, M.},
  journal = {Phys. Rev. Lett.},
  volume = {123},
  issue = {13},
  pages = {136801},
  numpages = {6},
  year = {2019},
  month = {Sep},
  publisher = {American Physical Society},
  doi = {10.1103/PhysRevLett.123.136801},
  url = {https://link.aps.org/doi/10.1103/PhysRevLett.123.136801}
}

@article{kormanyos2015,
	doi = {10.1088/2053-1583/2/2/022001},
	url = {https://doi.org/10.1088/2053-1583/2/2/022001},
	year = 2015,
	month = {apr},
	publisher = {{IOP} Publishing},
	volume = {2},
	number = {2},
	pages = {022001},
	author = {Andor Korm{\'{a}}nyos and Guido Burkard and Martin Gmitra and Jaroslav Fabian and Viktor Z{\'{o}}lyomi and Neil D Drummond and Vladimir Fal'ko},
	title = {$k\cdot p$ theory for two-dimensional transition metal dichalcogenide semiconductors},
	journal = {2D Materials}
}

@article{Stier2017,
  title = {Magnetooptics of Exciton Rydberg States in a Monolayer Semiconductor},
  author = {Stier, A. V. and Wilson, N. P. and Velizhanin, K. A. and Kono, J. and Xu, X. and Crooker, S. A.},
  journal = {Phys. Rev. Lett.},
  volume = {120},
  issue = {5},
  pages = {057405},
  numpages = {6},
  year = {2018},
  month = {Feb},
  publisher = {American Physical Society},
  doi = {10.1103/PhysRevLett.120.057405},
  url = {https://link.aps.org/doi/10.1103/PhysRevLett.120.057405}
}

@article{Smolenski2019,
  title = {Interaction-Induced Shubnikov--de Haas Oscillations in Optical Conductivity of Monolayer ${\mathrm{MoSe}}_{2}$},
  author = {Smole\ifmmode \acute{n}\else \'{n}\fi{}ski, T. and Cotlet, O. and Popert, A. and Back, P. and Shimazaki, Y. and Kn\"uppel, P. and Dietler, N. and Taniguchi, T. and Watanabe, K. and Kroner, M. and Imamoglu, A.},
  journal = {Phys. Rev. Lett.},
  volume = {123},
  issue = {9},
  pages = {097403},
  numpages = {6},
  year = {2019},
  month = {Aug},
  publisher = {American Physical Society},
  doi = {10.1103/PhysRevLett.123.097403},
  url = {https://link.aps.org/doi/10.1103/PhysRevLett.123.097403}
}

@article{Liu2019,
  title = {Gate Tunable Dark Trions in Monolayer ${\mathrm{WSe}}_{2}$},
  author = {Liu, Erfu and van Baren, Jeremiah and Lu, Zhengguang and Altaiary, Mashael M. and Taniguchi, Takashi and Watanabe, Kenji and Smirnov, Dmitry and Lui, Chun Hung},
  journal = {Phys. Rev. Lett.},
  volume = {123},
  issue = {2},
  pages = {027401},
  numpages = {6},
  year = {2019},
  month = {Jul},
  publisher = {American Physical Society},
  doi = {10.1103/PhysRevLett.123.027401},
  url = {https://link.aps.org/doi/10.1103/PhysRevLett.123.027401}
}

@article{Mak2010,
  title = {Atomically Thin ${\mathrm{MoS}}_{2}$: A New Direct-Gap Semiconductor},
  author = {Mak, Kin Fai and Lee, Changgu and Hone, James and Shan, Jie and Heinz, Tony F.},
  journal = {Phys. Rev. Lett.},
  volume = {105},
  issue = {13},
  pages = {136805},
  numpages = {4},
  year = {2010},
  month = {Sep},
  publisher = {American Physical Society},
  doi = {10.1103/PhysRevLett.105.136805},
  url = {https://link.aps.org/doi/10.1103/PhysRevLett.105.136805}
}

@article{Lezama2015,
author = {Lezama, Ignacio Gutiérrez and Arora, Ashish and Ubaldini, Alberto and Barreteau, Céline and Giannini, Enrico and Potemski, Marek and Morpurgo, Alberto F.},
title = {Indirect-to-Direct Band Gap Crossover in Few-Layer MoTe2},
journal = {Nano Letters},
volume = {15},
number = {4},
pages = {2336-2342},
year = {2015},
doi = {10.1021/nl5045007},
URL = {https://doi.org/10.1021/nl5045007 }
}

@Article{Arora_MoSe2,
author ="Arora, Ashish and Nogajewski, Karol and Molas, Maciej and Koperski, Maciej and Potemski, Marek",
title  ="Exciton band structure in layered MoSe2: from a monolayer to the bulk limit",
journal  ="Nanoscale",
year  ="2015",
volume  ="7",
issue  ="48",
pages  ="20769-20775",
publisher  ="The Royal Society of Chemistry",
doi  ="10.1039/C5NR06782K",
url  ="http://dx.doi.org/10.1039/C5NR06782K",
}

@Article{Mak2012,
author={Mak, Kin Fai and He, Keliang and Shan, Jie and Heinz, Tony F.},
title={Control of valley polarization in monolayer MoS2 by optical helicity},
journal={Nature Nanotechnology},
year={2012},
month={Aug},
day={01},
volume={7},
number={8},
pages={494-498},
issn={1748-3395},
doi={10.1038/nnano.2012.96},
url={https://doi.org/10.1038/nnano.2012.96}
}

@article{Xiao2012,
  title = {Coupled Spin and Valley Physics in Monolayers of ${\mathrm{MoS}}_{2}$ and Other Group-VI Dichalcogenides},
  author = {Xiao, Di and Liu, Gui-Bin and Feng, Wanxiang and Xu, Xiaodong and Yao, Wang},
  journal = {Phys. Rev. Lett.},
  volume = {108},
  issue = {19},
  pages = {196802},
  numpages = {5},
  year = {2012},
  month = {May},
  publisher = {American Physical Society},
  doi = {10.1103/PhysRevLett.108.196802},
  url = {https://link.aps.org/doi/10.1103/PhysRevLett.108.196802}
}

@Article{Cao2012,
author={Cao, Ting and Wang, Gang and Han, Wenpeng and Ye, Huiqi and Zhu, Chuanrui and Shi, Junren and Niu, Qian and Tan, Pingheng and Wang, Enge
and Liu, Baoli and Feng, Ji},
title={Valley-selective circular dichroism of monolayer molybdenum disulphide},
journal={Nature Communications},
year={2012},
month={Jun},
day={06},
volume={3},
number={1},
pages={887},
issn={2041-1723},
doi={10.1038/ncomms1882},
url={https://doi.org/10.1038/ncomms1882}
}

@Article{Jones2013,
author={Jones, Aaron M. and Yu, Hongyi and Ghimire, Nirmal J. and Wu, Sanfeng and Aivazian, Grant and Ross, Jason S. and Zhao, Bo and Yan, Jiaqiang and Mandrus, David G. and Xiao, Di and Yao, Wang and Xu, Xiaodong},
title={Optical generation of excitonic valley coherence in monolayer WSe2},
journal={Nature Nanotechnology},
year={2013},
month={Sep},
day={01},
volume={8},
number={9},
pages={634-638},
issn={1748-3395},
doi={10.1038/nnano.2013.151},
url={https://doi.org/10.1038/nnano.2013.151}
}

@Article{Mak2013,
author={Mak, Kin Fai and He, Keliang and Lee, Changgu and Lee, Gwan Hyoung
and Hone, James and Heinz, Tony F. and Shan, Jie},
title={Tightly bound trions in monolayer MoS2},
journal={Nature Materials},
year={2013},
month={Mar},
day={01},
volume={12},
number={3},
pages={207-211},
issn={1476-4660},
doi={10.1038/nmat3505},
url={https://doi.org/10.1038/nmat3505}
}

@Article{Ross2013,
author={Ross, Jason S. and Wu, Sanfeng and Yu, Hongyi and Ghimire, Nirmal J. and Jones, Aaron M. and Aivazian, Grant and Yan, Jiaqiang and Mandrus, David G. and Xiao, Di and Yao, Wang and Xu, Xiaodong},
title={Electrical control of neutral and charged excitons in a monolayer semiconductor},
journal={Nature Communications},
year={2013},
month={Feb},
day={12},
volume={4},
number={1},
pages={1474},
issn={2041-1723},
doi={10.1038/ncomms2498},
url={https://doi.org/10.1038/ncomms2498}
}

@Article{Molas_Nanoscale,
author ="Molas, Maciej R. and Nogajewski, Karol and Slobodeniuk, Artur O. and Binder, Johannes and Bartos, Miroslav and Potemski, Marek",
title  ="The optical response of monolayer{,} few-layer and bulk tungsten disulfide",
journal  ="Nanoscale",
year  ="2017",
volume  ="9",
issue  ="35",
pages  ="13128-13141",
publisher  ="The Royal Society of Chemistry",
doi  ="10.1039/C7NR04672C",
url  ="http://dx.doi.org/10.1039/C7NR04672C",
}

@Article{Hao2017,
author={Hao, Kai and Specht, Judith F. and Nagler, Philipp and Xu, Lixiang
and Tran, Kha and Singh, Akshay and Dass, Chandriker Kavir and Sch{\"u}ller, Christian and Korn, Tobias and Richter, Marten and Knorr, Andreas and Li, Xiaoqin and Moody, Galan},
title={Neutral and charged inter-valley biexcitons in monolayer MoSe2},
journal={Nature Communications},
year={2017},
month={Jun},
day={28},
volume={8},
number={1},
pages={15552},
issn={2041-1723},
doi={10.1038/ncomms15552},
url={https://doi.org/10.1038/ncomms15552}
}

@Article{Wierzbowski2017,
author={Wierzbowski, Jakob and Klein, Julian and Sigger, Florian and Straubinger, Christian and Kremser, Malte
and Taniguchi, Takashi and Watanabe, Kenji and Wurstbauer, Ursula and Holleitner, Alexander W. and Kaniber, Michael
and M{\"u}ller, Kai and Finley, Jonathan J.},
title={Direct exciton emission from atomically thin transition metal dichalcogenide heterostructures near the lifetime limit},
journal={Scientific Reports},
year={2017},
month={Sep},
day={28},
volume={7},
number={1},
pages={12383},
issn={2045-2322},
doi={10.1038/s41598-017-09739-4},
url={https://doi.org/10.1038/s41598-017-09739-4}
}

@article{Cadiz2017,
  title = {Excitonic Linewidth Approaching the Homogeneous Limit in ${\mathrm{MoS}}_{2}$-Based van der Waals Heterostructures},
  author = {Cadiz, F. and Courtade, E. and Robert, C. and Wang, G. and Shen, Y. and Cai, H. and Taniguchi, T. and Watanabe, K. and Carrere, H. and Lagarde, D. and Manca, M. and Amand, T. and Renucci, P. and Tongay, S. and Marie, X. and Urbaszek, B.},
  journal = {Phys. Rev. X},
  volume = {7},
  issue = {2},
  pages = {021026},
  numpages = {12},
  year = {2017},
  month = {May},
  publisher = {American Physical Society},
  doi = {10.1103/PhysRevX.7.021026},
  url = {https://link.aps.org/doi/10.1103/PhysRevX.7.021026}
}

@article{Ajayi2017,
doi = {10.1088/2053-1583/aa6aa1},
url = {https://dx.doi.org/10.1088/2053-1583/aa6aa1},
year = {2017},
month = {jul},
publisher = {IOP Publishing},
volume = {4},
number = {3},
pages = {031011},
author = {Obafunso A Ajayi and Jenny V Ardelean and Gabriella D Shepard and Jue Wang and Abhinandan Antony and Takeshi Taniguchi and Kenji Watanabe and Tony F Heinz and Stefan Strauf and X-Y Zhu and James C Hone},
title = {Approaching the intrinsic photoluminescence linewidth in transition metal dichalcogenide monolayers},
journal = {2D Materials}
}

@article{Tuan2022,
  title = {Six-Body and Eight-Body Exciton States in Monolayer ${\mathrm{WSe}}_{2}$},
  author = {Van Tuan, Dinh and Shi, Su-Fei and Xu, Xiaodong and Crooker, Scott A. and Dery, Hanan},
  journal = {Phys. Rev. Lett.},
  volume = {129},
  issue = {7},
  pages = {076801},
  numpages = {7},
  year = {2022},
  month = {Aug},
  publisher = {American Physical Society},
  doi = {10.1103/PhysRevLett.129.076801},
  url = {https://link.aps.org/doi/10.1103/PhysRevLett.129.076801}
}

@article{Tuan2022theory,
  title = {Composite excitonic states in doped semiconductors},
  author = {Van Tuan, Dinh and Dery, Hanan},
  journal = {Phys. Rev. B},
  volume = {106},
  issue = {8},
  pages = {L081301},
  numpages = {6},
  year = {2022},
  month = {Aug},
  publisher = {American Physical Society},
  doi = {10.1103/PhysRevB.106.L081301},
  url = {https://link.aps.org/doi/10.1103/PhysRevB.106.L081301}
}

@article{Klein2021,
  title = {Controlling exciton many-body states by the electric-field effect in monolayer ${\mathrm{MoS}}_{2}$},
  author = {Klein, J. and H\"otger, A. and Florian, M. and Steinhoff, A. and Delhomme, A. and Taniguchi, T. and Watanabe, K. and Jahnke, F. and Holleitner, A. W. and Potemski, M. and Faugeras, C. and Finley, J. J. and Stier, A. V.},
  journal = {Phys. Rev. Res.},
  volume = {3},
  issue = {2},
  pages = {L022009},
  numpages = {7},
  year = {2021},
  month = {Apr},
  publisher = {American Physical Society},
  doi = {10.1103/PhysRevResearch.3.L022009},
  url = {https://link.aps.org/doi/10.1103/PhysRevResearch.3.L022009}
}

@article{Klein2022,
  title = {Trions in ${\mathrm{MoS}}_{2}$ are quantum superpositions of intra- and intervalley spin states},
  author = {Klein, J. and Florian, M. and H\"otger, A. and Steinhoff, A. and Delhomme, A. and Taniguchi, T. and Watanabe, K. and Jahnke, F. and Holleitner, A. W. and Potemski, M. and Faugeras, C. and Stier, A. V. and Finley, J. J.},
  journal = {Phys. Rev. B},
  volume = {105},
  issue = {4},
  pages = {L041302},
  numpages = {7},
  year = {2022},
  month = {Jan},
  publisher = {American Physical Society},
  doi = {10.1103/PhysRevB.105.L041302},
  url = {https://link.aps.org/doi/10.1103/PhysRevB.105.L041302}
}

@Article{Li2021,
  author  = {Li, Jing and Goryca, Mateusz and Choi, Junho and Xu, Xiaodong and Crooker, Scott A.},
  journal = {Nano Letters},
  title   = {Many-Body Exciton and Intervalley Correlations in Heavily Electron-Doped WSe2 Monolayers},
  year    = {2022},
  note    = {PMID: 34918936},
  number  = {1},
  pages   = {426-432},
  volume  = {22},
  doi     = {10.1021/acs.nanolett.1c04217},
  eprint  = {https://doi.org/10.1021/acs.nanolett.1c04217},
  url     = {https://doi.org/10.1021/acs.nanolett.1c04217},
}

@Article{Li2019replica,
author={Li, Zhipeng and Wang, Tianmeng and Jin, Chenhao and Lu, Zhengguang
and Lian, Zhen and Meng, Yuze and Blei, Mark and Gao, Shiyuan and Taniguchi, Takashi and Watanabe, Kenji and Ren, Tianhui and Tongay, Sefaattin and Yang, Li and Smirnov, Dmitry and Cao, Ting and Shi, Su-Fei},
title={Emerging photoluminescence from the dark-exciton phonon replica in monolayer WSe2},
journal={Nature Communications},
year={2019},
month={Jun},
day={06},
volume={10},
number={1},
pages={2469},
issn={2041-1723},
doi={10.1038/s41467-019-10477-6},
url={https://doi.org/10.1038/s41467-019-10477-6}
}

@article{Li2019momentum,
author = {Li, Zhipeng and Wang, Tianmeng and Jin, Chenhao and Lu, Zhengguang and Lian, Zhen and Meng, Yuze and Blei, Mark and Gao, Mengnan and Taniguchi, Takashi and Watanabe, Kenji and Ren, Tianhui and Cao, Ting and Tongay, Sefaattin and Smirnov, Dmitry and Zhang, Lifa and Shi, Su-Fei},
title = {Momentum-Dark Intervalley Exciton in Monolayer Tungsten Diselenide Brightened via Chiral Phonon},
journal = {ACS Nano},
volume = {13},
number = {12},
pages = {14107-14113},
year = {2019},
URL = {https://doi.org/10.1021/acsnano.9b06682}
}

@article{Robert2016,
  title = {Excitonic properties of semiconducting monolayer and bilayer $\mathrm{MoT}{\mathrm{e}}_{2}$},
  author = {Robert, C. and Picard, R. and Lagarde, D. and Wang, G. and Echeverry, J. P. and Cadiz, F. and Renucci, P. and H\"ogele, A. and Amand, T. and Marie, X. and Gerber, I. C. and Urbaszek, B.},
  journal = {Phys. Rev. B},
  volume = {94},
  issue = {15},
  pages = {155425},
  numpages = {8},
  year = {2016},
  month = {Oct},
  publisher = {American Physical Society},
  doi = {10.1103/PhysRevB.94.155425},
  url = {https://link.aps.org/doi/10.1103/PhysRevB.94.155425}
}

@Article{Lu2019,
author={Lu, Zhengguang and Rhodes, Daniel and Li, Zhipeng and Van Tuan, Dinh
and Jiang, Yuxuan and Ludwig, Jonathan and Jiang, Zhigang and Lian, Zhen
and Shi, Su-Fei and Hone, James and Dery, Hanan and Smirnov, Dmitry},
title={Magnetic field mixing and splitting of bright and dark excitons in monolayer MoSe2},
journal={2D Materials},
year={2019},
month={Nov},
day={21},
publisher={IOP Publishing},
volume={7},
number={1},
pages={015017},
issn={2053-1583},
doi={10.1088/2053-1583/ab5614},
url={https://dx.doi.org/10.1088/2053-1583/ab5614},
url={https://doi.org/10.1088/2053-1583/ab5614}
}

@article{Molas2019spec,
  title = {Energy Spectrum of Two-Dimensional Excitons in a Nonuniform Dielectric Medium},
  author = {Molas, M. R. and Slobodeniuk, A. O. and Nogajewski, K. and Bartos, M. and Bala, \L{}. and Babi\ifmmode \acute{n}\else \'{n}\fi{}ski, A. and Watanabe, K. and Taniguchi, T. and Faugeras, C. and Potemski, M.},
  journal = {Phys. Rev. Lett.},
  volume = {123},
  issue = {13},
  pages = {136801},
  numpages = {6},
  year = {2019},
  month = {Sep},
  publisher = {American Physical Society},
  doi = {10.1103/PhysRevLett.123.136801},
  url = {https://link.aps.org/doi/10.1103/PhysRevLett.123.136801}
}

@Article{Li2020phonon,
author={Li, Zhipeng and Wang, Tianmeng and Miao, Shengnan and Li, Yunmei and Lu, Zhenguang
and Jin, Chenhao and Lian, Zhen and Meng, Yuze and Blei, Mark and Taniguchi, Takashi
and Watanabe, Kenji and Tongay, Sefaattin and Yao, Wang and Smirnov, Dmitry
and Zhang, Chuanwei and Shi, Su-Fei},
title={Phonon-exciton Interactions in WSe2 under a quantizing magnetic field},
journal={Nature Communications},
year={2020},
month={Jun},
day={19},
volume={11},
number={1},
pages={3104},
issn={2041-1723},
doi={10.1038/s41467-020-16934-x},
url={https://doi.org/10.1038/s41467-020-16934-x}
}

@article{Kipczak2024,
url = {https://doi.org/10.1515/nanoph-2024-0385},
title = {Impact of temperature on the brightening of neutral and charged dark excitons in WSe2 monolayer},
author = {Łucja Kipczak and Natalia Zawadzka and Dipankar Jana and Igor Antoniazzi and Magdalena Grzeszczyk and Małgorzata Zinkiewicz and Kenji Watanabe and Takashi Taniguchi and Marek Potemski and Clément Faugeras and Adam Babiński and Maciej R. Molas},
journal = {Nanophotonics},
doi = {doi:10.1515/nanoph-2024-0385},
year = {2024},
lastchecked = {2024-11-24}
}

@Article{Kipczak2023,
author={Kipczak, {\L}ucja and Slobodeniuk, Artur O. and Wo{\'{z}}niak, Tomasz and Bhatnagar, Mukul and Zawadzka, Natalia and Olkowska-Pucko, Katarzyna and Grzeszczyk, Magdalena
and Watanabe, Kenji and Taniguchi, Takashi and Babi{\'{n}}ski, Adam and Molas, Maciej R.},
title={Analogy and dissimilarity of excitons in monolayer and bilayer of MoSe2},
journal={2D Materials},
year={2023},
month={Feb},
day={28},
publisher={IOP Publishing},
volume={10},
number={2},
pages={025014},
issn={2053-1583},
doi={10.1088/2053-1583/acbc8b},
url={https://dx.doi.org/10.1088/2053-1583/acbc8b}
}

@Article{Pucko2022,
author={Pucko, Katarzyna Olkowska and Blundo, Elena and Zawadzka, Natalia
and Cianci, Salvatore and Vaclavkova, Diana and Kapu{\'{s}}ci{\'{n}}ski, Piotr
and Jana, Dipankar and Pettinari, Giorgio and Felici, Marco and Nogajewski, Karol
and Barto{\v{s}}, Miroslav and Watanabe, Kenji and Taniguchi, Takashi
and Faugeras, Clement and Potemski, Marek and Babi{\'{n}}ski, Adam and Polimeni, Antonio
and Molas, Maciej R.},
title={Excitons and trions in WSSe monolayers},
journal={2D Materials},
year={2022},
month={Dec},
day={19},
publisher={IOP Publishing},
volume={10},
number={1},
pages={015018},
issn={2053-1583},
doi={10.1088/2053-1583/aca915},
url={https://dx.doi.org/10.1088/2053-1583/aca915}
}

@article{Wozniak2020,
  title = {Exciton $g$ factors of van der Waals heterostructures from first-principles calculations},
  author = {Wo\ifmmode \acute{z}\else \'{z}\fi{}niak, Tomasz and Faria Junior, Paulo E. and Seifert, Gotthard and Chaves, Andrey and Kunstmann, Jens},
  journal = {Phys. Rev. B},
  volume = {101},
  issue = {23},
  pages = {235408},
  numpages = {11},
  year = {2020},
  month = {Jun},
  publisher = {American Physical Society},
  doi = {10.1103/PhysRevB.101.235408},
  url = {https://link.aps.org/doi/10.1103/PhysRevB.101.235408}
}

@Article{Förste2020,
author={F{\"o}rste, Jonathan and Tepliakov, Nikita V. and Kruchinin, Stanislav Yu.
and Lindlau, Jessica and Funk, Victor and F{\"o}rg, Michael and Watanabe, Kenji
and Taniguchi, Takashi and Baimuratov, Anvar S. and H{\"o}gele, Alexander},
title={Exciton g-factors in monolayer and bilayer WSe2 from experiment and theory},
journal={Nature Communications},
year={2020},
month={Sep},
day={10},
volume={11},
number={1},
pages={4539},
issn={2041-1723},
doi={10.1038/s41467-020-18019-1},
url={https://doi.org/10.1038/s41467-020-18019-1}
}

@Article{Ye2018, 
author={Ye, Ziliang and Waldecker, Lutz and Ma, Eric Yue and Rhodes, Daniel and Antony, Abhinandan and Kim, Bumho and Zhang, Xiao-Xiao and Deng, Minda
and Jiang, Yuxuan and Lu, Zhengguang and Smirnov, Dmitry and Watanabe, Kenji and Taniguchi, Takashi and Hone, James and Heinz, Tony F.},
title={Efficient generation of neutral and charged biexcitons in encapsulated WSe2 monolayers},
journal={Nature Communications},
year={2018},
month={Sep},
day={13},
volume={9},
number={1},
pages={3718},
issn={2041-1723},
doi={10.1038/s41467-018-05917-8},
url={https://doi.org/10.1038/s41467-018-05917-8}
}

@article{OlkowskaPucko2026,
  author  = {Katarzyna Olkowska-Pucko and Tomasz Wo{\'z}niak and Elena Blundo and Natalia Zawadzka and {\L}ucja Kipczak and Paulo E. Faria Junior and Jan Szpakowski and Grzegorz Krasucki and Salvatore Cianci and Diana Vaclavkova and Dipankar Jana and Piotr Kapu{\'s}ci{\'n}ski and Amit Pawbake and Shalini Badola and Magdalena Grzeszczyk and Daniele Cecchetti and Giorgio Pettinari and Igor Antoniazzi and Zden{\v e}k Sofer and Iva Plutnarov{\'a} and Kenji Watanabe and Takashi Taniguchi and Cl{\'e}ment Faugeras and Marek Potemski and Adam Babi{\'n}ski and Antonio Polimeni and Maciej R. Molas},
  title   = {Extremely High Excitonic $g$ Factors in 2D Crystals by Alloy-Induced Admixing of Band States},
  journal = {Physical Review Letters},
  volume  = {136},
  pages   = {076901},
  year    = {2026},
  doi     = {10.1103/lx4n-7bb7}
}

@article{Chen2023Excitonic,
  title = {Excitonic Complexes in Two-Dimensional Transition Metal Dichalcogenides},
  author = {Chen, Xiaotong and Lian, Zhen and Meng, Yuze and Ma, Lei
            and Shi, Su-Fei},
  journal = {Nature Communications},
  volume = {14},
  pages = {8233},
  year = {2023},
  doi = {10.1038/s41467-023-44119-9}
}

@article{Xu2014,
  author  = {Xu, Xiaodong and Yao, Wang and Xiao, Di and Heinz, Tony F.},
  title   = {Spin and pseudospins in layered transition metal dichalcogenides},
  journal = {Nature Physics},
  volume  = {10},
  pages   = {343--350},
  year    = {2014},
  doi     = {10.1038/nphys2942}
}

@article{Shree2018,
  title = {Observation of exciton-phonon coupling in ${\mathrm{MoSe}}_{2}$ monolayers},
  author = {Shree, S. and Semina, M. and Robert, C. and Han, B. and Amand, T. 
            and Balocchi, A. and Manca, M. and Courtade, E. and Marie, X. 
            and Taniguchi, T. and Watanabe, K. and Glazov, M. M. 
            and Urbaszek, B.},
  journal = {Physical Review B},
  volume = {98},
  pages = {035302},
  year = {2018},
  doi = {10.1103/PhysRevB.98.035302}
}

@article{Martin2020,
  title = {Encapsulation Narrows and Preserves the Excitonic Homogeneous 
           Linewidth of Exfoliated Monolayer ${\mathrm{MoSe}}_{2}$},
  author = {Martin, Eric W. and Horng, Jason and Ruth, Hanna G. and Paik, Eunice
            and Wentzel, Michael-Henr and Deng, Hui and Cundiff, Steven T.},
  journal = {Physical Review Applied},
  volume = {14},
  pages = {021002},
  year = {2020},
  doi = {10.1103/PhysRevApplied.14.021002}
}

@article{Singh2020,
  author  = {Singh, Akshay and Lee, Hae Yeon and Grade{\v{c}}ak, Silvija},
  title   = {Direct optical-structure correlation in atomically thin 
             dichalcogenides and heterostructures},
  journal = {Nano Research},
  volume  = {13},
  pages   = {1363--1368},
  year    = {2020},
  doi     = {10.1007/s12274-019-2601-7}
}

@article{Wu2021,
  author  = {Wu, Lishu and Cong, Chunxiao and Yang, Weihuang and Chen, Yu 
             and Shao, Yan and Do, T. Thu Ha and Wen, Wen and Feng, Shun 
             and Zou, Chenji and Zhang, Hongbo and Du, Bowen and Cao, Bingchen 
             and Shang, Jingzhi and Xiong, Qihua and Loh, Kian Ping and Yu, Ting},
  title   = {Observation of Strong Valley Magnetic Response in Monolayer 
             Transition Metal Dichalcogenide Alloys of 
             Mo$_{0.5}$W$_{0.5}$Se$_2$ and 
             Mo$_{0.5}$W$_{0.5}$Se$_2$/WS$_2$ Heterostructures},
  journal = {ACS Nano},
  volume  = {15},
  number  = {5},
  pages   = {8397--8406},
  year    = {2021},
  doi     = {10.1021/acsnano.0c10478}
}

@article{Selig2020Sidebands,
  author  = {Roberto Rosati and Koloman Wagner and Samuel Brem and Ra{\"u}l Perea-Caus{\'i}n and Edith Wietek and Jonas Zipfel and Jonas D. Ziegler and Malte Selig and Takashi Taniguchi and Kenji Watanabe and Andreas Knorr and Alexey Chernikov and Ermin Malic},
  title   = {Temporal Evolution of Low-Temperature Phonon Sidebands in Transition Metal Dichalcogenides},
  journal = {ACS Photonics},
  volume  = {7},
  number  = {10},
  pages   = {2756--2764},
  year    = {2020},
  doi     = {10.1021/acsphotonics.0c00866}
}

@article{Echeverry2016,
  author  = {Echeverry, J. P. and Urbaszek, B. and Amand, T. and Marie, X. and Gerber, I. C.},
  title   = {Splitting between bright and dark excitons in transition metal dichalcogenide monolayers},
  journal = {Physical Review B},
  volume  = {93},
  pages   = {121107},
  year    = {2016},
  doi     = {10.1103/PhysRevB.93.121107}
}

@article{Berkelbach2018,
  author  = {Berkelbach, Timothy C. and Reichman, David R.},
  title   = {Optical and Excitonic Properties of Atomically Thin Transition-Metal Dichalcogenides},
  journal = {Annual Review of Condensed Matter Physics},
  volume  = {9},
  pages   = {379--396},
  year    = {2018},
  doi     = {10.1146/annurev-conmatphys-033117-054009}
}

@article{Mueller2018,
  author  = {Mueller, Thomas and Malic, Ermin},
  title   = {Exciton physics and device application of two-dimensional transition metal dichalcogenide semiconductors},
  journal = {npj 2D Materials and Applications},
  volume  = {2},
  pages   = {29},
  year    = {2018},
  doi     = {10.1038/s41699-018-0074-2}
}

@article{Arora2021,
  author  = {Arora, Ashish},
  title   = {Magneto-optics of layered two-dimensional semiconductors and heterostructures: Progress and prospects},
  journal = {Journal of Applied Physics},
  volume  = {129},
  number  = {12},
  pages   = {120902},
  year    = {2021},
  doi     = {10.1063/5.0042683}
}

@article{Han2018,
  author  = {Han, B. and Robert, C. and Courtade, E. and Manca, M. and Shree, S.
             and Amand, T. and Renucci, P. and Taniguchi, T. and Watanabe, K.
             and Marie, X. and Golub, L. E. and Glazov, M. M. and Urbaszek, B.},
  title   = {Exciton States in Monolayer MoSe$_2$ and MoTe$_2$ Probed by
             Upconversion Spectroscopy},
  journal = {Physical Review X},
  volume  = {8},
  pages   = {031073},
  year    = {2018},
  doi     = {10.1103/PhysRevX.8.031073}
}

@article{Courtade2018,
  author  = {Courtade, E. and Han, B. and Nakhaie, S. and Robert, C. and Marie, X.
             and Renucci, P. and Taniguchi, T. and Watanabe, K. and Geelhaar, L.
             and Lopes, J. M. J. and Urbaszek, B.},
  title   = {Spectrally narrow exciton luminescence from monolayer MoS$_2$
             and MoSe$_2$ exfoliated onto epitaxially grown hexagonal BN},
  journal = {Applied Physics Letters},
  volume  = {113},
  pages   = {032106},
  year    = {2018},
  doi     = {10.1063/1.5033554}
}

@article{Zuberek2022,
  author  = {{\.Z}uberek, E. and Majak, M. and Lubczy{\'n}ski, J. and Debus, J.
             and Watanabe, K. and Taniguchi, T. and Ho, C.-H. and Bryja, L.
             and Jadczak, J.},
  title   = {Upconversion photoluminescence excitation reveals exciton--trion
             and exciton--biexciton coupling in hBN/WS$_2$/hBN van der Waals
             heterostructures},
  journal = {Scientific Reports},
  volume  = {12},
  pages   = {13699},
  year    = {2022},
  doi     = {10.1038/s41598-022-18104-z}
}

@article{Vaquero2020,
  author  = {Vaquero, Daniel and Cleric{\`o}, Vito and Salvador-S{\'a}nchez, Juan
             and Mart{\'i}n-Ramos, Adri{\'a}n and D{\'i}az, Elena and
             Dom{\'i}nguez-Adame, Francisco and Meziani, Yahya M. and
             Diez, Enrique and Quereda, Jorge},
  title   = {Excitons, trions and Rydberg states in monolayer MoS$_2$
             revealed by low-temperature photocurrent spectroscopy},
  journal = {Communications Physics},
  volume  = {3},
  pages   = {194},
  year    = {2020},
  doi     = {10.1038/s42005-020-00460-9}
}

@article{Golyshkov2024,
  author  = {Golyshkov, G. M. and Brichkin, A. S. and Bisti, V. E. and
             Chernenko, A. V.},
  title   = {Excited States of Excitons in MoSe$_2$ and WSe$_2$ Monolayers},
  journal = {JETP Letters},
  volume  = {120},
  pages   = {270--276},
  year    = {2024},
  doi     = {10.1134/S0021364024602690}
}

@article{Druppel2017,
  author  = {Dr{\"u}ppel, Matthias and Deilmann, Thorsten and Kr{\"u}ger, Peter
             and Rohlfing, Michael},
  title   = {Diversity of trion states and substrate effects in the optical
             properties of an MoS$_2$ monolayer},
  journal = {Nature Communications},
  volume  = {8},
  pages   = {2117},
  year    = {2017},
  doi     = {10.1038/s41467-017-02286-6}
}

@article{Rytova1967,
  author  = {Rytova, N. S.},
  title   = {The Screened Potential of a Point Charge in a Thin Film},
  journal = {Moscow University Physics Bulletin},
  volume  = {22},
  number  = {3},
  pages   = {18},
  year    = {1967}
}

@article{Keldysh1979,
  author  = {Keldysh, L. V.},
  title   = {Coulomb Interaction in Thin Semiconductor and Semimetal Films},
  journal = {JETP Letters},
  volume  = {29},
  pages   = {658},
  year    = {1979}
}

@article{Cudazzo2011,
  author  = {Cudazzo, Pierluigi and Tokatly, Ilya V. and Rubio, Angel},
  title   = {Dielectric Screening in Two-Dimensional Insulators:
             Implications for Excitonic and Impurity States in Graphane},
  journal = {Physical Review B},
  volume  = {84},
  pages   = {085406},
  year    = {2011},
  doi     = {10.1103/PhysRevB.84.085406}
}

@article{Berkelbach2013,
  author  = {Berkelbach, Timothy C. and Hybertsen, Mark S. and
             Reichman, David R.},
  title   = {Theory of Neutral and Charged Excitons in Monolayer
             Transition Metal Dichalcogenides},
  journal = {Physical Review B},
  volume  = {88},
  pages   = {045318},
  year    = {2013},
  doi     = {10.1103/PhysRevB.88.045318}
}

@article{Borghardt2017,
  author  = {Borghardt, Sven and Tu, Jhih-Sian and Winkler, Florian and
             Schubert, J{\"u}rgen and Zander, Willi and Leosson, Kristjan and
             Kardyna{\l}, Beata E.},
  title   = {Engineering of optical and electronic band gaps in transition
             metal dichalcogenide monolayers through external dielectric screening},
  journal = {Physical Review Materials},
  volume  = {1},
  pages   = {054001},
  year    = {2017},
  doi     = {10.1103/PhysRevMaterials.1.054001}
}

@article{Aslan2018,
  author  = {Aslan, Ozgur Burak and Deng, Minda and Heinz, Tony F.},
  title   = {Strain tuning of excitons in monolayer WSe$_2$},
  journal = {Physical Review B},
  volume  = {98},
  pages   = {115308},
  year    = {2018},
  doi     = {10.1103/PhysRevB.98.115308}
}

@article{VanTuan2019,
  author  = {Van Tuan, Dinh and Scharf, Benedikt and Wang, Zefang and
             Shan, Jie and Mak, Kin Fai and {\v{Z}}uti{\'c}, Igor and
             Dery, Hanan},
  title   = {Probing many-body interactions in monolayer transition-metal
             dichalcogenides},
  journal = {Physical Review B},
  volume  = {99},
  pages   = {085301},
  year    = {2019},
  doi     = {10.1103/PhysRevB.99.085301}
}

@article{Slobodeniuk2023,
  author  = {Slobodeniuk, Artur O. and Molas, Maciej R.},
  title   = {Exciton spectrum in atomically thin monolayers: The role of hBN encapsulation},
  journal = {Physical Review B},
  volume  = {108},
  pages   = {035427},
  year    = {2023},
  doi     = {10.1103/PhysRevB.108.035427}
}

@article{Robert2018,
  author  = {Robert, C. and Semina, M. A. and Cadiz, F. and Manca, M. and
             Courtade, E. and Taniguchi, T. and Watanabe, K. and Cai, H. and
             Tongay, S. and Lassagne, B. and Renucci, P. and Amand, T. and
             Marie, X. and Glazov, M. M. and Urbaszek, B.},
  title   = {Optical spectroscopy of excited exciton states in MoS$_2$
             monolayers in van der Waals heterostructures},
  journal = {Physical Review Materials},
  volume  = {2},
  pages   = {011001},
  year    = {2018},
  doi     = {10.1103/PhysRevMaterials.2.011001}
}

@article{Takahashi2024,
  author  = {Takahashi, S. and Kusaba, S. and Watanabe, K. and
             Taniguchi, T. and Yanagi, K. and Tanaka, K.},
  title   = {3D hydrogen-like screening effect on excitons in
             hBN-encapsulated monolayer transition metal dichalcogenides},
  journal = {Scientific Reports},
  volume  = {14},
  pages   = {27286},
  year    = {2024},
  doi     = {10.1038/s41598-024-77625-x}
}

@article{Schneider2020,
  author  = {Schneider, Lorenz Maximilian and Esdaille, Shanece S. and Rhodes, Daniel A. and Barmak, Katayun and Hone, James C. and Rahimi-Iman, Arash},
  title   = {Direct Measurement of the Radiative Pattern of Bright and Dark Excitons and Exciton Complexes in Encapsulated Tungsten Diselenide},
  journal = {Scientific Reports},
  volume  = {10},
  pages   = {8091},
  year    = {2020},
  doi     = {10.1038/s41598-020-64838-z}
}

@article{Jadczak2021Upconversion,
  author  = {Jadczak, Joanna and Glazov, Mikhail and Kutrowska-Girzycka, Joanna and Schindler, Janina J. and Debus, Joerg and Ho, Ching-Hwa and Watanabe, Kenji and Taniguchi, Takashi and Bayer, Manfred and Bryja, Leszek},
  title   = {Upconversion of Light into Bright Intravalley Excitons via Dark Intervalley Excitons in hBN-Encapsulated {WSe$_2$} Monolayers},
  journal = {ACS Nano},
  volume  = {15},
  number  = {12},
  pages   = {19165--19174},
  year    = {2021},
  doi     = {10.1021/acsnano.1c08286}
}

@article{Zinkiewicz2022,
  author  = {Zinkiewicz, Ma{\l}gorzata and Grzeszczyk, Magdalena and Kipczak, {\L}ucja and Kazimierczuk, Tomasz and Watanabe, Kenji and Taniguchi, Takashi and Kossacki, Piotr and Babi{\'n}ski, Adam and Molas, Maciej R.},
  title   = {The effect of dielectric environment on the brightening of neutral and charged dark excitons in {WSe$_2$} monolayer},
  journal = {Applied Physics Letters},
  volume  = {120},
  number  = {16},
  pages   = {163101},
  year    = {2022},
  doi     = {10.1063/5.0085950}
}

@article{Li2023Screening,
  author  = {Li, Wei-Hua and Lin, Jhen-Dong and Lo, Ping-Yuan and Peng, Guan-Hao and Hei, Ching-Yu and Chen, Shao-Yu and Cheng, Shun-Jen},
  title   = {The Key Role of Non-Local Screening in the Environment-Insensitive Exciton Fine Structures of Transition-Metal Dichalcogenide Monolayers},
  journal = {Nanomaterials},
  volume  = {13},
  number  = {11},
  pages   = {1739},
  year    = {2023},
  doi     = {10.3390/nano13111739}
}

@article{Ren2023,
  author  = {Ren, L. and Robert, C. and Glazov, M. and Semina, M. and Amand, T. and Lombez, L. and Lagarde, D. and Taniguchi, T. and Watanabe, K. and Marie, X.},
  title   = {Control of the Bright-Dark Exciton Splitting Using the Lamb Shift in a Two-Dimensional Semiconductor},
  journal = {Physical Review Letters},
  volume  = {131},
  number  = {11},
  pages   = {116901},
  year    = {2023},
  doi     = {10.1103/PhysRevLett.131.116901}
}

@article{Selig2016,
  author  = {Selig, Malte and Bergh{\"a}user, Gunnar and Raja, Archana and Nagler, Philipp and Sch{\"u}ller, Christian and Heinz, Tony F. and Korn, Tobias and Chernikov, Alexey and Malic, Ermin and Knorr, Andreas},
  title   = {Excitonic linewidth and coherence lifetime in monolayer transition metal dichalcogenides},
  journal = {Nature Communications},
  volume  = {7},
  pages   = {13279},
  year    = {2016},
  doi     = {10.1038/ncomms13279}
}

@article{Selig2020,
  author  = {Selig, Malte and Katsch, Florian and Brem, Samuel and Mkrtchian, Garnik F. and Malic, Ermin and Knorr, Andreas},
  title   = {Suppression of intervalley exchange coupling in the presence of momentum-dark states in transition metal dichalcogenides},
  journal = {Physical Review Research},
  volume  = {2},
  pages   = {023322},
  year    = {2020},
  doi     = {10.1103/PhysRevResearch.2.023322}
}

@article{Li2025ExcitonDynamics,
  author  = {Li, Jingang and Yang, Rundi and Li, Runxuan and Grigoropoulos, Costas P.},
  title   = {Exciton Dynamics in 2D Transition Metal Dichalcogenides},
  journal = {Advanced Optical Materials},
  volume  = {13},
  number  = {11},
  pages   = {2403137},
  year    = {2025},
  doi     = {10.1002/adom.202403137}
}

@article{Roadmap2025,
  author  = {Garc{\'i}a de Abajo, F. Javier and Basov, D. N. and Koppens, Frank H. L. and others},
  title   = {Roadmap for Photonics with 2D Materials},
  journal = {ACS Photonics},
  volume  = {12},
  number  = {8},
  pages   = {3961--4095},
  year    = {2025},
  doi     = {10.1021/acsphotonics.5c00353}
}

@article{Meshulam2026,
  author  = {Meshulam, Matan and Kahlon, Anabel Atash and Gershuni, Yonatan and Poirier, Thomas and Edgar, James H. and Tongay, Sefaattin A. and Epstein, Itai},
  title   = {Temperature-Dependent Optical and Polaritonic Properties of Excitons in hBN-Encapsulated Monolayer TMDs},
  journal = {Advanced Optical Materials},
  volume  = {14},
  number  = {7},
  pages   = {e02535},
  year    = {2026},
  doi     = {10.1002/adom.202502535}
}

@article{Ryu2024,
  author  = {Ryu, Huije and Hong, Seong Chul and Kim, Kangwon and Jung, Yeonjoon and Lee, Yangjin and Lee, Kihyun and Kim, Youngbum and Kim, Hyunjun and Watanabe, Kenji and Taniguchi, Takashi and Kim, Jeongyong and Kim, Kwanpyo and Cheong, Hyeonsik and Lee, Gwan-Hyoung},
  title   = {Optical grade transformation of monolayer transition metal dichalcogenides via encapsulation annealing},
  journal = {Nanoscale},
  volume  = {16},
  pages   = {5836--5844},
  year    = {2024},
  doi     = {10.1039/D3NR06641J}
}

@article{Dijkstra2025,
  author  = {Dijkstra, Alain and Ben Mhenni, Amine and Van Tuan, Dinh and Çetiner, Elif and Schur-Wilkens, Muriel and Kim, Junghwan and Steiner, Laurin and Watanabe, Kenji and Taniguchi, Takashi and Barbone, Matteo and Wilson, Nathan P. and Dery, Hanan and Finley, Jonathan J.},
  title   = {Ten-valley excitonic complexes in charge-tunable monolayer WSe$_2$},
  journal = {Nature Communications},
  volume  = {16},
  pages   = {9743},
  year    = {2025},
  doi     = {10.1038/s41467-025-65731-x}
}

@article{Krasucki2026Doping,
  author        = {Krasucki, Grzegorz and Slobodeniuk, Artur O. and Walczyk, Kacper and Olkowska-Pucko, Katarzyna and Watanabe, Kenji and Taniguchi, Takashi and Babi{\'n}ski, Adam and Molas, Maciej R.},
  title         = {Doping-Induced Brightening of Dark Excitons and Trions in a WSe$_2$ Monolayer},
  journal       = {arXiv preprint},
  year          = {2026},
  eprint        = {2604.25553},
  archivePrefix = {arXiv},
  doi = {10.48550/arXiv.2604.25553},
  primaryClass  = {cond-mat.mes-hall}
}

@article{Sidler2017,
  author  = {Sidler, Meinrad and Back, Patrick and Cotlet, Ovidiu and Srivastava, Ajit and Fink, Thomas and Kroner, Martin and Demler, Eugene and Imamoglu, Atac},
  title   = {Fermi polaron-polaritons in charge-tunable atomically thin semiconductors},
  journal = {Nature Physics},
  volume  = {13},
  pages   = {255--261},
  year    = {2017},
  doi     = {10.1038/nphys3949}
}

@article{Efimkin2017,
  author  = {Efimkin, Dmitry K. and MacDonald, Allan H.},
  title   = {Many-body theory of trion absorption features in two-dimensional semiconductors},
  journal = {Physical Review B},
  volume  = {95},
  number  = {3},
  pages   = {035417},
  year    = {2017},
  doi     = {10.1103/PhysRevB.95.035417}
}

@article{Zhu2025,
  author  = {Zhu, Xing and Bacon, David R. and Pareek, Vivek and Mad{\'e}o, Julien and Taniguchi, Takashi and Watanabe, Kenji and Man, Michael K. L. and Dani, Keshav M.},
  title   = {A holistic view of the dynamics of long-lived valley polarized dark excitonic states in monolayer WS$_2$},
  journal = {Nature Communications},
  volume  = {16},
  pages   = {6385},
  year    = {2025},
  doi     = {10.1038/s41467-025-61677-2}
}

@article{Wragg2025,
  author  = {Wragg, Joseph and Bolzonello, Luca and Donati, Ludovica and Soundarapandian, Karuppasamy Pandian and Bertini, Riccardo and Tongay, Seth Ariel and Watanabe, Kenji and Taniguchi, Takashi and Koppens, Frank H. L. and van Hulst, Niek F.},
  title   = {Dual Action Spectroscopy Exposes the Bright and Dark Excitons of Room-Temperature WSe$_2$},
  journal = {Nano Letters},
  volume  = {25},
  number  = {19},
  pages   = {7658--7664},
  year    = {2025},
  doi     = {10.1021/acs.nanolett.4c06349}
}

@article{Glazov2020,
  author  = {Glazov, M. M.},
  title   = {Optical properties of charged excitons in two-dimensional semiconductors},
  journal = {The Journal of Chemical Physics},
  volume  = {153},
  pages   = {034703},
  year    = {2020},
  doi     = {10.1063/5.0012475}
}

@article{Chang2018,
  author  = {Chang, Yia-Chung and Shiau, Shiue-Yuan and Combescot, Monique},
  title   = {Crossover from trion-hole complex to exciton-polaron in $n$-doped two-dimensional semiconductor quantum wells},
  journal = {Physical Review B},
  volume  = {98},
  pages   = {235203},
  year    = {2018},
  doi     = {10.1103/PhysRevB.98.235203}
}

@article{Goldstein2020,
  author  = {Goldstein, Thomas and Wu, Yueh-Chun and Chen, Shao-Yu and Taniguchi, Takashi and Watanabe, Kenji and Varga, Kalman and Yan, Jun},
  title   = {Ground and excited state exciton polarons in monolayer MoSe$_2$},
  journal = {The Journal of Chemical Physics},
  volume  = {153},
  pages   = {071101},
  year    = {2020},
  doi     = {10.1063/5.0013092}
}

@article{Wagner2020,
  author  = {Wagner, Koloman and Wietek, Edith and Ziegler, Jonas D. and Semina, Marina A. and Taniguchi, Takashi and Watanabe, Kenji and Zipfel, Jonas and Glazov, Mikhail M. and Chernikov, Alexey},
  title   = {Autoionization and Dressing of Excited Excitons by Free Carriers in Monolayer WSe$_2$},
  journal = {Physical Review Letters},
  volume  = {125},
  pages   = {267401},
  year    = {2020},
  doi     = {10.1103/PhysRevLett.125.267401}
}

@article{Liu2021Polaron,
  author  = {Liu, Erfu and van Baren, Jeremiah and Lu, Zhengguang and Taniguchi, Takashi and Watanabe, Kenji and Smirnov, Dmitry and Chang, Yia-Chung and Lui, Chun Hung},
  title   = {Exciton-polaron Rydberg states in monolayer MoSe$_2$ and WSe$_2$},
  journal = {Nature Communications},
  volume  = {12},
  pages   = {6131},
  year    = {2021},
  doi     = {10.1038/s41467-021-26304-w}
}

@article{Genco2025,
  author  = {Genco, Armando and Trovatello, Chiara and Shahnazaryan, Vanik A. and Dogadov, Oleg and Cadore, Alisson R. and Rosa, Barbara L. T. and Kerfoot, James A. and Ahmed, Tanweer and Balci, Osman and Alexeev, Evgeny M. and Rostami, Habib and Watanabe, Kenji and Taniguchi, Takashi and Tongay, Seth Ariel and Ferrari, Andrea C. and Cerullo, Giulio and Dal Conte, Stefano},
  title   = {Ultrafast Dynamics of Rydberg Excitons and Their Optically Induced Charged Complexes in Encapsulated WSe$_2$ Monolayers},
  journal = {Nano Letters},
  volume  = {25},
  number  = {19},
  pages   = {7673--7681},
  year    = {2025},
  doi     = {10.1021/acs.nanolett.4c06428}
}

@article{Arora2019Trion,
  author  = {Arora, Ashish and Deilmann, Thorsten and Reichenauer, Till and Kern, Johannes and Michaelis de Vasconcellos, Steffen and Rohlfing, Michael and Bratschitsch, Rudolf},
  title   = {Excited-State Trions in Monolayer WS$_2$},
  journal = {Physical Review Letters},
  volume  = {123},
  pages   = {167401},
  year    = {2019},
  doi     = {10.1103/PhysRevLett.123.167401}
}

@article{Slobodeniuk2016,
  author  = {Slobodeniuk, A. O. and Basko, D. M.},
  title   = {Spin-flip processes and radiative decay of dark intravalley excitons in transition metal dichalcogenide monolayers},
  journal = {2D Materials},
  volume  = {3},
  number  = {3},
  pages   = {035009},
  year    = {2016},
  doi     = {10.1088/2053-1583/3/3/035009}
}

@article{Nagler2018,
  author  = {Nagler, Philipp and Ballottin, Mariana V. and Mitioglu, Anatolie A. and Durnev, Mikhail V. and Taniguchi, Takashi and Watanabe, Kenji and Chernikov, Alexey and Sch{\"u}ller, Christian and Glazov, Mikhail M. and Christianen, Peter C. M. and Korn, Tobias},
  title   = {Zeeman Splitting and Inverted Polarization of Biexciton Emission in Monolayer WS$_2$},
  journal = {Physical Review Letters},
  volume  = {121},
  number  = {5},
  pages   = {057402},
  year    = {2018},
  doi     = {10.1103/PhysRevLett.121.057402}
}

@article{Steinhoff2018,
  author  = {Steinhoff, Alexander and Florian, Matthias and Singh, Akshay and Tran, Kha and Kolarczik, Mirco and Helmrich, Sophia and Achtstein, Alexander W. and Woggon, Ulrike and Owschimikow, Nina and Jahnke, Frank and Li, Xiaoqin},
  title   = {Biexciton fine structure in monolayer transition metal dichalcogenides},
  journal = {Nature Physics},
  volume  = {14},
  pages   = {1199--1204},
  year    = {2018},
  doi     = {10.1038/s41567-018-0282-x}
}

@article{Zhang2026Biexciton,
  author  = {Zhang, Xiu and Huang, Haiyun and Qin, Tingxiao and Zhu, Junzhi and Zhang, Mengdi and Xiong, Qihua and Liu, Haiyun},
  title   = {Coherent Dynamics of Neutral and Charged Biexcitons in Monolayer WSe$_2$},
  journal = {Nano Letters},
  volume  = {26},
  number  = {37},
  pages   = {12290--12298},
  year    = {2026},
  doi     = {10.1021/acs.nanolett.6c02450}
}

@article{Bao2020,
  author  = {Bao, Di and Granados del {\'A}guila, Andr{\'e}s and Do, T. Thu Ha and Liu, Sheng and Pei, Jiajie and Xiong, Qihua},
  title   = {Probing momentum-indirect excitons by near-resonance photoluminescence excitation spectroscopy in WS$_2$ monolayer},
  journal = {2D Materials},
  volume  = {7},
  number  = {3},
  pages   = {031002},
  year    = {2020},
  doi     = {10.1088/2053-1583/ab817a}
}

@article{Adler2024,
  author  = {Adler, Eli R. and Le, Thy Doan Mai and Boulares, Ibrahim and Boyd, Robert and He, Yangchen and Rhodes, Daniel and Van Keuren, Edward and Barbara, Paola and Najmaei, Sina},
  title   = {Observation of Multi-Phonon Emission in Monolayer WS$_2$ on Various Substrates},
  journal = {Nanomaterials},
  volume  = {14},
  number  = {1},
  pages   = {37},
  year    = {2024},
  doi     = {10.3390/nano14010037}
}

@article{Nielsen2025Q,
  author  = {M{\o}rch Nielsen, Carl Emil and Fischer, Franz and Bester, Gabriel},
  title   = {Beyond the K-valley: exploring unique trion states in indirect band gap monolayer WSe$_2$},
  journal = {npj 2D Materials and Applications},
  volume  = {9},
  pages   = {11},
  year    = {2025},
  doi     = {10.1038/s41699-025-00532-w}
}

@article{Dogadov2026,
  author  = {Dogadov, Oleg and Mittenzwey, Henry and Bertolotti, Micol and Olsen, Nicholas and Deckert, Thomas and Trovatello, Chiara and Zhu, Xiaoyang and Brida, Daniele and Cerullo, Giulio and Knorr, Andreas and Dal Conte, Stefano},
  title   = {Dissecting intervalley coupling mechanisms in monolayer transition metal dichalcogenides},
  journal = {npj 2D Materials and Applications},
  volume  = {10},
  pages   = {21},
  year    = {2026},
  doi     = {10.1038/s41699-025-00653-2}
}

@article{Dery2025,
  title = {Energy Shifts and Broadening of Excitonic Resonances in Electrostatically Doped Semiconductors},
  author = {Dery, Hanan and Robert, Cedric and Crooker, Scott A. and Marie, Xavier and Van Tuan, Dinh},
  journal = {Phys. Rev. X},
  volume = {15},
  issue = {3},
  pages = {031049},
  year = {2025},
  publisher = {American Physical Society},
  doi = {10.1103/ddn8-d8bs},
  url = {https://doi.org/10.1103/ddn8-d8bs}
}

@article{Tongay2014,
  author  = {Tongay, Sefaattin and Narang, Deepa S. and Kang, Jun and Fan, Wen and Ko, Changhyun and Luce, Alexander V. and Wang, Kevin X. and Suh, Joonki and Patel, K. D. and Pathak, V. M. and Li, Jingbo and Wu, Junqiao},
  title   = {Two-Dimensional Semiconductor Alloys: Monolayer Mo$_{1-x}$W$_x$Se$_2$},
  journal = {Applied Physics Letters},
  volume  = {104},
  number  = {1},
  pages   = {012101},
  year    = {2014},
  doi     = {10.1063/1.4834358}
}

@article{Zhang2014,
  author  = {Zhang, Mei and Wu, Juanxia and Zhu, Yiming and Dumcenco, Dumitru O. and Hong, Jinhua and Mao, Nannan and Deng, Shibin and Chen, Yanfeng and Yang, Yanlian and Jin, Chuanhong and Chaki, Sunil H. and Huang, Ying-Sheng and Zhang, Jin and Xie, Liming},
  title   = {Two-Dimensional Molybdenum Tungsten Diselenide Alloys: Photoluminescence, Raman Scattering, and Electrical Transport},
  journal = {ACS Nano},
  volume  = {8},
  number  = {7},
  pages   = {7130--7137},
  year    = {2014},
  doi     = {10.1021/nn5020566}
}

@article{Mann2014,
  author  = {Mann, John and Ma, Quan and Odenthal, Patrick M. and Isarraraz, Miguel and Le, Duy and Preciado, Edwin and Barroso, David and Yamaguchi, Koichi and von Son Palacio, Gretel and Nguyen, Andrew and Tran, Tai and Wurch, Michelle and Nguyen, Ariana and Klee, Velveth and Bobek, Sarah and Sun, Dezheng and Heinz, Tony F. and Rahman, Talat S. and Kawakami, Roland and Bartels, Ludwig},
  title   = {2-Dimensional Transition Metal Dichalcogenides with Tunable Direct Band Gaps: MoS$_{2(1-x)}$Se$_{2x}$ Monolayers},
  journal = {Advanced Materials},
  volume  = {26},
  number  = {9},
  pages   = {1399--1404},
  year    = {2014},
  doi     = {10.1002/adma.201304389}
}

@article{Li2014,
  author  = {Li, Honglai and Duan, Xidong and Wu, Xueping and Zhuang, Xiujuan and Zhou, Hong and Zhang, Qinglin and Zhu, Xiaoli and Hu, Wei and Ren, Pinyun and Guo, Pengfei and Ma, Liang and Fan, Xiaopeng and Wang, Xiaoxia and Xu, Jinyou and Pan, Anlian and Duan, Xiangfeng},
  title   = {Growth of Alloy MoS$_{2x}$Se$_{2(1-x)}$ Nanosheets with Fully Tunable Chemical Compositions and Optical Properties},
  journal = {Journal of the American Chemical Society},
  volume  = {136},
  number  = {10},
  pages   = {3756--3759},
  year    = {2014},
  doi     = {10.1021/ja500069b}
}

@article{Feng2014,
  author  = {Feng, Qingliang and Zhu, Yiming and Hong, Jinhua and Zhang, Mei and Duan, Wenjie and Mao, Nannan and Wu, Juanxia and Xu, Hua and Dong, Fengliang and Lin, Fang and Jin, Chuanhong and Wang, Chunming and Zhang, Jin and Xie, Liming},
  title   = {Growth of Large-Area 2D MoS$_{2(1-x)}$Se$_{2x}$ Semiconductor Alloys},
  journal = {Advanced Materials},
  volume  = {26},
  number  = {17},
  pages   = {2648--2653},
  year    = {2014},
  doi     = {10.1002/adma.201306095}
}

@article{Wang2015Alloy,
  author  = {Wang, Gang and Robert, Cedric and Suslu, Aslihan and Chen, Bin and Yang, Sijie and Alamdari, Sarah and Gerber, Iann C. and Amand, Thierry and Marie, Xavier and Tongay, Sefaattin and Urbaszek, Bernhard},
  title   = {Spin-Orbit Engineering in Transition Metal Dichalcogenide Alloy Monolayers},
  journal = {Nature Communications},
  volume  = {6},
  pages   = {10110},
  year    = {2015},
  doi     = {10.1038/ncomms10110}
}

@article{Masenda2021,
  author  = {Masenda, Hilary and Schneider, Lucas M. and Adel Aly, M. and Machchhar, S. J. and Usman, A. and Meerholz, Klaus and Gebhard, Florian and Baranovskii, Sergei D. and Koch, Martin},
  title   = {Energy Scaling of Compositional Disorder in Ternary Transition-Metal Dichalcogenide Monolayers},
  journal = {Advanced Electronic Materials},
  volume  = {7},
  number  = {8},
  pages   = {2100196},
  year    = {2021},
  doi     = {10.1002/aelm.202100196}
}

@article{Braganca2019,
  author  = {Braganca, Helena and Riche, Flavio and Qu, Fanyao and Lopez-Richard, Victor and Marques, Gilmar Eugenio},
  title   = {Dark-Exciton Valley Dynamics in Transition Metal Dichalcogenide Alloy Monolayers},
  journal = {Scientific Reports},
  volume  = {9},
  pages   = {4575},
  year    = {2019},
  doi     = {10.1038/s41598-019-40932-9}
}

@article{Ciesiolkiewicz2026,
  author  = {Ciesio{\l}kiewicz-Klepek, Karolina and Kopaczek, Jan and Serafi{\'n}czuk, Jaros{\l}aw and Kudrawiec, Robert},
  title   = {Sensitivity of Excitonic Transitions to Temperature in Monolayers of TMD Alloys},
  journal = {The Journal of Physical Chemistry C},
  volume  = {130},
  number  = {2},
  pages   = {1014--1022},
  year    = {2026},
  doi     = {10.1021/acs.jpcc.5c06399}
}

@article{Jeong2026,
  author  = {Jeong, Won Uk and Ahn, Jung Bin and Jeong, Tae Jin and Ryu, Yeonhee and Kim, Sung},
  title   = {Composition-Tunable Valley Polarization in Monolayer Mo$_{1-x}$W$_x$S$_2$ Alloys},
  journal = {Current Applied Physics},
  volume  = {84},
  pages   = {12--18},
  year    = {2026},
  doi     = {10.1016/j.cap.2025.12.009}
}

@article{Chen2013,
  author  = {Chen, Yanfeng and Xi, Jinyang and Dumcenco, Dumitru O. and Liu, Zheng and Suenaga, Kazu and Wang, Dong and Shuai, Zhigang and Huang, Ying-Sheng and Xie, Liming},
  title   = {Tunable Band Gap Photoluminescence from Atomically Thin Transition-Metal Dichalcogenide Alloys},
  journal = {ACS Nano},
  volume  = {7},
  number  = {5},
  pages   = {4610--4616},
  year    = {2013},
  doi     = {10.1021/nn401420h}
}

@article{Chand2026,
  author  = {Saroj B. Chand and John M. Woods and Jiamin Quan and Enrique Mejia and Ashok Adhikari and Wei Wang and Takashi Taniguchi and Kenji Watanabe and Andrea Al{\`u} and Gabriele Grosso},
  title   = {Phonon-Dressed States of Dark Excitons in Transition Metal Dichalcogenides},
  journal = {Advanced Optical Materials},
  volume  = {14},
  number  = {13},
  pages   = {e03833},
  year    = {2026},
  doi     = {10.1002/adom.202503833}
}

@article{PauloPRB2026,
  title = {Generalized many-body exciton $g$ factors: Magnetic hybridization and nonmonotonic Rydberg series in monolayer ${\mathrm{WSe}}_{2}$},
  author = {Faria Junior, Paulo E. and Hernang{\'o}mez-P{\'e}rez, Daniel and Amit, Tomer and Fabian, Jaroslav and Refaely-Abramson, Sivan},
  journal = {Phys. Rev. B},
  volume = {112},
  issue = {24},
  pages = {L241404},
  numpages = {9},
  year = {2025},
  month = {Dec},
  publisher = {American Physical Society},
  doi = {10.1103/c5y7-w9tk},
  url = {https://link.aps.org/doi/10.1103/c5y7-w9tk}
}

@article{Danovich2017,
  author = {Danovich, M. and Z{\'o}lyomi, V. and Fal'ko, V. I.},
  title = {Dark trions and biexcitons in WS2 and WSe2 made bright by e-e scattering},
  journal = {Scientific Reports},
  volume = {7},
  pages = {45998},
  year = {2017},
  doi = {10.1038/srep45998}
}

@article{Jindal2025Dark,
  title = {Brightened emission of dark trions in transition metal dichalcogenide monolayers},
  author = {Jindal, V. and Mourzidis, K. and Balocchi, A. and Robert, C. and Li, P. and Van Tuan, D. and Lombez, L. and Lagarde, D. and Renucci, P. and Taniguchi, T. and Watanabe, K. and Dery, H. and Marie, X.},
  journal = {Phys. Rev. B},
  volume = {111},
  pages = {155409},
  year = {2025},
  doi = {10.1103/PhysRevB.111.155409}
}

\end{document}